\documentclass[twocolumn]{aastex63}

\usepackage{subfigure}
\usepackage{url}
\usepackage{hyperref}
\usepackage{epsfig}
\usepackage{graphicx}
\usepackage{sidecap}
\usepackage{hanging}
\usepackage{color}
\usepackage{multirow}
\usepackage{enumerate}
\usepackage{natbib}
\usepackage{amssymb}
\usepackage{amsmath}
\usepackage[bottom]{footmisc}
\usepackage{apjfonts}

\newcommand{\sci}{Science}
\newcommand{\jatis}{JATIS}

\newcommand{\kepler}{{\it Kepler}}
\newcommand{\tess}{{TESS}}

\newcommand{\jwst}{{\it JWST}}

\newcommand{\gaia}{{Gaia}}

\providecommand{\e}[1]{\ensuremath{\, \times \, 10^{#1}}}

\newcommand{\koi}{KOI-375.01}
\newcommand{\planet}{{Kepler-1704~b}}
\newcommand{\host}{{Kepler-1704}}

\providecommand{\bjdtdb}{\ensuremath{\rm {BJD_{TDB}}}}

\providecommand{\msun}{\ensuremath{\,M_\Sun}}
\providecommand{\rsun}{\ensuremath{\,R_\Sun}}
\providecommand{\lsun}{\ensuremath{\,L_\Sun}}
\providecommand{\mj}{\ensuremath{\,M_{\rm J}}}
\providecommand{\rj}{\ensuremath{\,R_{\rm J}}}

\providecommand{\fave}{\langle F \rangle}
\providecommand{\fluxcgs}{10$^9$ erg s$^{-1}$ cm$^{-2}$}

\shorttitle{GOT `EM II. A Highly Eccentric, Failed Hot Jupiter}
\shortauthors{Dalba et al.}

\begin{document}

\title{Giant Outer Transiting Exoplanet Mass (GOT `EM) Survey. II. \\ Discovery of a Failed Hot Jupiter on a 2.7 Year, Highly Eccentric Orbit\footnote{Some of the data presented herein were obtained at the W. M. Keck Observatory, which is operated as a scientific partnership among the California Institute of Technology, the University of California and the National Aeronautics and Space Administration. The Observatory was made possible by the generous financial support of the W. M. Keck Foundation.}}

\correspondingauthor{Paul A. Dalba}
\email{pdalba@ucr.edu}


\author[0000-0002-4297-5506]{Paul A.\ Dalba}
\altaffiliation{NSF Astronomy and Astrophysics Postdoctoral Fellow}
\affiliation{Department of Astronomy and Astrophysics, University of California Santa Cruz, 1156 High St., Santa Cruz, CA, USA}
\affiliation{Department of Earth and Planetary Sciences, University of California Riverside, 900 University Ave., Riverside, CA 92521, USA}

\author[0000-0002-7084-0529]{Stephen R.\ Kane}
\affiliation{Department of Earth and Planetary Sciences, University of California Riverside, 900 University Ave., Riverside, CA 92521, USA}

\author[0000-0002-4860-7667]{Zhexing Li}
\affiliation{Department of Earth and Planetary Sciences, University of California Riverside, 900 University Ave., Riverside, CA 92521, USA}

\author[0000-0003-2562-9043]{Mason G.\ MacDougall}
\affiliation{Department of Physics \& Astronomy, University of California Los Angeles, Los Angeles, CA 90095, USA}

\author[0000-0001-8391-5182]{Lee J.\ Rosenthal}
\affiliation{Department of Astronomy, California Institute of Technology, Pasadena, CA 91125, USA}

\author[0000-0002-8466-5469]{Collin Cherubim}
\affiliation{Department of Astronomy, California Institute of Technology, Pasadena, CA 91125, USA}

\author[0000-0002-0531-1073]{Howard Isaacson}
\affiliation{{Department of Astronomy,  University of California Berkeley, Berkeley CA 94720, USA}}
\affiliation{Centre for Astrophysics, University of Southern Queensland, Toowoomba, QLD, Australia}

\author[0000-0002-5113-8558]{Daniel P.\ Thorngren}
\affiliation{Institute for Research on Exoplanets (iREx), Universit\'e de Montr\'eal, Canada}

\author[0000-0003-3504-5316]{Benjamin Fulton}
\affiliation{NASA Exoplanet Science Institute/Caltech-IPAC, MC 314-6, 1200 E. California Blvd., Pasadena, CA 91125, USA}

\author[0000-0001-8638-0320]{Andrew W.\ Howard}
\affiliation{Department of Astronomy, California Institute of Technology, Pasadena, CA 91125, USA}

\author[0000-0003-0967-2893]{Erik A.\ Petigura}
\affiliation{Department of Physics \& Astronomy, University of California Los Angeles, Los Angeles, CA 90095, USA}

\author[0000-0002-2949-2163]{Edward W.\ Schwieterman}
\affiliation{Department of Earth and Planetary Sciences, University of California Riverside, 900 University Ave, Riverside, CA 92521, USA}
\affiliation{Blue Marble Space Institute of Science, Seattle, WA, 98115}

\author[0000-0002-9427-0014]{Dan O.\ Peluso}
\affiliation{Centre for Astrophysics, University of Southern Queensland, Toowoomba, QLD, Australia}
\affiliation{SETI Institute, Carl Sagan Center, 189 Bernardo Avenue, Mountain View, CA, USA}

\author[0000-0002-0792-3719]{Thomas M.\ Esposito}
\affiliation{SETI Institute, Carl Sagan Center, 189 Bernardo Avenue, Mountain View, CA, USA}
\affiliation{Astronomy Department, University of California, Berkeley, CA 94720, USA}

\author[0000-0001-7016-7277]{Franck Marchis}
\affiliation{SETI Institute, Carl Sagan Center, 189 Bernardo Avenue, Mountain View, CA, USA}
\affiliation{Unistellar, 198 Alabama Street, San Francisco CA 94110 USA}

\author[0000-0001-5133-6303]{Matthew J.\ Payne}
\affiliation{Harvard-Smithsonian Center for Astrophysics, 60 Garden St., MS 51, Cambridge, MA 02138, USA}


\begin{abstract}

Radial velocity (RV) surveys have discovered giant exoplanets on au-scale orbits with a broad distribution of eccentricities. Those with the most eccentric orbits are valuable laboratories for testing theories of high eccentricity migration. However, few such exoplanets transit their host stars thus removing the ability to apply constraints on formation from their bulk internal compositions. We report the discovery of Kepler-1704 b, a transiting 4.15 $M_{\rm J}$ giant planet on a 988.88 day orbit with the extreme eccentricity of $0.921^{+0.010}_{-0.015}$. Our decade-long RV baseline from the Keck I telescope allows us to measure the orbit and bulk heavy element composition of Kepler-1704 b and place limits on the existence of undiscovered companions. Kepler-1704 b is a failed hot Jupiter that was likely excited to high eccentricity by scattering events that possibly began during its gas accretion phase. Its final periastron distance was too large to allow for tidal circularization, so now it orbits it host from distances spanning 0.16--3.9 au. The maximum difference in planetary equilibrium temperature resulting from this elongated orbit is over 700~K. A simulation of the thermal phase curve of Kepler-1704 b during periastron passage demonstrates that it is a remarkable target for atmospheric characterization from the \textit{James Webb Space Telescope}, which could potentially also measure the planet's rotational period as the hot spot from periastron rotates in and out of view. Continued characterization of the Kepler-1704 system promises to refine theories explaining the formation of hot Jupiters and cool giant planets like those in the solar system. \\

\end{abstract}

 
\section{Introduction}\label{sec:intro}

Giant planet migration is typically invoked to explain the present day architecture of exoplanetary systems. Theories of planetary migration abound but can broadly be categorized as disk-driven migration, caused by torques from the protoplanetary disk \citep[e.g.,][]{Goldreich1980,Lin1986,Ward1997,Baruteau2014}, or high-eccentricity migration (HEM), whereby a giant planet exchanges orbital energy and angular momentum with one or more other objects in its system and subsequently experiences tidal circularization during close periastron passages \citep[e.g.,][]{Rasio1996,Wu2003,Nagasawa2008,Wu2011}. The characterization of giant planets and their orbits offers a window into which mechanisms might have been at play. 

There are multiple pathways for generating high eccentricities including Kozai-Lidov oscillations \citep{Kozai1962,Lidov1962} induced by a stellar companion \citep[e.g.,][]{Wu2003,Fabrycky2007,Naoz2012} or planetary companion \citep[e.g.,][]{Naoz2011,Lithwick2011b}, planet-planet scattering \citep[e.g.,][]{Rasio1996,Ford2006a,Chatterjee2008,Juric2008,Raymond2010,Nagasawa2011}, and secular chaos \citep[e.g.,][]{Wu2011,Hamers2017a}. Each mechanism can excite the eccentricity of a giant planet and, in doing so, imprints identifying (although not necessarily unique) clues in the present day system. Disentangling all the possible migration pathways for a single system or even determining the fraction of systems that migrated through various channels is challenging, though \citep[e.g.,][]{Fabrycky2009,Socrates2012,Dawson2015,Dawson2013}. 

HEM theories are readily tested in systems containing hot Jupiters, giant planets on orbits shorter than $\sim$10 days that are thought to have formed at greater distances from their host stars \citep[for a recent review, see][]{Dawson2018}. In addressing the mysteries of giant planet HEM, it is beneficial not only to investigate hot giant planets themselves but also \emph{proto-} and \emph{failed} hot Jupiters, objects in the process of becoming hot Jupiters or those that followed a similar evolutionary pathway but will not become hot Jupiters, respectively. HD~80606~b \citep[e.g.,][]{Naef2001,Moutou2009,Winn2009} is possibly a proto-hot Jupiter caught in the act of tidal circularization \citep[e.g.,][]{Wu2003,Fabrycky2007,Socrates2012}. Motivated by this planet, \citet{Socrates2012} theorized that if HEM is the preferred pathway of hot Jupiter migration, then the \kepler\ mission \citep{Borucki2010} should detect a population of highly eccentric ($e>0.9$) giant planets and their current orbital periods ($P$) should distinguish which are likely to be proto-hot Jupiters ($P \lesssim 2$~years) or failed hot Jupiters ($P \gtrsim 2$~years). This theory was supported by the detection of highly eccentric eclipsing binaries by \kepler\ \citep{Dong2013}. However, similar support was not offered by \kepler's planet discoveries. Based on analyses of the photoeccentric effect \citep{Ford2008a,Dawson2012a}, \citet{Dawson2015} reported a paucity of proto-hot Jupiters on highly eccentric orbits in the \kepler\ sample even after considering the limited sensitivity of transit surveys to planets with orbital distances of a few au. This work instead suggested that the dominant pathway of hot Jupiter formation is either disk migration or interaction with a planetary rather than stellar companion causing HEM to begin interior to 1~au. Only one proto-hot Jupiter candidate was identified, Kepler-419~b \citep{Dawson2012b}, which was later labeled as a failed hot Jupiter after subsequent (RV) observations \citep{Dawson2014}. 

Although RV surveys have detected a handful of failed hot Jupiter exoplanets, Kepler-419~b stands alone owing to its transiting geometry. By definition, a failed hot Jupiter must have a sufficiently wide orbit such that its periastron distance (despite its high eccentricity) is too large for tidal forces to efficiently circularize its orbit. By the observational biases of the transit method \citep[e.g.,][]{Beatty2008}, such long-period planets are unlikely to be observed in transit \citep[e.g.,][]{Dalba2019a}, although eccentricity can increase this probability \citep[e.g.,][]{Kane2007}. According to the NASA Exoplanet Archive\footnote{Accessed 2021 February 2 (\url{https://exoplanetarchive.ipac.caltech.edu/}).}, of the 16 non-controversial exoplanets with measured eccentricity above 0.8, only Kepler-419~b, HD~80606~b, and Kepler-1656~b \citep{Brady2018} have measured radii. Increasing this threshold to $e>0.9$ leaves only HD~80606~b. 

Those rare few eccentric, long-period giant exoplanets that do transit their hosts are exceptionally valuable because their radii and bulk compositions provide new windows into their formation and migration. Metal-rich stars preferentially host eccentric hot Jupiters \citep{Dawson2013,Buchhave2018}, lending credence to theories of planet-planet scattering since host star metallicity is known to correlate with giant planet occurrence \citep[e.g.,][]{Gonzalez1997,Santos2004,Fischer2005}. Furthermore, empirical trends in giant planet metal enrichment (relative to stellar) with planet mass hint at a fundamental and expected connection between the metal content of stars and their planets \citep{Miller2011,Thorngren2016,Teske2019}. With this in mind, giant planet bulk metallicity is likely a key piece of information for understanding migration history \citep[e.g.,][]{Alibert2005,Ginzburg2020,Shibata2020}. 

As the second discovery of the Giant Outer Transiting Exoplanet Mass (GOT `EM) survey \citep{Dalba2021a}, we present a new failed hot Jupiter from the \kepler\ sample: \koi\ (hereafter \planet\ as we will confirm its planetary nature). In Section~\ref{sec:obs}, we show the observations of this system including photometry from the \kepler\ spacecraft that detected two transits spaced by 989~days, follow-up adaptive optics (AO) imaging, and a follow-up Doppler spectroscopy spanning a decade. In Section~\ref{sec:model}, we combine these data sets through the comprehensive modeling of system parameters using \textsf{EXOFASTv2} \citep{Eastman2013,Eastman2019}. In Section~\ref{sec:results}, we conduct a thorough analysis to explore the plausibility of planetary or stellar companions across a wide swath of orbital separation space.  We also retrieve this planet's bulk metallicity and simulate its reflected light phase curve, the detection of which would be an unprecedented discovery that is within the anticipated capability of the \textit{James Webb Space Telescope} (\jwst). In Section~\ref{sec:disc}, we offer our interpretation of all of the analyses of the \host\ system in regards to the formation history of \planet\ and motivate a campaign to measure the stellar obliquity during a future transit. In Section~\ref{sec:concl}, we summarize our findings.


\section{Observations}\label{sec:obs}

We employ photometric, spectroscopic, and imaging observations in our analysis of the \host\ system. In the following sections, we describe how each of these data sets was collected and processed.


\subsection{Photometric Data from Kepler}\label{sec:kepler}

The \kepler\ spacecraft observed \host\ at 30-minute cadence in all 18 quarters of its primary mission. We accessed the Pre-search Data Conditioning Simple Aperture Photometry \citep[PDCSAP;][]{Jenkins2010a,Smith2012,Stumpe2012} through the Mikulski Archive for Space Telescopes (MAST), stitching together the light curves from individual quarters into one time series with a common baseline flux using \textsf{lightkurve} \citep{Lightkurve2018}. We further cleaned the photometry by removing all data points flagged for ``bad quality'' and dividing out the median background flux to produce a normalized light curve. We then measured a preliminary time of conjunction, duration, and period for the transiting planet using a box least squares transit search \citep[BLS;][]{Kovacs2002}, identifying only two transit events in Quarters 2 and 13. The time separating these two transits was 989~days, the suspected orbital period of \planet. However, the data gap between Quarters 7 and 8 occurred precisely in between these transits, introducing a $\sim$494-day orbital period alias.

We used the BLS results to mask out both transits and detrend any variability in the light curve without risk of obscuring the signal. Interpolating over the masked transit events, we fit a smoothed curve to systematics in the photometry using a Savitzky--Golay filter \citep{SciPy2020} and then subtracted out this additional structure to produce our final data product. Before unmasking the transit events, we clipped any remaining individual outliers with residuals to the smoothed fit that were greater than 3-$\sigma$ discrepant. 

\begin{figure}
  \begin{center}
  \includegraphics[width=\columnwidth]{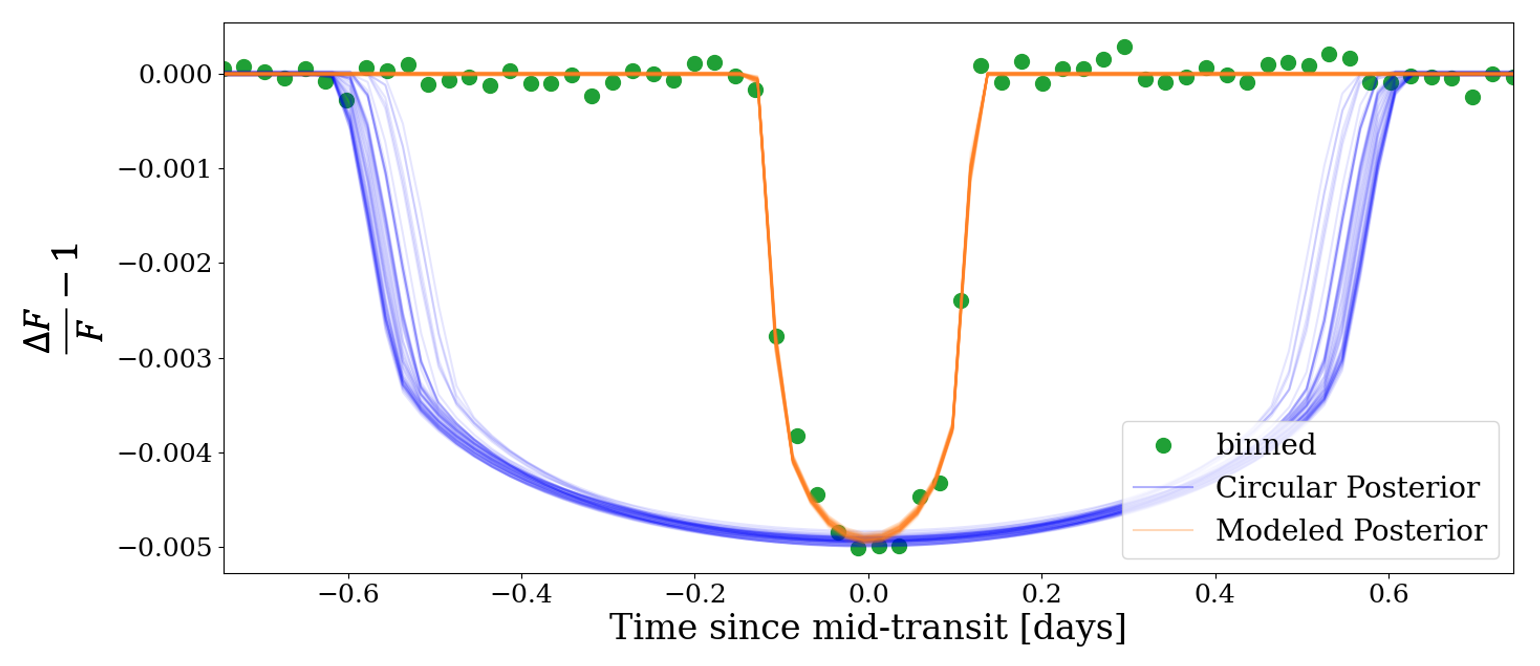}  
  \end{center}
  \caption{Phase-folded, binned \kepler\ data for \planet\ (green dots). The transit duration is substantially shorter than expected for a circular orbit assuming the stellar density listed in Table~\ref{tab:stellar} even when fit for impact parameter (blue models). The data are better reproduced by models with high eccentricity (orange lines).}
  \label{fig:photoecc_tran}
\end{figure}

We present the binned, detrended \kepler\ transits of \planet\ in Figure \ref{fig:photoecc_tran}. Under the assumption of a circular edge-on orbit, the mean transit duration of \planet\ and stellar properties reported by the NASA Exoplanet Archive suggest an orbital period of approximately 11~days. This scenario is thoroughly ruled out by the extensive \kepler\ data set. Instead, we explored the possibility that orbital eccentricity affected the duration of the transit.

\subsubsection{Photoeccentric Transit Modeling}\label{sec:photo}

The observed transits of \planet\ have a duration of $\sim$6 hours, which is nearly 5~times shorter than would be expected for a Jovian-size planet with a 989~day orbital period. The two plausible sources of this discrepancy are high impact parameter ($b$) or high eccentricity ($e$), but a preliminary transit fit reveals that high $b$ alone cannot account for the anomalously short transit duration. We instead developed a model to account for both of these properties through a photometric transit fit that takes into consideration the \emph{photoeccentric} framework of \citet{Dawson2012a}, as shown in Figure~\ref{fig:photoecc_tran}.

We modeled the standard transit parameters, including orbital period ($P$), time of conjunction ($T_C$), planet-star radius ratio ($R_p/R_{\star}$), and $b$, along with the expected stellar density assuming a circular orbit, $\rho_{\star,{\rm circ}}$, to obtain a model that encodes information about the true orbital eccentricity of the planet according to \citet{Kipping2012a}. We derived this dynamical information from our results by comparing our modeled $\rho_{\star,{\rm circ}}$ to the true stellar density, $\rho_{\star}$, represented by the median of our \textsf{EXOFASTv2} $\rho_{\star}$ posterior (Section~\ref{sec:model}). A value of $\rho_{\star,{\rm circ}}$ greater than $\rho_{\star}$ would imply that the planet transited faster than expected and vice versa, given an initial assumption of $e = 0$. Breaking from this assumption, however, we calculated which values of $e$ and the argument of periastron ($\omega$) were necessary to account for the unusually fast transit, subsequently bringing $\rho_{\star,{\rm circ}}$ into agreement with $\rho_{\star}$. For both parameters, we calculated posterior probability distributions using the log-likelihood function \citep{Dawson2012a}

\begin{figure}
    \centering
    \includegraphics[width=0.9\columnwidth]{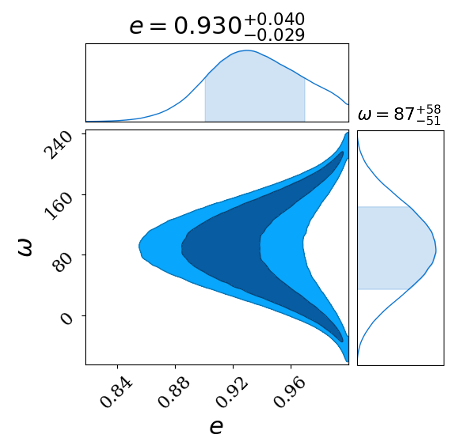}
    \caption{Posterior probabilities distributions for orbital eccentricity ($e$) and argument of periastron ($\omega$, in degrees) from the photoeccentric modeling. The shaded regions in the 1D histograms are 68\% credible intervals, and the shaded contours in the 2D histogram are the 68\% and 95\% credible regions. Values reported are the median and 68\% credible intervals.}
    \label{fig:photoecc_post}
\end{figure}

\begin{equation}\label{eq:photoecc}
\log P(e,\omega | \rho_{\star}, \rho_{\star,{\rm circ}}) = -\frac{1}{2}  \left(\frac{g(e,\omega)^3 \rho_{\star} - \rho_{\star,{\rm circ}}}{g(e,\omega)^3 \sigma_{\rho_\star}}\right)^2
\end{equation}
where
\begin{equation}
    g(e,\omega) = \frac{1 + e\sin{\omega}}{\sqrt{1 - e^2}} \;,
\end{equation}
following the notation of \citet{Kipping2010a} and \citet{Kipping2012a}.

Constraints on $\omega$ using this method tend to be broad, but they are sufficient to determine if a transit occurs closer to periastron (as is the case for \planet) or apastron. On the other hand, we were able to constrain the eccentricity of \planet\ here with high certainty. We found that the 68\% credible interval for eccentricity is 0.901--0.970 (Figure~\ref{fig:photoecc_post}).

In a previous analysis of the photoeccentric effect in \kepler\ transit data, \citet{Dawson2015} found $0^{+1}_{-0}$ giant planets on highly eccentric orbits that are likely undergoing tidal circularization. This non-detection refuted the hypothesis of \citet{Socrates2012} that approximately four such planets should be detected if HEM is the dominant hot Jupiter migration mechanism. \citet{Dawson2015} only considered planet candidates with three or more transits, to more accurately account for the completeness of \kepler\ pipeline detections \citep[e.g.,][]{Christiansen2020}, so \host\ was not included in their analysis.

Assuming tidal decay at constant angular momentum, the highest allowed values of eccentricity from our photoeccentric modeling would produce a final orbital period below 10~days, the canonical threshold for hot Jupiters. Therefore, based on just this photoeccentric effect analysis, \planet\ is a candidate proto-hot Jupiter. However, additional orbital characterization via RV monitoring of the host is needed to refine the eccentricity and the nature of \planet.


\subsection{Spectroscopic Data from HIRES}\label{sec:hires}

We acquired 15 high resolution spectra of \host\ with the High Resolution Echelle Spectrometer \citep[HIRES;][]{Vogt1994} on the Keck~I telescope in support of our Doppler monitoring of the \host\ system. The baseline of these observations spans nearly a decade. For each observation, the starlight passed through a heated iodine cell before reaching the slit to enable the precise wavelength calibration of each RV measurement. 

We did not acquire a high signal-to-noise (S/N) template spectrum as is typical for HIRES RV observations \citep[e.g.,][]{Howard2010}. Instead, we identified a pre-existing, ``best match'' template spectrum in the HIRES spectral library following \citet{Dalba2020b}. The best match star was HD~203473, a brighter G6V star with similar spectroscopic properties to \host\ according to a \textsf{SpecMatch--Emp}\footnote{\url{https://github.com/samuelyeewl/specmatch-emp}} analysis \citep{Yee2017}. The use of a best match template incurs extra uncertainty in addition to internal RV errors. Following conservative estimations by \citet[][their Table~2]{Dalba2020b}, we added 6.2~m~s$^{-1}$ to our internal RV errors in quadrature to account for this method. After swapping in the template of HD~203473, the RV extraction proceeded following the standard forwarding techniques employed the by California Planet Search \citep[e.g.,][]{Howard2010,Howard2016}. 

We provide the full RV data set for \host\ in Table \ref{tab:rvs}. The uncertainties listed include the additional uncertainty incurred by the matched-template method of RV extraction \citep{Dalba2020b}. We also include corresponding $S_{\rm HK}$ activity indicators derived from the Ca II H and K spectral lines \citep{Wright2004,Isaacson2010}. 

We note that the first RV measurement (from ${\rm BJD}=2455669$) is the least precise observation in the series. Its uncertainty is three standard deviations above the mean. This larger error is not surprising as the exposure time for the spectrum used to measure that RV was substantially shorter than the others. The resulting best fit velocity in each two-angstrom chunk of spectrum, which typically only contain one stellar and one iodine line, was less precise, leading to the larger error in RV. When folded on the ephemeris of \planet, this data point occupies a non-critical phase in the orbit. However, this data point extends the baseline of RVs observations by 826 days and is critical to our consideration of acceleration in the \host\ system (Section~\ref{sec:companions}). Although there is no obvious reason to exclude this data point from our analysis besides its larger uncertainty, we will treat this data point with skepticism moving forward.

\begin{deluxetable}{ccc}
\tablecaption{RV Measurements of \host \label{tab:rvs}}
\tablehead{
  \colhead{BJD$_{\rm TDB}$} & 
  \colhead{RV (m s$^{-1}$)} &
  \colhead{$S_{\rm HK}$}}
  \startdata
    2455669.111196 & 25.3$\pm$8.5 & 0.0966$\pm$0.0010\\
    2456495.013178 & 28.9$\pm$6.8 & 0.1220$\pm$0.0010\\
    2456532.811313 & 31.3$\pm$6.8 & 0.1330$\pm$0.0010\\
    2458383.894210 & 16.2$\pm$7.5 & 0.1609$\pm$0.0010\\
    2458593.029972 & 38.6$\pm$6.8 & 0.1172$\pm$0.0010\\
    2458679.811045 & 63.2$\pm$6.8 & 0.1260$\pm$0.0010\\
    2458765.877254 & 68.1$\pm$6.8 & 0.1311$\pm$0.0010\\
    2458815.758493 & 90.0$\pm$7.2 & 0.1267$\pm$0.0010\\
    2459006.997818 & 195.5$\pm$6.8 & 0.1222$\pm$0.0010\\
    2459038.992753 & $-118.9\pm$6.9 & 0.1222$\pm$0.0010\\
    2459041.035816 & $-119.9\pm$7.1 & 0.1205$\pm$0.0010\\
    2459051.874260 & $-93.1\pm$6.7 & 0.1265$\pm$0.0010\\
    2459070.992339 & $-72.1\pm$7.2 & 0.0964$\pm$0.0010\\
    2459189.758826 & $-31.5\pm$7.6 & 0.1183$\pm$0.0010\\
  \enddata
\end{deluxetable}

In Section~\ref{sec:model}, we model the RVs and transits simultaneously, confirming that the orbital period of \planet\ is accurately represented by the time elapsed between the two \kepler\ transits ({988.88~days}) and not half of that value. Visual inspection of the RV data folded on an orbital period of 494.44~days suggests no Keplerian signal at this periodicity. Therefore, we hereafter do not consider the possibility that another transit occurred during the gap in observations between \kepler\ quarters 7 and 8.


\subsection{Archival AO Imaging}\label{sec:imaging}

The \host\ system has been observed in several imaging campaigns previously \citep[for a summary, see][]{Furlan2017a}. To explore the existence of bound or background stellar neighbors, we present three data sets acquired from the Exoplanet Follow-up Observing Program\footnote{ExoFOP, accessed 2021 February 5 (\url{https://exofop.ipac.caltech.edu/}).}. 

\begin{figure*}
  \centering
    \begin{tabular}{cc}
      \includegraphics[width=0.5\textwidth]{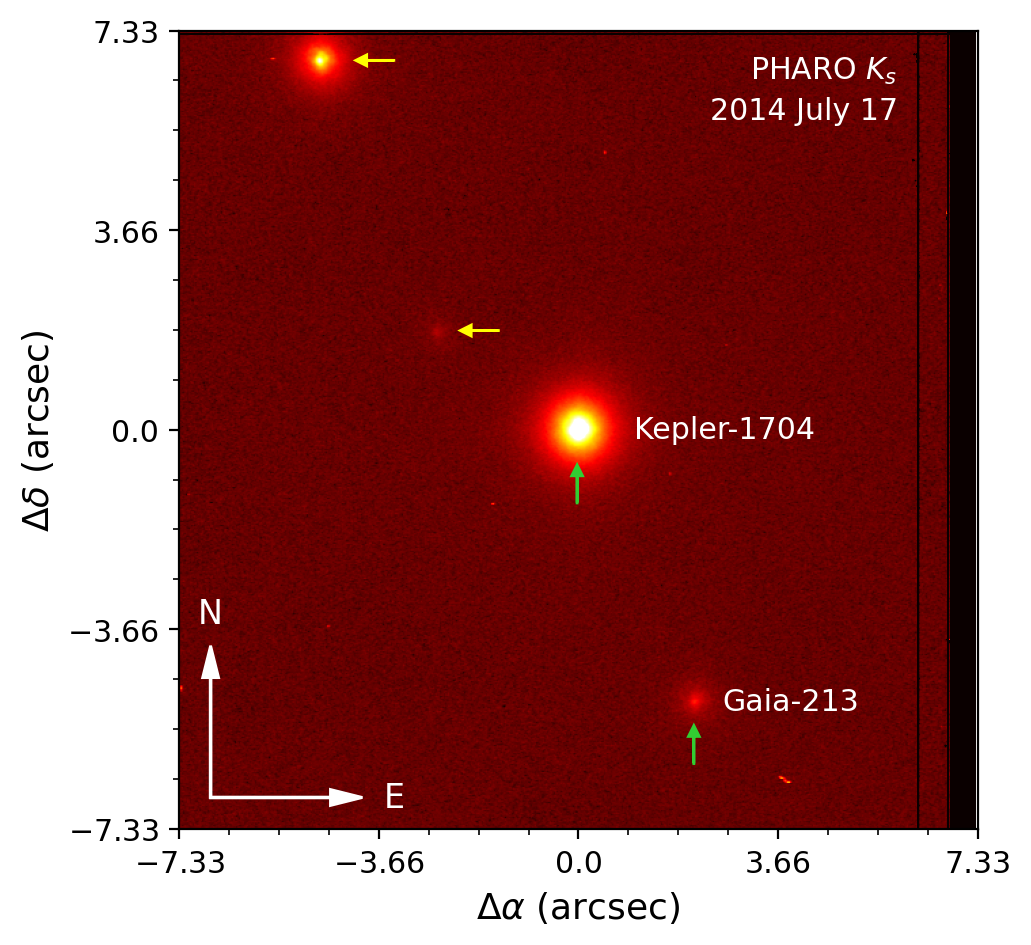} &
      \includegraphics[width=0.5\textwidth]{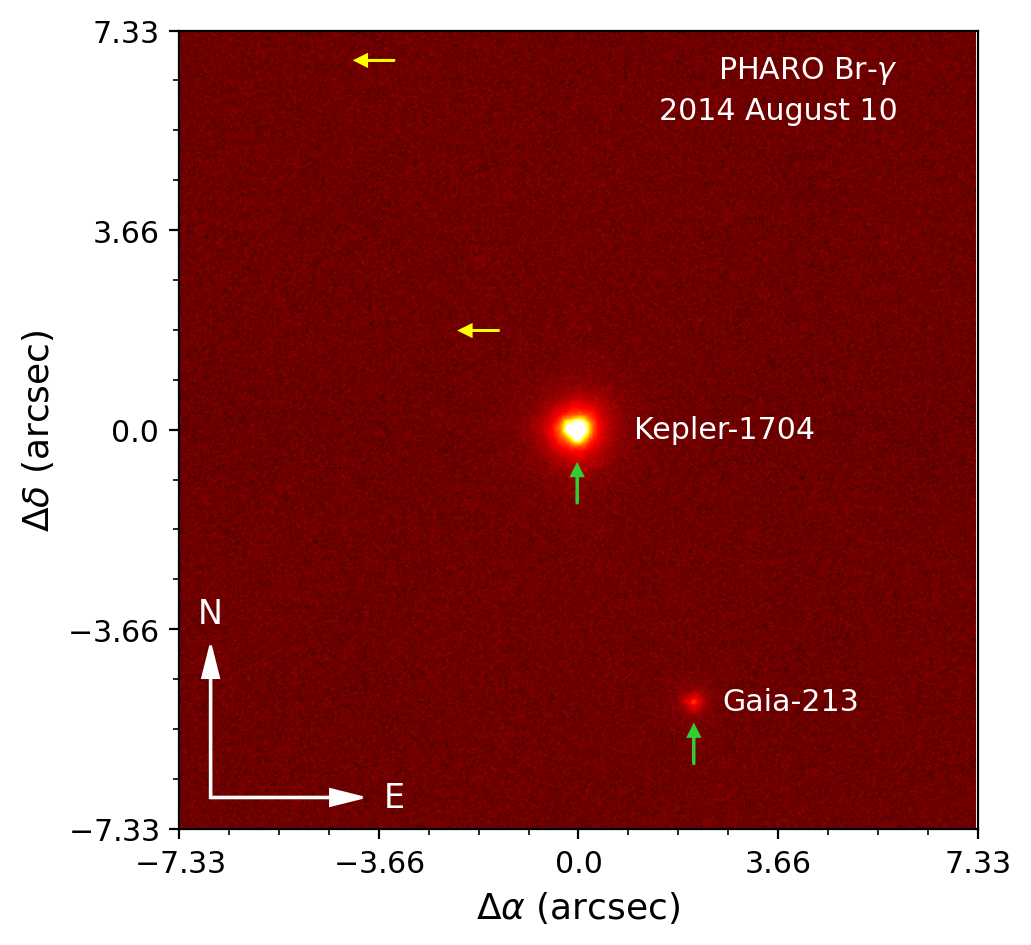} 
    \end{tabular}
  \caption{AO images of \host\ taken with the PHARO instrument on the 200~inch telescope at Palomar Observatory and acquired from ExoFOP. \textit{Left:} Observation from \citet{Wang2015a} showing \host\ and three other sources. Green, vertical arrows identify \host\ (at center) and \emph{Gaia-213} (see text), as resolved by \gaia. Yellow, horizontal arrows identify two additional sources not resolved by \gaia, the fainter of which was claimed as a detection by \citet{Wang2015a}. The white stripes on the eastern edge of the image are mosaicking artifacts. \textit{Right:} PHARO observation from \citet{Furlan2017a} showing \host\ at center and \emph{Gaia-213}. In both images, the scales and locations of the arrows are identical. The two sources present in the left panel that are absent in the right panel are spurious duplications of \host\ and \emph{Gaia-213} caused by an alignment error. According to \gaia\ astrometry, \emph{Gaia-213} is not gravitationally bound to \host.}
  \label{fig:pharo}
\end{figure*}

The first imaging data set comprises AO images from the PHARO instrument \citep{Hayward2001} at the 200~inch telescope at Palomar Observatory as published by \citet{Wang2015a}. This work used a 3-point dither pattern to obtain a set of images in the $K_s$ band that were combined and searched for stellar companions (Figure~\ref{fig:pharo}, left panel). \citet{Wang2015a} claimed two detections: one source with $\Delta K_s = 3.3$ with a separation and position angle (PA) of $5\farcs47$ and 157.0$^{\circ}$, respectively, and another source with $\Delta K_s = 4.6$ with a separation and PA of $3\farcs19$ and $305.5^{\circ}$, respectively. Both detections are visible in the left panel of Figure~\ref{fig:pharo}. The source with PA=$157.0^{\circ}$ (indicated by a green, vertical arrow) is resolved by \gaia\ \citep{Gaia2016,Gaia2020} and has the EDR3 source ID of 2136191732305041920 (hereafter \emph{Gaia-213} for simplicity). The parallax and proper motion of \host\ and \emph{Gaia-213} as measured by \gaia\ definitively show that these two stars are not gravitationally associated. The other source identified by \citet{Wang2015a}, as well as a brighter source near the upper edge of the image that was not identified by \citet{Wang2015a}, are not present in the \gaia\ EDR3 catalog. In the left panel of Figure~\ref{fig:pharo}, these unidentified sources are shown with horizontal yellow arrows.

The second imaging data set also comprises AO images from the PHARO instrument but in the Br-$\gamma$ filter as published by \citet{Furlan2017a}. Surprisingly, only \host\ and \emph{Gaia-213} (at PA=$157.0^{\circ}$) are visible despite deeper magnitude limits near $3\farcs19$: $\Delta K_s = 4.9$ versus $\Delta$Br-$\gamma = 7.0$ \citep{Wang2015a,Furlan2017a}. The time elapsed between the epochs of imaging, roughly one month, is too short to explain the discrepancy. 

The solution to this conundrum lies in the relative positioning of the two sources in question with respect to the positioning of \host\ and \emph{Gaia-213}. The separation and PA between these two pairs are identical. Visual inspection suggests that the contrast between the stars in each pair is also similar. Thus, our conclusion is that the two sources identified by yellow, horizontal arrows in the left panel Figure~\ref{fig:pharo} are spurious duplications of \host\ and \emph{Gaia-213} caused by an accidental image alignment error. 

\begin{figure}
    \centering
    \includegraphics[width=\columnwidth]{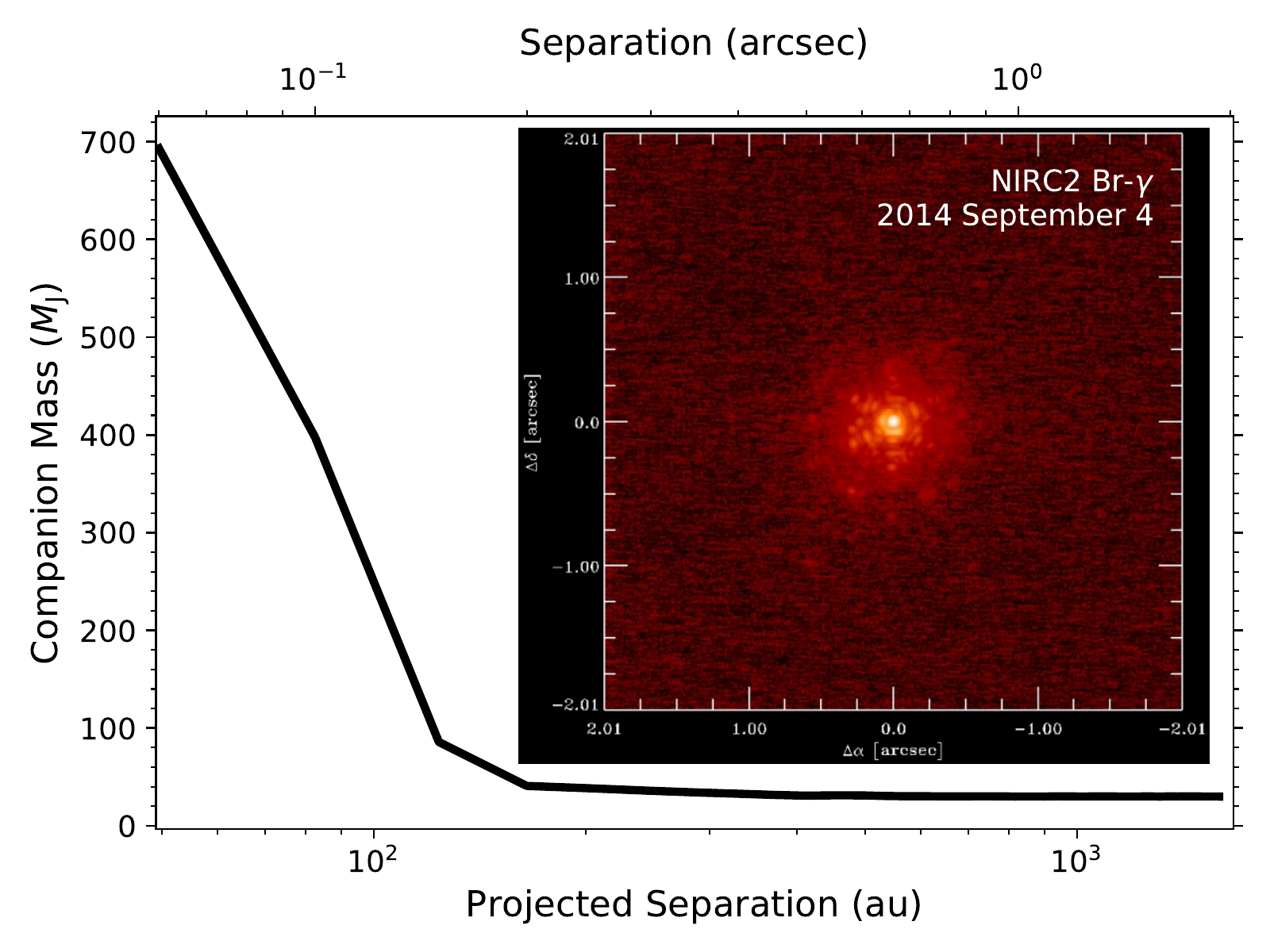}
    \caption{Upper limit on companion mass in the \host\ system based on the contrast curve measured from NIRC2 AO images. The masses were estimated by interpolating a MIST isochrone (in the stellar regime) and a brown dwarf isochrone (in the substellar regime). The inset is the NIRC2 image of \host\ published by \citet{Furlan2017a}.}
    \label{fig:mass_limit}
\end{figure}

The third imaging data set comprises AO images from the NIRC2 instrument \citep{Wizinowich2000} at the Keck~II telescope as published by \citet{Furlan2017a}. Observations were taken in the Br-$\gamma$ filter and the field of view was too small to include any of the other sources (astrophysical or spurious) mentioned previously (Figure~\ref{fig:mass_limit}). The NIRC2 data yield a nondetection of a stellar neighbor within 2$\arcsec$ with delta-magnitude limits of 8.4 and 8.7 at separations of $0\farcs$5 and $1\farcs$0, respectively \citep{Furlan2017a}. Since the NIRC2 observations of \host\ provide the strongest constraints on neighboring stars, we continue our analysis using only these data.

We used the NIRC2 contrast curve (i.e., 5$\sigma$ limiting delta-magnitude as a function of separation) to derive the corresponding limiting mass for a bound companion. First, we downloaded a MESA Isochrones and Stellar Tracks (MIST) isochrone \citep{Paxton2011,Paxton2013,Paxton2015,Dotter2016,Choi2016} from the MIST web interpolator\footnote{Accessed 2020 December 17 (\url{http://waps.cfa.harvard.edu/MIST/}).}. We provided values of initial stellar metallicity, extinction, and age based on the system modeling described in Section~\ref{sec:model}. This isochrone provide a numerical relationship between stellar mass and absolute $K_s$ magnitude, which we treated interchangeably with Br-$\gamma$. After converting absolute magnitude to apparent magnitude (using the distance from Section~\ref{sec:model}), we interpolated the $\Delta K_s$ values with those measured by NIRC2 to calculate an upper limit of companion mass as a function of projected separation (Figure~\ref{fig:mass_limit}). At wider separations, the delta-magnitude values exceeded those in the MIST isochrone. For those separations we instead interpolated a 5~Gyr brown dwarf isochrone from \citet{Baraffe2003}. Beyond a projected separation of $\sim$200~au, we find that any companion in the \host\ system must have a mass below $\sim$32~$M_{\rm J}$.


\section{Modeling the Stellar and Planetary Parameters}\label{sec:model}

We simultaneously fit models to the transit and RV data for \host\ while also modeling the stellar spectral energy distribution (SED) using the \textsf{EXOFASTv2} suite. The result was a set of precise, consistent stellar (Table~\ref{tab:stellar}) and planetary (Table~\ref{tab:planet}) parameters. 

\begin{deluxetable}{lcc}
\tabletypesize{\scriptsize}
\tablecaption{Median values and 68\% confidence intervals for the stellar parameters for \host. \label{tab:stellar}}
\tablehead{\colhead{~~~Parameter} & \colhead{Units} & \colhead{Values}}
\startdata
\multicolumn{2}{l}{Informative Priors:}& \smallskip\\
~~~~$T_{\rm eff}$\dotfill &Effective Temperature (K)\dotfill & $\mathcal{N}(5772,115)$\\
~~~~$[{\rm Fe/H}]$\dotfill &Metallicity (dex)\dotfill & $\mathcal{N}(0.2,0.06)$\\
~~~~$\varpi$\dotfill &Parallax (mas)\dotfill & $\mathcal{N}(1.213,0.016)$\\
~~~~$A_V$\dotfill &V-band extinction (mag)\dotfill & $\mathcal{U}(0,0.2902)$\\
\smallskip\\\multicolumn{2}{l}{Stellar Parameters:}&\smallskip\\
~~~~$M_*$ & Mass (\msun) & $1.132^{+0.040}_{-0.050}$\\
~~~~$R_*$ & Radius (\rsun) & $1.697^{+0.059}_{-0.058}$\\
~~~~$L_*$ & Luminosity (\lsun) & $2.83^{+0.17}_{-0.19}$\\
~~~~$F_{Bol}$ & Bolometric Flux (cgs) & $1.336\times10^{-10}$$^{+7.1\times10^{-12}}_{-8.6\times10^{-12}}$\\
~~~~$\rho_*$ & Density (g cm$^{-3}$) & $0.325^{+0.036}_{-0.032}$\\
~~~~$\log{g}$ & Surface gravity (cgs) & $4.031^{+0.031}_{-0.032}$\\
~~~~$T_{\rm eff}$ & Effective Temperature (K) & $5746^{+87}_{-88}$\\
~~~~$[{\rm Fe/H}]$ & Metallicity (dex) & $0.196\pm0.058$\\
~~~~$[{\rm Fe/H}]_{0}$ & Initial Metallicity$^{a}$  & $0.219^{+0.053}_{-0.056}$\\
~~~~Age & Age (Gyr) & $7.4^{+1.5}_{-1.0}$\\
~~~~EEP & Equal Evolutionary Phase$^{b}$  & $452.9^{+4.5}_{-5.7}$\\
~~~~$A_V$ & $V$-band extinction (mag) & $0.190^{+0.067}_{-0.091}$\\
~~~~$\sigma_{SED}$ & SED photometry error scaling  & $1.05^{+0.43}_{-0.26}$\\
~~~~$\varpi$ & Parallax (mas) & $1.213\pm0.016$\\
~~~~$d$ & Distance (pc) & $824\pm11$\\
\enddata
\tablenotetext{}{See Table~3 in \citet{Eastman2019} for a detailed description of all parameters and all default (non-informative) priors beyond those specified here. $\mathcal{N}(a,b)$ denotes a normal distribution with mean $a$ and variance $b^2$. $\mathcal{U}(a,b)$ denotes a uniform distribution over the interval [$a$,$b$].}
\tablenotetext{a}{Initial metallicity is that of the star when it formed.}
\tablenotetext{b}{Corresponds to static points in a star's evolutionary history. See Section~2 of \citet{Dotter2016}.}
\end{deluxetable}

\begin{deluxetable}{lcc}
\tabletypesize{\scriptsize}
\tablecaption{Median values and 68\% confidence interval of the planet parameters for \planet. \label{tab:planet}}
\tablehead{\colhead{~~~Parameter} & \colhead{Units} & \colhead{Values}}
\startdata
~~~~$P$ & Period (days) & $988.88112\pm0.00091$\\
~~~~$R_p$ & Radius (\rj) & $1.066^{+0.044}_{-0.042}$\\
~~~~$M_p$ & Mass$^{a}$ (\mj) & $4.16^{+0.29}_{-0.28}$\\
~~~~$T_C$ & Time of conjunction (\bjdtdb) & $2, 455, 072.22337^{+0.00063}_{-0.00064}$\\
~~~~$a$ & Semimajor axis (au) & $2.027^{+0.024}_{-0.030}$\\
~~~~$i$ & Inclination (deg) & $89.00^{+0.56}_{-0.26}$\\
~~~~$e$ & Eccentricity  & $0.920^{+0.010}_{-0.016}$\\
~~~~$\omega_*$ & Argument of periastron$^{b}$ (deg) & $82.4^{+4.5}_{-5.1}$\\
~~~~$T_{eq}$ & Equilibrium temperature$^{c}$ (K) & $253.8^{+3.7}_{-4.1}$\\
~~~~$\tau_{\rm circ}$ & Tidal circularization time-scale$^{d}$ (Gyr) & $80, 000^{+160, 000}_{-48, 000}$\\
~~~~$K$ & RV semiamplitude (m s$^{-1}$) & $190^{+17}_{-16}$\\
~~~~$\dot{\gamma}$ & RV slope$^{e}$ (m s$^{-1}$ day$^{-1}$) & $0.0031^{+0.0029}_{-0.0027}$\\
~~~~$R_p/R_*$ & Radius of planet in stellar radii  & $0.0644^{+0.0016}_{-0.0011}$\\
~~~~$a/R_*$ & Semimajor axis in stellar radii  & $256.4^{+9.1}_{-8.7}$\\
~~~~$\tau$ & Ingress / egress transit duration (days) & $0.0173^{+0.0041}_{-0.0022}$\\
~~~~$T_{14}$ & Total transit duration (days) & $0.2503^{+0.0035}_{-0.0026}$\\
~~~~$T_{FWHM}$ & FWHM transit duration (days) & $0.2325\pm0.0017$\\
~~~~$b$ & Transit Impact parameter  & $0.37^{+0.16}_{-0.22}$\\
~~~~$b_S$ & Eclipse impact parameter  & $7.6^{+2.4}_{-4.5}$\\
~~~~$\rho_p$ & Density (g cm$^{-3}$) & $4.07^{+0.55}_{-0.49}$\\
~~~~$\log g_p$ & Surface gravity (cgs)  & $3.938\pm0.040$\\
~~~~$\fave$ & Incident Flux (\fluxcgs) & $0.000465^{+0.000027}_{-0.000028}$\\
~~~~$T_P$ & Time of periastron (\bjdtdb) & $2, 455, 071.88\pm0.19$\\
~~~~$T_S$ & Time of eclipse (\bjdtdb) & $2, 454, 760^{+100}_{-110}$\\
\smallskip\\\multicolumn{2}{l}{Wavelength Parameters}&Kepler\smallskip\\
~~~~$u_{1}$ & Linear limb-darkening coefficient  & $0.453^{+0.039}_{-0.040}$\\
~~~~$u_{2}$ & Quadratic limb-darkening coefficient  & $0.264\pm0.048$\\
\smallskip\\\multicolumn{2}{l}{Telescope Parameters}&Keck-HIRES\smallskip\\
~~~~$\gamma_{\rm rel}$ & Relative RV Offset$^{e}$ (m s$^{-1}$) & $34.1^{+3.4}_{-3.6}$\\
~~~~$\sigma_J$ & RV jitter (m s$^{-1}$) & $6.7^{+4.3}_{-4.2}$\\
\enddata
\tablenotetext{}{See Table 3 in \citet{Eastman2019} for a detailed description of all parameters and all default (non-informative) priors. The coordinates of the planet are barycentric.}
\tablenotetext{a}{The value and uncertainty for $M_P$ were determined using the full posterior distribution.}
\tablenotetext{b}{$\omega$ is the argument of periastron of the star's orbit due to the planet.}
\tablenotetext{c}{Calculated with Equation~\ref{eq:teq}, which assumes no albedo and perfect redistribution. Between apastron and periastron, $T_{\rm eq}$ varies from {180--900~K}. See the text for a discussion.}
\tablenotetext{d}{The tidal circularization timescales is calculated from Equation~\ref{eq:tides}.}
\tablenotetext{e}{The reference epoch is BJD$_{\rm TDB} =$ 2457429.435011.}
\end{deluxetable}

We began by defining informative priors on several stellar parameters, which are listed at the top of Table~\ref{tab:stellar}. We constrained stellar effective temperature ($T_{\rm eff}$) and metallicity (as described by [Fe/H]) based on a \textsf{SpecMatch}\footnote{\url{https://github.com/petigura/specmatch-syn}} analysis \citep{Petigura2015,Petigura2017b} of a moderate S/N ($\sim$40) spectrum of \host\ acquired with Keck-HIRES without the iodine cell. This analysis produced an uncertainty on $T_{\rm eff}$ of 100~K, which we inflated to 115~K, in line with the systematic uncertainty floor reported by \citet{Tayar2020}. The \textsf{SpecMatch} analysis also suggested that the stellar radius is $\sim$1.7~$R_{\sun}$, hinting that this G2 star has evolved off of the main sequence (see Section~\ref{sec:bimod}). In addition to $T_{\rm eff}$ and [Fe/H], we constrained the upper limit on $V$-band extinction using the galactic reddening maps of \citet{Schlafly2011}. Lastly, we constrained the parallax of \host\ as measured by \gaia\ in EDR3 \citep{Gaia2016,Gaia2020}. Following the astrometric solution of \citet{Lindegren2020}\footnote{We calculated the astrometric solution using the software described at \url{https://www.cosmos.esa.int/web/gaia/edr3-code}.}, we subtracted $-0.026\pm0.013$~mas from the EDR3 value. 

For the SED portion of the \textsf{EXOFASTv2} fit, we modeled broadband photometry from 2MASS \citep{Cutri2003}, ALLWISE \citep{Cutri2014}, and \gaia\ \citep{Gaia2018} with inflated uncertainties as recommended by \citet{Eastman2019}. In doing so, we employed the MIST stellar evolution models \citep{Paxton2011,Paxton2013,Paxton2015,Dotter2016,Choi2016} packaged within \textsf{EXOFASTv2}. We imposed a noise floor of 2\% on the bolometric flux used in the SED modeling following \citet{Tayar2020}.

\begin{figure*}
    \centering
    \includegraphics[width=0.9\textwidth]{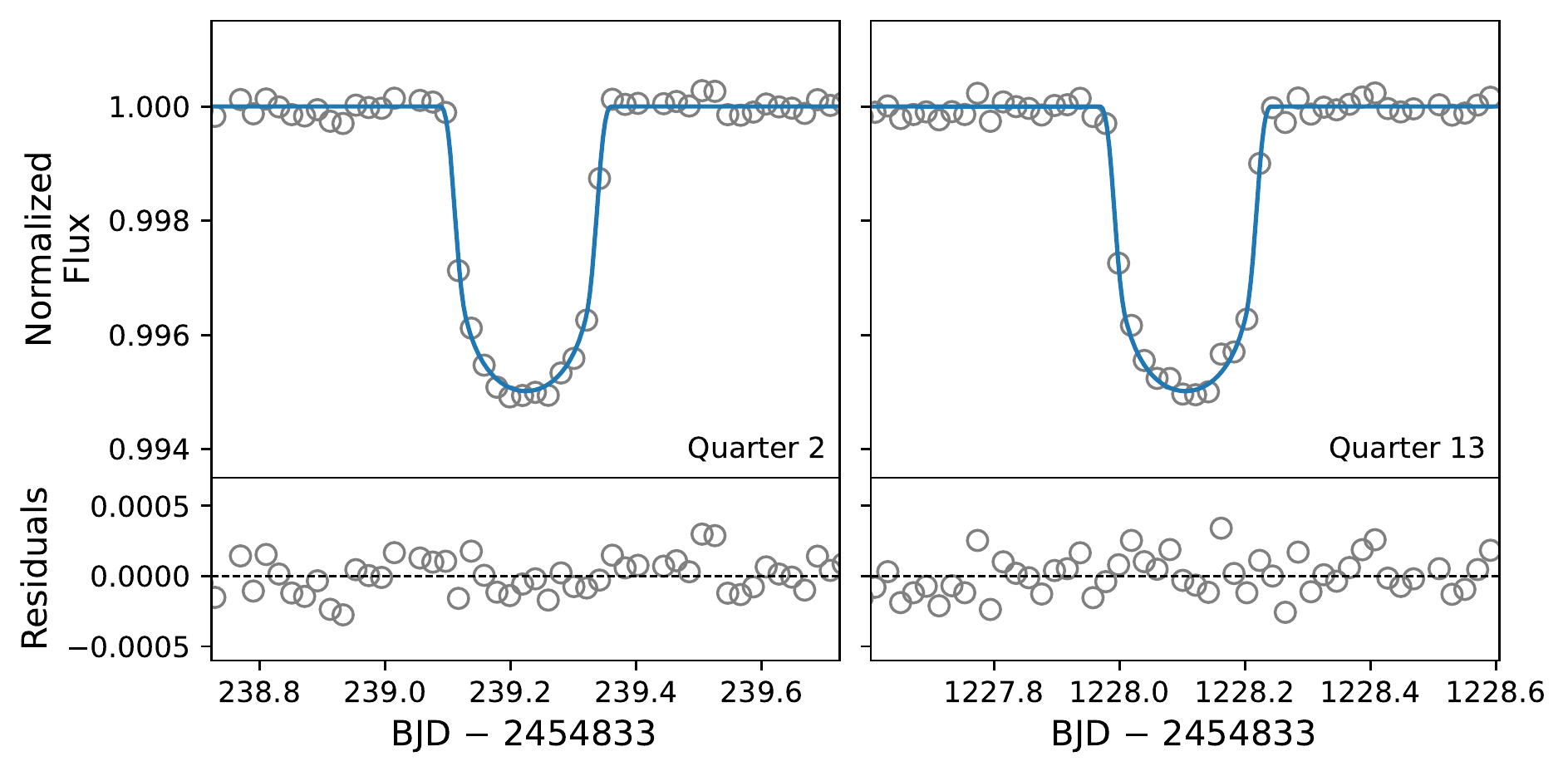}
    \caption{Detrended \kepler\ photometry of both transits (gray circles) and the best fit \textsf{EXOFASTv2} model (blue line).}
    \label{fig:transits}
\end{figure*}

\begin{figure}
    \centering
    \includegraphics[width=\columnwidth]{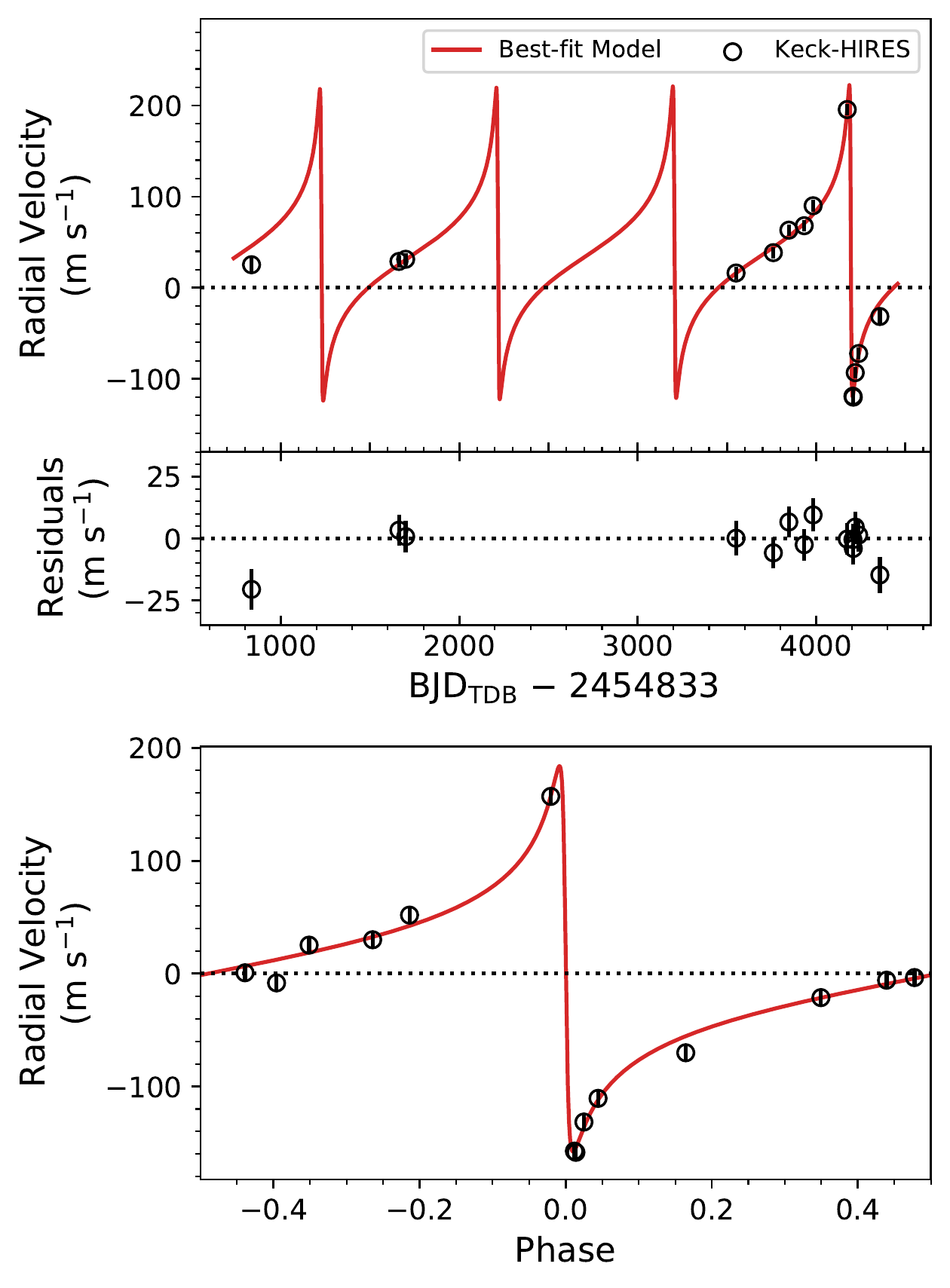}
    \caption{RV measurements of \host\ from Keck-HIRES with the best fit \textsf{EXOFASTv2} model. The top panel shows the time series and the bottom panel shows the data phase-folded on the best-fit ephemeris with $P=988.88$~days.}
    \label{fig:rv}
\end{figure}

The \textsf{EXOFASTv2} fit progressed until the number of independent draws of the underlying posterior probability distribution of each parameter exceeded 1000 and the Gelman--Rubin statistic for each parameter decreased below 1.01 \citep{Gelman1992,Ford2006b}. We show the resulting best fit models with the transit and RV data in Figures \ref{fig:transits} and \ref{fig:rv}, respectively.

\subsection{The Bimodality of Stellar Mass and Age}\label{sec:bimod}

\begin{figure}
    \centering
    \includegraphics[width=\columnwidth]{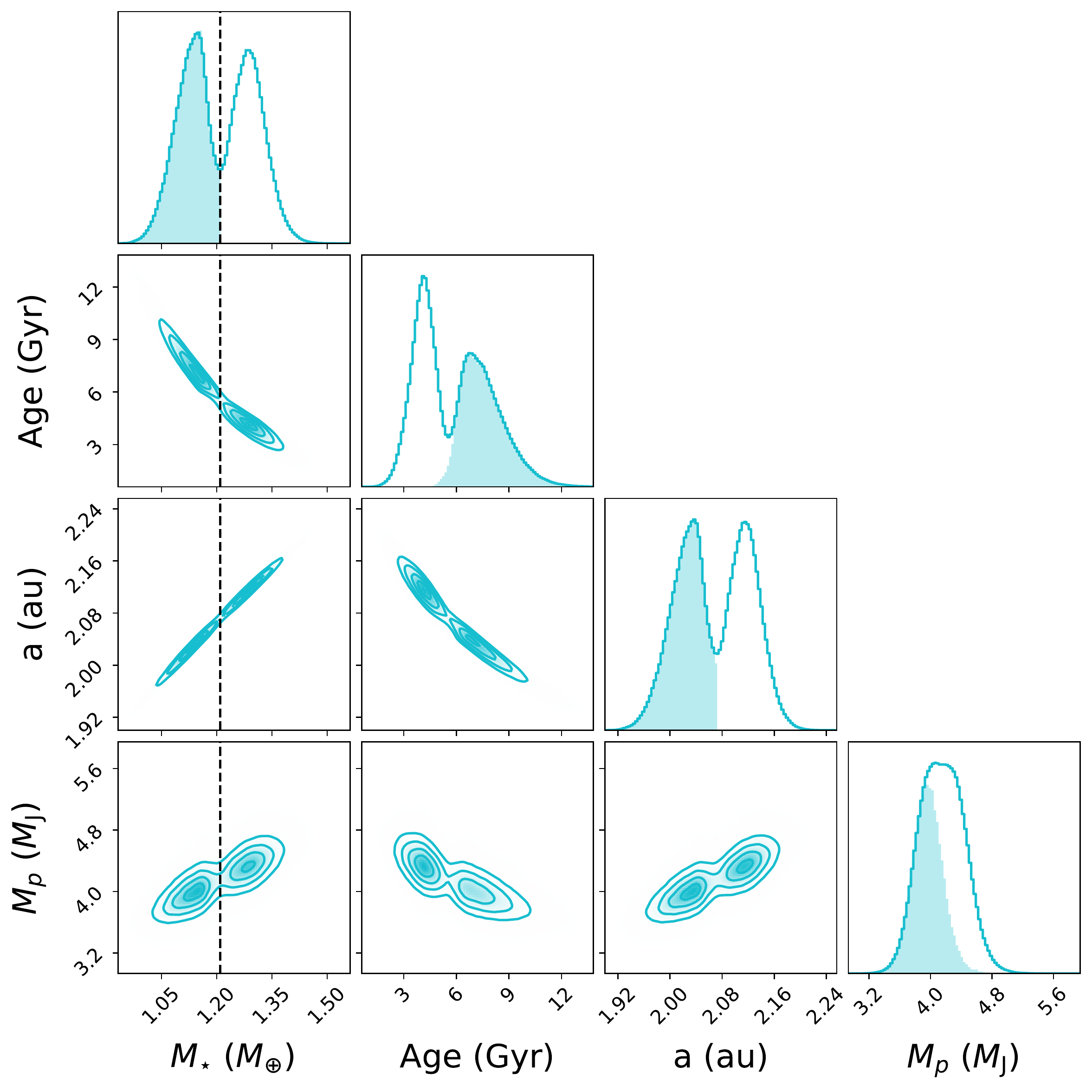}
    \caption{Posterior probability distributions showing the bimodality in stellar properties and its effect on the inferred properties of \planet. The dashed vertical line at 1.21~$M_{\sun}$ shows where we separated the low and high mass solutions, the former of which is slightly preferred (51.8\% to 48.2\%) and is shown as the shaded portion of each distribution.}
    \label{fig:bimod}
\end{figure}

The converged \textsf{EXOFASTv2} fit yielded bimodal posterior probability distributions for the stellar mass ($M_{\star}$) and age (Figure~\ref{fig:bimod}). The region of parameter space preferred by the MIST stellar evolution models, as influenced by all of the \host\ data, exists near the subgiant branch as we suspected based on the \textsf{SpecMatch} radius estimation. \textsf{EXOFASTv2} found that multiple stellar ages and surface gravity values ($\log{g}$) correspond to the $T_{\rm eff}$ prior, meaning that the bimodality is astrophysical and not due to inadequate posterior sampling. The bimodality propagates to the semi-major axis ($a$) of \planet\ and, to a lesser extent, its mass ($M_p$; Figure~\ref{fig:bimod}). 

Since we could not distinguish between the two families of solutions with the data of the \host\ system in hand, we adopted the strategy of \citet{IkwutUkwa2020} and divided the solutions at a fiducial $M_{\star}$ value of {1.21~$M_{\sun}$}, which corresponds to the trough between the posterior probability peaks in Figure~\ref{fig:bimod}. The lower stellar mass, older age solution contains 51.8\% of the posterior samples, which we treated as a slight preference over the higher stellar mass, younger solution. Therefore, in Tables~\ref{tab:stellar} and \ref{tab:planet}, we only publish the parameters for the preferred, lower stellar mass solution. The only exception is the planet mass, $M_p$, for which we determine the 68\% confidence interval using the full posterior distribution. None of our interpretations of the nature or formation history of \planet\ are changed by considering the alternate solution.


\section{Results}\label{sec:results}

\subsection{Confirming \planet\ as a Genuine Planet}

A photometric dimming event with a depth corresponding to a giant planet transit can be created by substellar or stellar objects or various systematic signals \citep[e.g.,][]{Brown2003,Torres2005,Cameron2012,ForemanMackey2016b,Dalba2020b}. False-positive signals can be harder to identify for longer (compared to shorter) orbital periods owing to the difficulty in quantifying the reliability of genuine transit events from similarly long-period exoplanets \citep[e.g.,][]{Thompson2018}. Indeed, \citet{Santerne2016} measured a 55\% false-positive rate for \kepler\ giant planets within 400~days of orbital period. For these reasons, long-period giant planet candidates like \koi\ must be vetted with Doppler spectroscopy before any weight is placed upon their standing as a genuine planet.

\begin{figure}
    \centering
    \includegraphics[width=0.9\columnwidth]{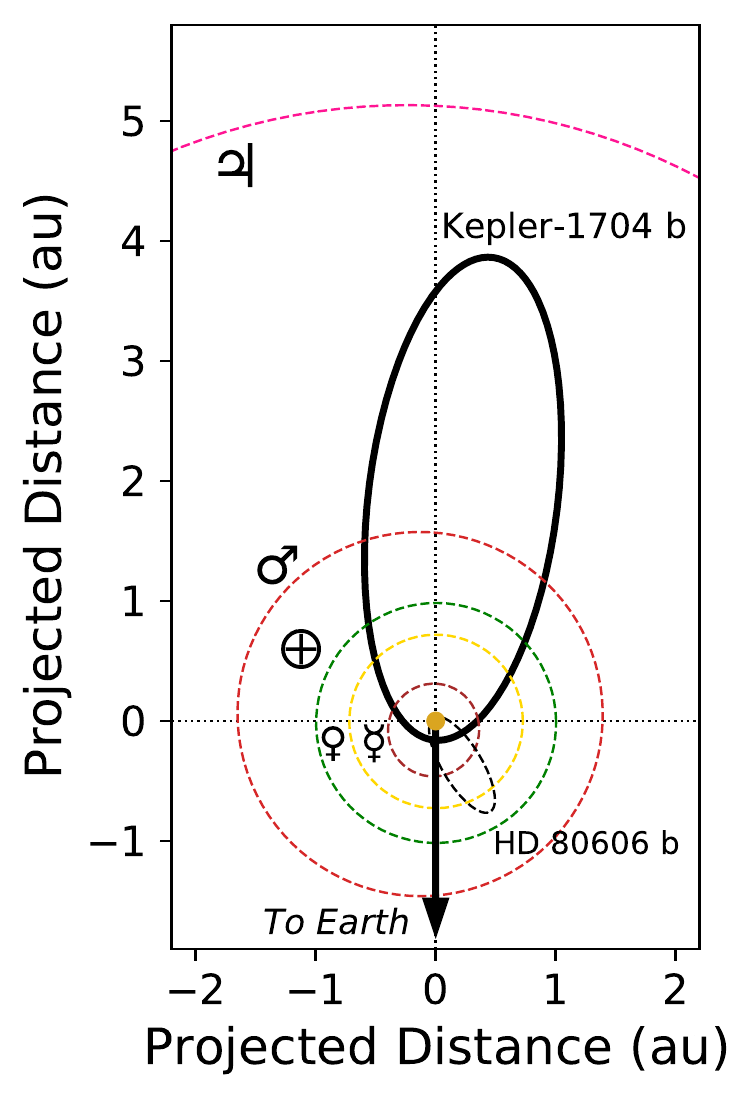}
    \caption{Face-on view of the orbit of \planet. The orbits of five Solar System planets and HD~80606~b (dashed black line) are included for reference. All orbits are drawn to scale, although the size of \host\ is not.}
    \label{fig:orbit_view}
\end{figure}

Our 10-year baseline of RV measurements for \host\ confirmed the genuine planetary nature of \planet. It also confirmed the 988.88~day orbital period, placing \planet\ among the top five longest-period (non-controversial) transiting exoplanets with precisely measured periods known to date\footnote{According to the NASA Exoplanet Archive, accessed 2021 June 23.}. With a semi-major axis of {2.03~au} and an orbital eccentricity of {0.92}, its elongated orbit brings it within {0.16~au} of its host star and then slingshots it out to {3.9~au}---the largest apastron distance of any transiting exoplanet with known orbital period and eccentricity. Figure~\ref{fig:orbit_view} is a diagram showing the orbit of \planet\ relative to those of Jupiter, the Solar System terrestrial planets, and HD~80606~b. The RV data also contain a slight, although tentative, acceleration ({$0.0031^{+0.0029}_{-0.0027}$~m~s$^{-1}$~day$^{-1}$}) that possibly indicates the existence of an outer companion.

The equilibrium temperature ($T_{\rm eq}$) for \planet\ as shown in Table~\ref{tab:planet} is calculated following
\begin{equation}\label{eq:teq}
    T_{\rm eq} = T_{\rm eff}\sqrt{\frac{R_{\star}}{2a}} \;,
\end{equation}
which assumes no albedo and perfect heat redistribution \citep{Hansen2007}. However, including a factor of 1/($\sqrt{1\pm e}$) in this equation suggests that $T_{\rm eq}$ varies from {$\sim$180~K} at apastron to {$\sim$900~K} at periastron.  This substantial {$\sim$700~K} swing in temperature likely affects the atmosphere on \planet.

In the following sections, we will investigate the possibility of companions, migration history, interior composition, and atmospheric characterization prospects for \planet. We take advantage of the fact that this planet's orbital period, eccentricity, and radius are known precisely, which is remarkable for an exoplanet with its orbital properties.


\subsection{Outer Companions in the \host\ System}\label{sec:companions}

As described in Section~\ref{sec:intro} and extensively in the broader orbital dynamics literature \citep[e.g.,][]{Naoz2016}, the presence of an outer planetary or stellar companion may have direct consequences on the migration history of a giant planet. For \host, archival AO imaging data yield a nondetection of stellar companions beyond $\sim$100~au and upper mass limits on such a companion down to $\sim$50~au (see Figure~\ref{fig:mass_limit} and Section~\ref{sec:imaging}). In the following sections, we exploit our long baseline of RV observations to improve upon these limits with an injection-recovery test (Section~\ref{sec:inj-rec}), a RV trend analysis (Section~\ref{sec:trend}), and a chaos indicator analysis (Section~\ref{sec:megno}).

\subsubsection{RV Injection-Recovery Test}\label{sec:inj-rec}

We characterized the sensitivity of our RV data set to additional bound companions by running injection-recovery tests, in which we added synthetic signals to our RV data and converted the signal recovery rate into a map of search completeness. We used \textsf{RVSearch} \citep{Rosenthal2021}, an iterative periodogram search algorithm, to search for evidence of additional companions to \planet\ in the RV data and perform these tests. We initialized \textsf{RVSearch} with the best-fit Keplerian model for \planet\ and searched for additional companions with orbital periods spanning 2--10,000 days. We found no evidence for additional companions in this period range. Once the search was completed, \textsf{RVSearch} injected synthetic planets into the data and repeated the additional iteration to determine whether it recovered these synthetic planets. We ran 3,000 injection tests for \host. We drew the injected planet $a$ and $M_p \sin{i}$ from log-uniform distributions, and drew eccentricity from the Beta distribution with shape parameters $\alpha = 0.867$ and $\beta = 3.03$, which \citet{Kipping2013a} found represented the sample of RV-observed exoplanets. After \textsf{RVSearch} performed the injection-recovery tests, we measured search completeness across a wide range of $a$ and $M_p \sin{i}$ by determining the fraction of recovered synthetic signals in localized regions of $a$ and $M_p \sin{i}$.

\begin{figure*}
  \centering
    \begin{tabular}{cc}
      \includegraphics[width=0.5\textwidth]{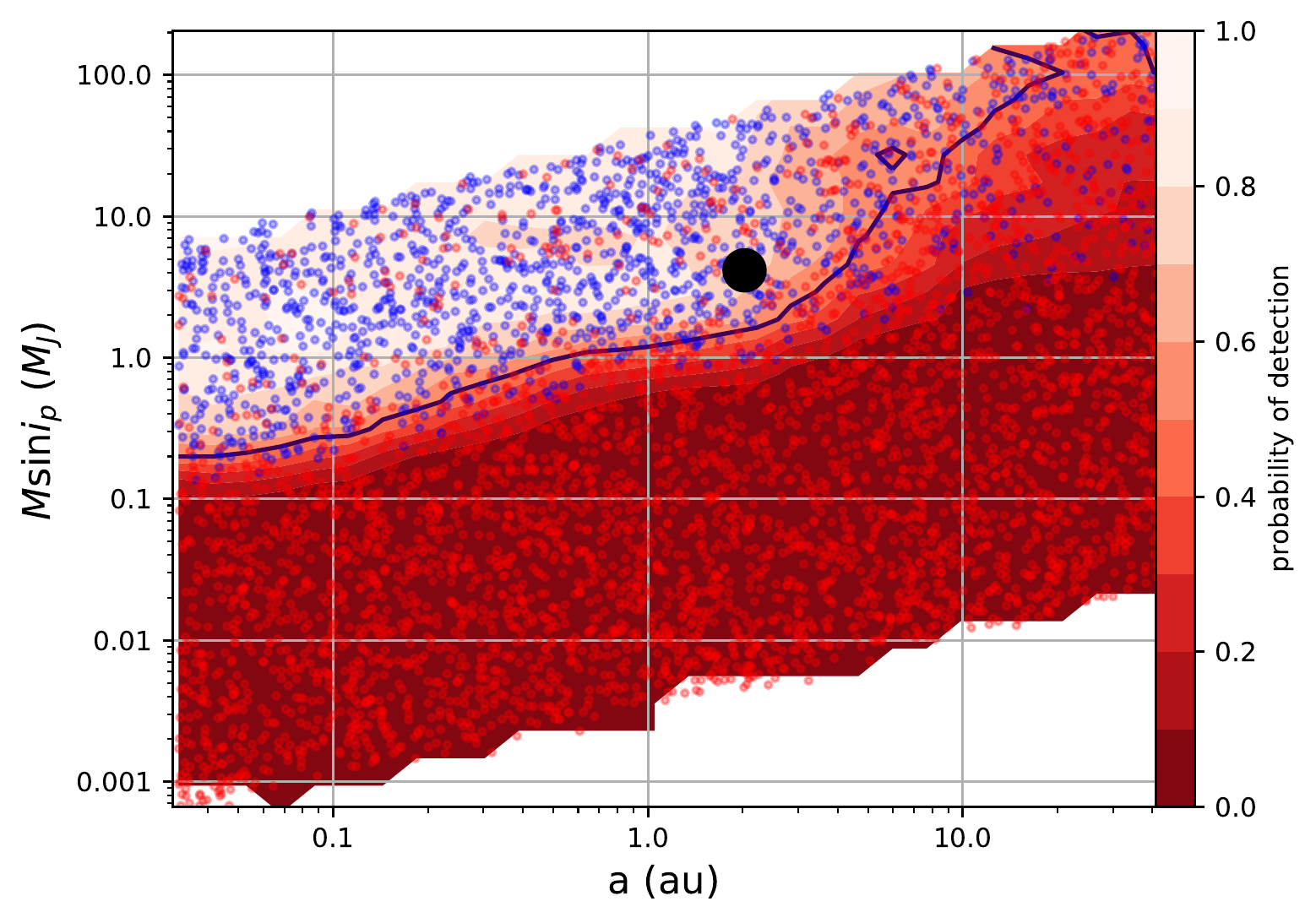} &
      \includegraphics[width=0.5\textwidth]{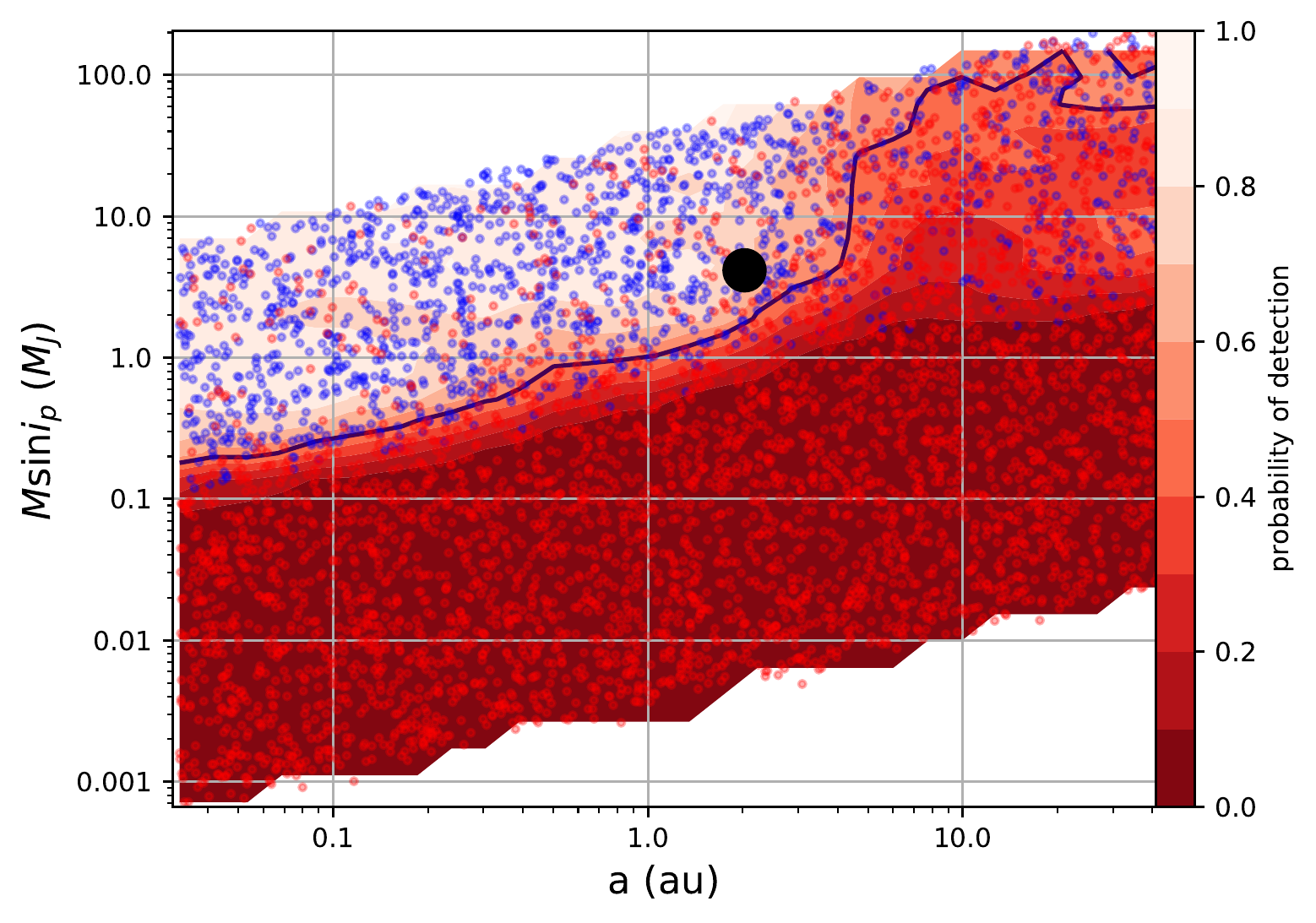} 
    \end{tabular}
  \caption{\textsf{RVSearch} injection and recovery to search for other signals in the RV data set. The left panel shows completeness contours for all RV data, while the right panel shows contours with the earliest RV data point removed (see Section~\ref{sec:hires}). Red dots represent injected signals that were not recovered as opposed to blue dots that show recovered signals. The black dot is \planet, and the black line shows the 50\% recovery contour.}
  \label{fig:rvsearch}
\end{figure*}

Figure \ref{fig:rvsearch} shows a pair of search completeness results, one of which includes the first low-S/N RV data point (left panel) and one of which excludes it (right panel). In both cases, our RV sensitivity to companions beyond the orbital separation of \planet\ is limited, dropping below 50$\%$ completeness at 4~$M_{\rm J}$ beyond 4~au. The sparsity and high RMS of the RV data set drive the high lower limit on detectability in $M_p \sin{i}$, and the nearly 10-year observational baseline sets the sharp change in completeness around 3~au.

\subsubsection{RV Trend Analysis}\label{sec:trend}

To build upon the injection-recovery test, we conducted a complementary analysis of acceleration (i.e., a long-term RV trend) in the Keck-HIRES RVs. This analysis focused specifically on partially sampled signals from giant planets, substellar objects, or stars that could be lurking undetected in the outer reaches of the \host\ system. When combined with a nondetection from the AO imaging, RV trends can greatly reduce the parameter space that a possible undetected companion could occupy \citep[e.g.,][]{Crepp2012,Kane2019b,Dalba2021b}.

The \textsf{EXOFASTv2} fit to the transit, RV, and SED (Section~\ref{sec:model}) included a parameter for ``RV slope'' ($\dot\gamma$), which quantified any acceleration measured from the RVs. As shown in Table~\ref{tab:planet}, we made a low significance detection of acceleration: {$\dot\gamma = 0.0031^{+0.0029}_{-0.0027}$~m~s$^{-1}$~day$^{-1}$.} To refine the mass ($M_c$) and orbital distance ($a_c$) of the companion that could have caused this RV drift, we simulated RVs over a grid of scenarios broadly following the procedure of \citet{Montet2014}. 

Firstly, we subtracted the maximum likelihood \textsf{EXOFASTv2} solution for \planet\ from the Keck-HIRES RV data but without including the acceleration (i.e., we set $\dot\gamma = 0$). In doing so, we also inflated the RV uncertainties ($\sigma_{v_r}(t)$) to account for the fitted RV jitter (Table~\ref{tab:planet}). The resulting RV time series ($v_r(t)$) only contained the long-term trend. 

Next, we defined a logarithmically spaced 30x30 grid in companion mass ($1\; M_{\rm J} < M_c  <1\; M_{\sun}$) and semi-major axis ($4< a/{\rm au}<200$). The mass boundaries were chosen to complement the constraints from the \textsf{RVsearch} injection-recovery tests and the AO imaging (Section~\ref{sec:imaging}). The orbital distance boundaries were chosen to span the gap between the apastron distance of \planet\ and the stringent upper boundary from the AO imaging. 

At each point along the $M_c$--$a_c$ grid, we drew 500 sets of the orbital elements \{$\omega$, $e$, $i$\}, which are the argument of periastron, the eccentricity, and the inclination, respectively. We drew $\omega$ randomly from a uniform distribution over the interval [0, 2$\pi$], and we drew $i$ randomly from a uniform distribution in $\cos{i}$ over the interval [0, 1]. For $e$, we drew values from the Beta distribution from \citet{Kipping2013a} mentioned previously (Section~\ref{sec:inj-rec}). These random draws were meant to account for the variety of orbital configurations a massive companion could have.

Then, for each of the individual orbits, we simulated 50 sets of RV time series ($\hat v_r(t)$) with a cadence matching $v_r$. Each of the 50 sets started at a different orbital phase spaced evenly across the entire orbit. This accounted for the fact that the Keck-HIRES observations could have sampled any portion of the companion's orbit. 

\begin{figure*}
  \centering
    \begin{tabular}{cc}
      \includegraphics[width=0.5\textwidth]{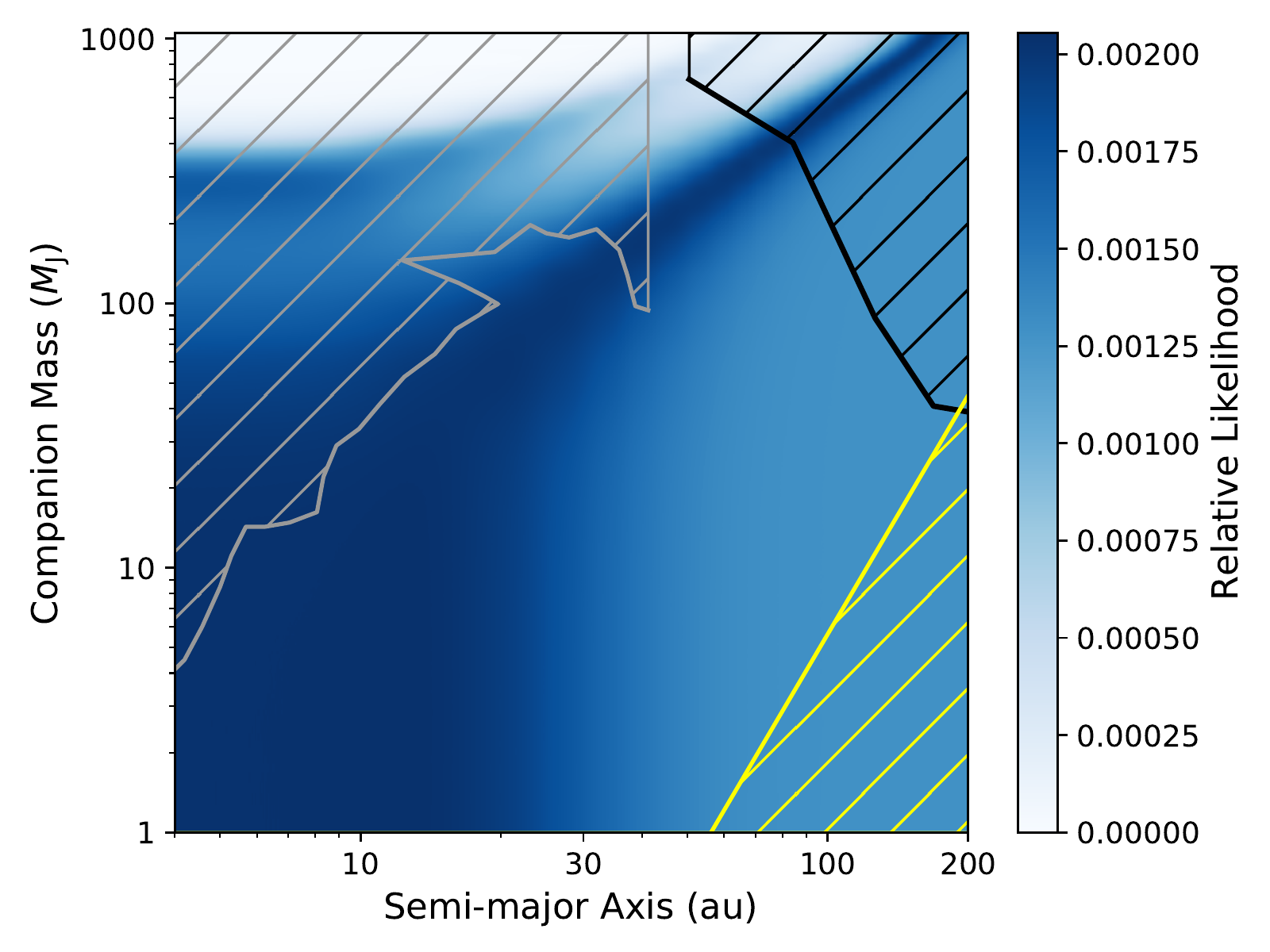} &
      \includegraphics[width=0.5\textwidth]{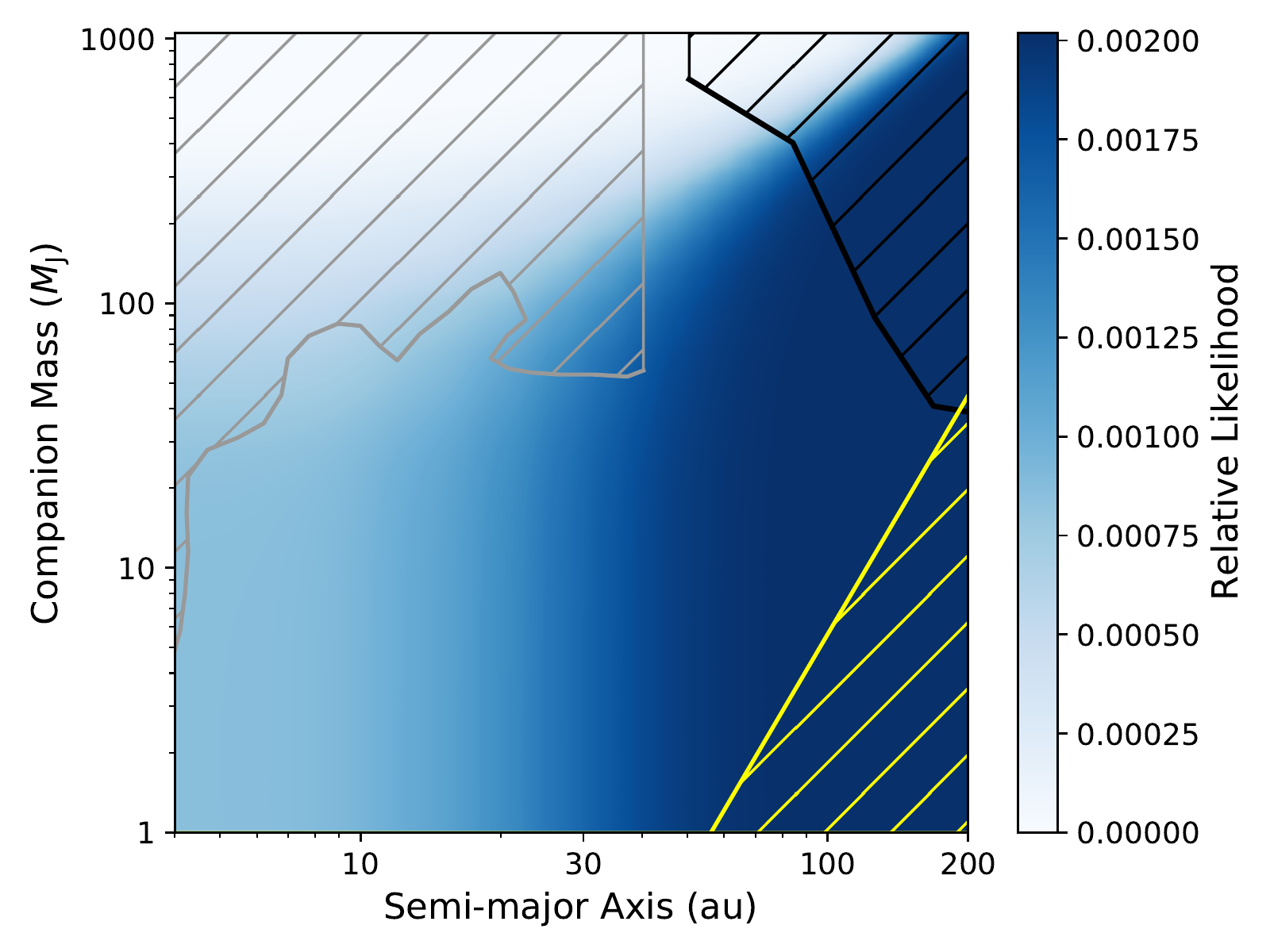} 
    \end{tabular}
  \caption{Relative likelihood of a companion in mass and semi-major axis space based on the acceleration in the Keck-HIRES RV residuals after the signal from \planet\ was subtracted. \textit{Left:} Likelihoods were calculated using all Keck-HIRES RV data points. \textit{Right:} Likelihoods were calculated after removing the first Keck-HIRES RV data point (Section~\ref{sec:hires}). In both panels, the black hatched region (upper, right) is ruled out to 5$\sigma$ by the AO imaging (Figure~\ref{fig:mass_limit}) and the yellow hatched regions (lower, right) cover companions that could not excite the observed eccentricity through Kozai--Lidov cycles \citep{Dong2014}. Any potential candidates in the gray hatched regions (upper, left) have greater than a 50\% recovery rate in the corresponding \textsf{RVsearch} analysis (Figure~\ref{fig:rvsearch}). }
  \label{fig:trend_map}
\end{figure*}

Finally, we used a least-squares regression routine to minimize the familiar statistic $\chi^2 = \sum_t [v_r(t) - \hat v_r(t)]^2/\sigma_{v_r}(t)^2$. This minimization was necessary because the Keck-HIRES RVs are relative, not absolute. Assuming uncorrelated errors, we converted the 50 $\chi^2$ values for each individual orbit to relative probabilities following $P \propto {\rm exp}(-\chi^2/2)$, and we summed the probabilities to effectively marginalize over the portion of the orbit captured by the data. We also summed the probabilities of the 500 sets of orbits at each grid point to effectively marginalize over all orbital properties other than $M_c$ and $a_c$. Lastly, we normalized the map of probabilities such that 22.5 million probability calculations summed to unity ultimately yielding relative likelihood values. Figure~\ref{fig:trend_map} (left panel) shows the resulting map. 

The slight acceleration detected in the full set of RVs prefers companions within roughly 30~au and less massive than a few hundred Jupiter masses although some probable solutions are still present at wide separation and high mass. Incorporating the upper mass limit from the AO imaging (Section~\ref{sec:imaging}) trims a correlated region of parameter space at the highest masses and largest orbital separations. Also, assuming that any companion with at least a 50\% \textsf{RVsearch} recovery rate should have been detected, the trend analysis further refines the likely parameter space of a companion.  

We repeated this entire analysis but after removing the first Keck-HIRES RV data point, as its timing and quality may have inaccurately affected the measured RV trend (Section~\ref{sec:hires}). The resulting map of relative likelihoods calculated without the first RV data point is shown in the right panel of Figure~\ref{fig:trend_map}. For context supporting the second map, we also conducted a second \textsf{EXOFASTv2} fit without the first Keck-HIRES data point that was otherwise identical to the fit described in Section~\ref{sec:model}. The only appreciable difference between the two \textsf{EXOFASTv2} fits was value of $\dot \gamma$, which decreased in significance to {$-0.0002\pm0.0029$~m~s$^{-1}$~day$^{-1}$} in the latter case. This difference manifests in the relative likelihood map as a preference toward larger orbits ($a\gtrsim$30~au) rather than smaller ones. The map is again complemented by the AO imaging upper limit and the region with over a 50\% \textsf{RVsearch} recovery rate. 

Within the parameter space we are exploring, it is also helpful to consider which companions would be capable of overcoming precession caused by general relativity and exciting the eccentricity of \planet\ through Kozai--Lidov cycles. \citet{Dong2014} calculated an approximate \emph{strength} criterion for warm Jupiters (their Equation~5) that we apply to \planet. In the limiting case of an initially circular orbit that is much longer-period than that of a hot Jupiter, we identify the region of $M_c$--$a$ parameter space with objects that are unable to  have excited the eccentricity of \planet\ (Figure~\ref{fig:trend_map}, yellow region). By all of our other analyses, we cannot rule out the existence of a companion at the lowest masses and largest separation we consider. However, such a companion is also too low mass and orbits too far from \planet\ to overcome GR precession through Kozai--Lidov interactions.

Considering all of the companion analyses together yields three conclusions for possible outer companions in the \host\ system. Firstly, for $M_c\gtrsim700$~$M_{\rm J}$, we should have either recovered the signal in the Keck-HIRES RVs ($\ge50$\% recovery rate) or detected the source directly in the AO imaging for nearly all values of $a$. Secondly, for $50\lesssim M_c/M_{\rm J} \lesssim 700$, those with $a \lesssim 40$~au should have been recovered by the RV data, and those with $a \gtrsim 150$~au should have been detected in the AO imaging. A companion with separation between these values could go undetected. Thirdly, we do not have sensitivity to companions with $M_c\lesssim$50~$M_{\rm J}$ within $\sim$150~au, so there could be substellar or planetary companions in this region. At these masses, our RV trend analysis reveals a preference for companions with $a\gtrsim30$~au. Although, some companions in this region of parameter space would be unable to excite the observed eccentricity of \planet.

\subsubsection{MEGNO Simulations}\label{sec:megno}

To test whether additional constraints can be placed on the orbital configurations of the potential outer companion, we ran a dynamical simulation using the Mean Exponential Growth of Nearby Orbits (MEGNO) chaos indicator \citep{Cincotta2000}. The MEGNO indicator demonstrates whether a specific system configuration would lead to chaos after a certain integration time by distinguishing between quasi-periodic and chaotic evolution of the bodies within the system \citep[e.g.,][]{Hinse2010}. The final MEGNO value returned for a specific orbital configuration is useful for determining the stochasticity of the configuration, where chaos is more likely to result in unstable orbits for planetary bodies. With a grid of orbital parameters, a MEGNO simulation can provide valuable information on the orbital configurations that are favored by dynamical simulations, and reject configurations that return chaos results.

The MEGNO simulation to explore the dynamically viable locations for various outer companions was carried out within the N-body package \textsf{REBOUND} \citep{Rein2012} with the symplectic integrator WHFast \citep{Rein2015}. We used the stellar and planetary parameters from Table \ref{tab:stellar} and Table \ref{tab:planet}, respectively. We provided a linear-uniform grid in semi-major axis (20--60~au) and companion mass (1--100~$M_{\rm J}$) that aligned with the higher likelihood region in Figure~\ref{fig:trend_map} (right panel). The eccentricity of the outer companion was set to zero. The simulation was integrated for 20~million~years with a time step of 0.034~years ($\sim$12.4~days). This time step was chosen to be 1/80 of the orbital period of \planet, a fourth of the recommended value \citep{Duncan1998}, to increase the sampling near the periastron passage of this highly eccentric planet. The integration was set to stop and return chaos results if any of the planetary orbits started extending beyond 100~au.

\begin{figure}
    \centering
    \includegraphics[width=\columnwidth]{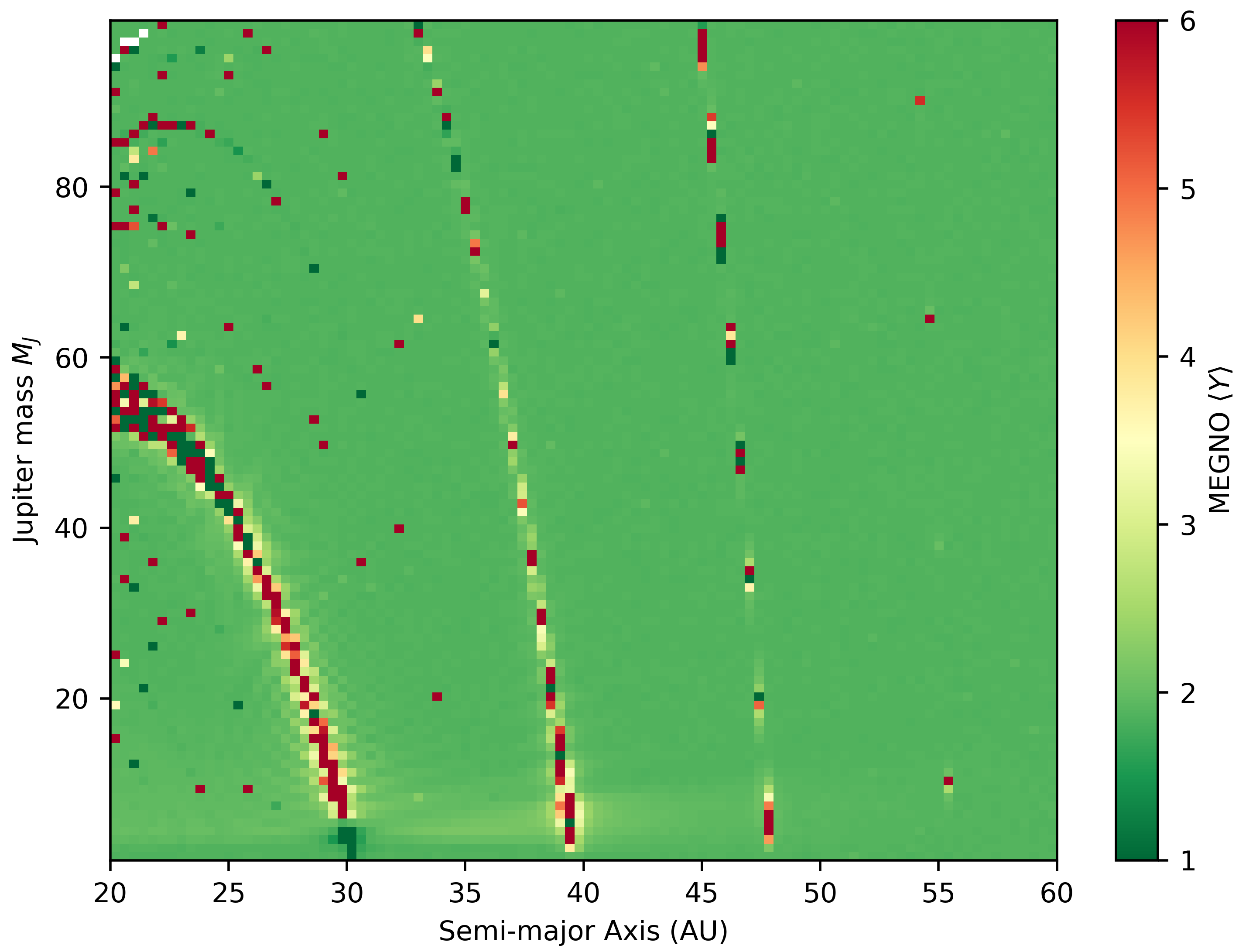}
    \caption{MEGNO simulation result with a grid of orbital configurations for the outer companion. Green regions (low values) are stable against chaos. The stripes identify chaos due to the overlap of secular resonances.}
    \label{fig:megno}
\end{figure}

Figure \ref{fig:megno} shows the grid of results of the MEGNO simulation. Each grid point is color-coded according to the final MEGNO value for the orbital configuration of that outer companion. A MEGNO value around 2 (green) is considered non-chaotic \citep{Hinse2010} and is thus a dynamically viable region where the outer companion could exist without making the system chaotic. Grid points in red indicate simulations that returned chaotic results, and those in white indicate irregular events such as close encounters and collisions, all of which are unfavorable configurations for an outer companion. 

Only a few stripes of parameter space contain orbital configurations that lead to chaos. The stripes indicate systems where nonlinear eccentricity secular resonances overlap leading to secular chaos \citep[e.g.,][]{Lithwick2011a,Wu2011}. The stripes become less defined at wider orbits because our simulation has not been run long enough to capture the resonance evolution. Besides these narrow regions, this analysis fails to rule out any extra substantial area of parameter space where a massive companion could exist.


\subsection{Bulk Metallicity Retrieval for \planet}\label{sec:zpzs}

Continuing our discussion of results, we now shift the attention from the outer reaches of the \host\ system back to \planet\ itself. 

With the measured mass and radius of \planet, along with other system properties, we retrieved the mass of its heavy elements or its bulk metals ($M_z$) and calculated its bulk metallicity ($Z_p \equiv M_z / M_p$) following \citet{Thorngren2019a}. Briefly, we modeled the thermal evolution of \planet\ using one-dimensional structure models with a core composed of a rock–ice mixture at equal amounts, a homogeneous convective envelope made of a H/He–rock–ice mixture, and a radiative atmosphere. The atmosphere models were interpolated from the grid of \citet{Fortney2007}. Samples were drawn from the posterior probability distributions for planet mass, radius, and age (Section~\ref{sec:model}), and the heavy element mass was adjusted in the structure models to recover the planet radius. 

This analysis relied on two assumptions. First, we assumed that the planet radius is not inflated \citep[e.g.,][]{Laughlin2018} because the average irradiation flux received by \planet\ (see Table~\ref{tab:planet}), which is well below the canonical 2$\times$10$^8$~erg~s$^{-1}$~cm$^{-2}$ empirical threshold for giant planets \citep{Miller2011,Demory2011b,Sestovic2018}. Second, we neglected any internal heating from circularization tides. We assumed that tides are an inefficient means of heating \planet\ as evidenced by its tidal circularization timescale ($\tau_{\rm circ}$) given by Equation~3 from \citet{Adams2006}: 
\begin{equation}\label{eq:tides}
    \tau_{\rm circ} = 1.6\;{\rm Gyr}\; \left ( \frac{Q}{10^6} \right ) \left ( \frac{M_p}{M_J} \right ) \left ( \frac{M_{\star}}{M_{\sun}} \right )^{-3/2} \left ( \frac{R_p}{R_J} \right )^{-5} \left ( \frac{a}{0.05\; {\rm au}} \right )^{13/2}
\end{equation}

\noindent where $Q$ is tidal quality factor that is assumed to be $10^6$ (similar to Jupiter). As listed in Table~\ref{tab:planet}, $\tau_{\rm circ}$ is {80,000}~Gyr, much longer than the age of the universe even considering error introduced by our estimate value for $Q$.

The metallicity retrieval was complicated slightly by the bimodal probability distribution for age that we inferred from the comprehensive system modeling (Figure~\ref{fig:bimod}). Instead of using separate normal priors for stellar mass and age, we used a bivariate Gaussian kernel-density estimate. Then, we sampled the posterior with a Markov chain Monte Carlo technique. 

\begin{figure}
    \centering
    \includegraphics[width=\columnwidth]{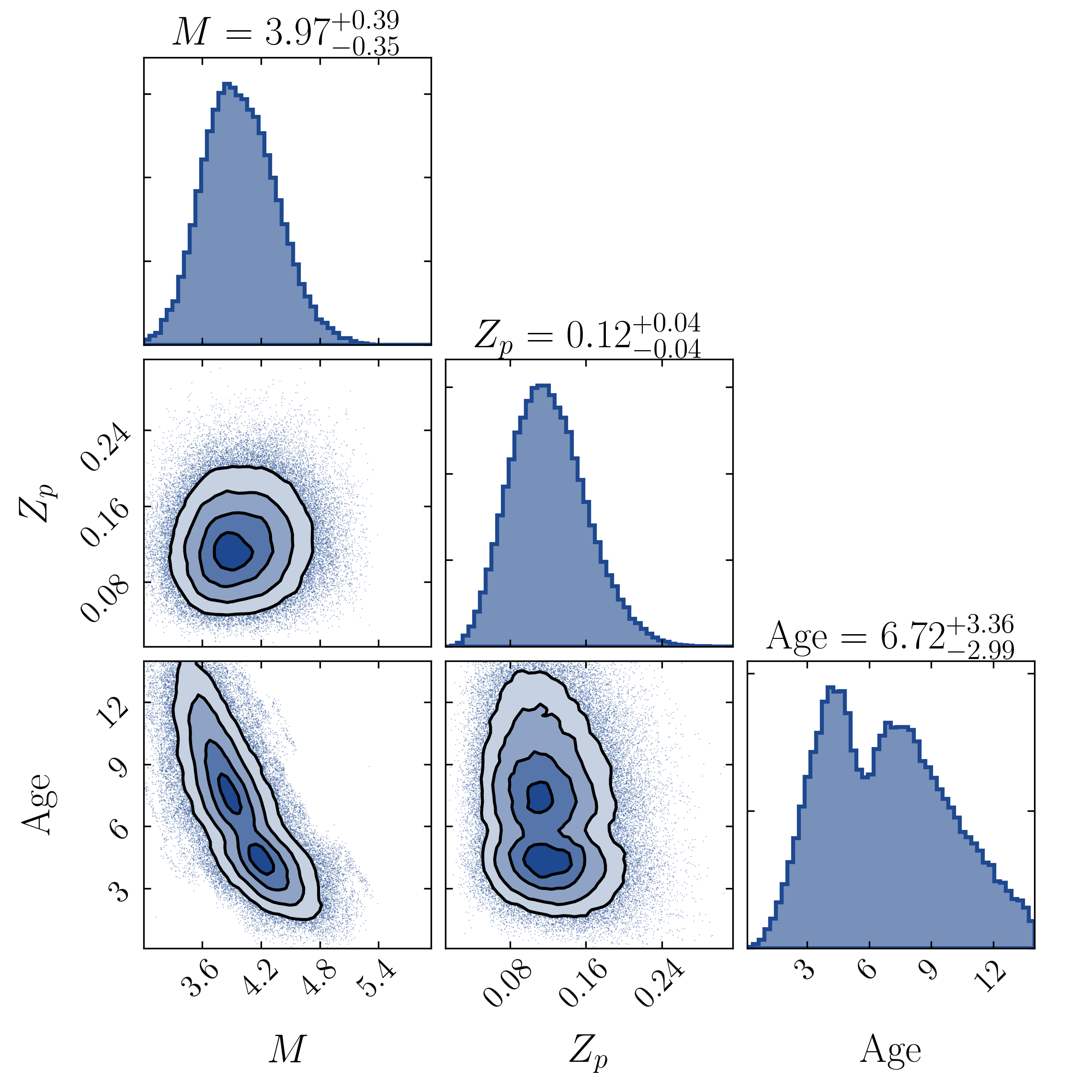}
    \caption{Posterior probability distributions from the heavy element mass retrieval for \planet. The symbols $M$ and $Z_p$ represent planet mass and bulk metallicity, respectively. Despite the bimodality in age (see Section~\ref{sec:model}), $Z_p$ is normal. The inferred bulk metallicity of \planet\ corresponds to a heavy element mass of {$\sim$150}~$M_{\earth}$ and an enrichment (relative to stellar) {of $\sim$5.}}
    \label{fig:metal_post}
\end{figure}

The results of the bulk metal mass retrieval are shown in Figure~\ref{fig:metal_post}. Despite the bimodality in age, the marginalized posterior probability distribution for bulk metallicity is a near-normal distribution at {$Z_p = 0.12\pm0.04$}, corresponding to {$M_z \approx 150$~$M_{\earth}$.} To calculate the stellar metallicity ($Z_{\star}$), we assumed that the iron abundance ([Fe/H]) scales with total heavy metal content such that $Z_{\star} \equiv 0.0142\times10^{\rm [Fe/H]}$ \citep{Asplund2009,Miller2011}, which yields {$Z_{\star} = 0.0229\pm0.0031$}. Finally, we calculated the bulk metallicity enrichment relative to stellar for \planet\ as {$Z_p/Z_{\star} = 5.2\pm1.9$}. 

\begin{figure*}
  \centering
    \begin{tabular}{cc}
      \includegraphics[width=0.47\textwidth]{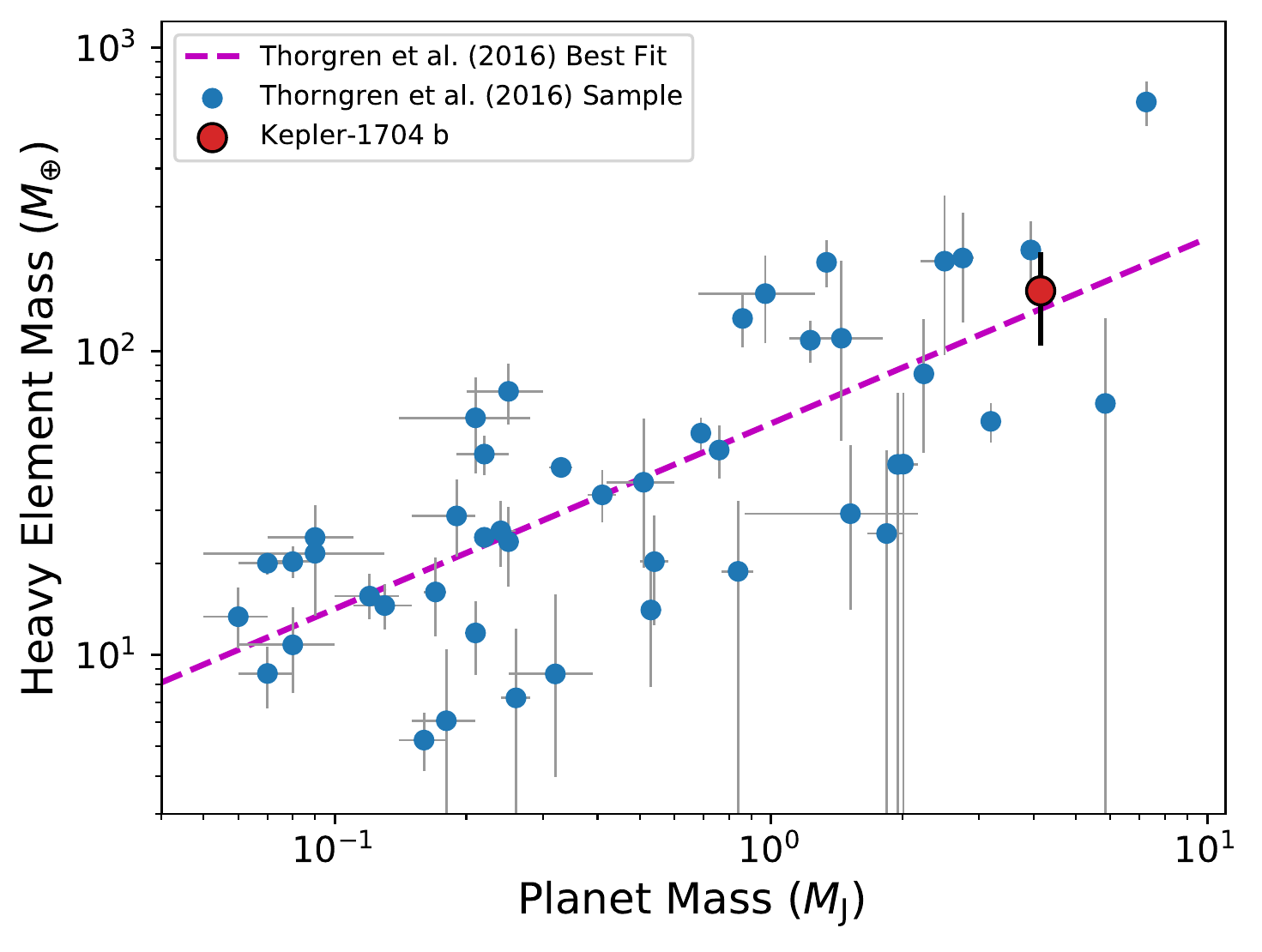} &
      \includegraphics[width=0.47\textwidth]{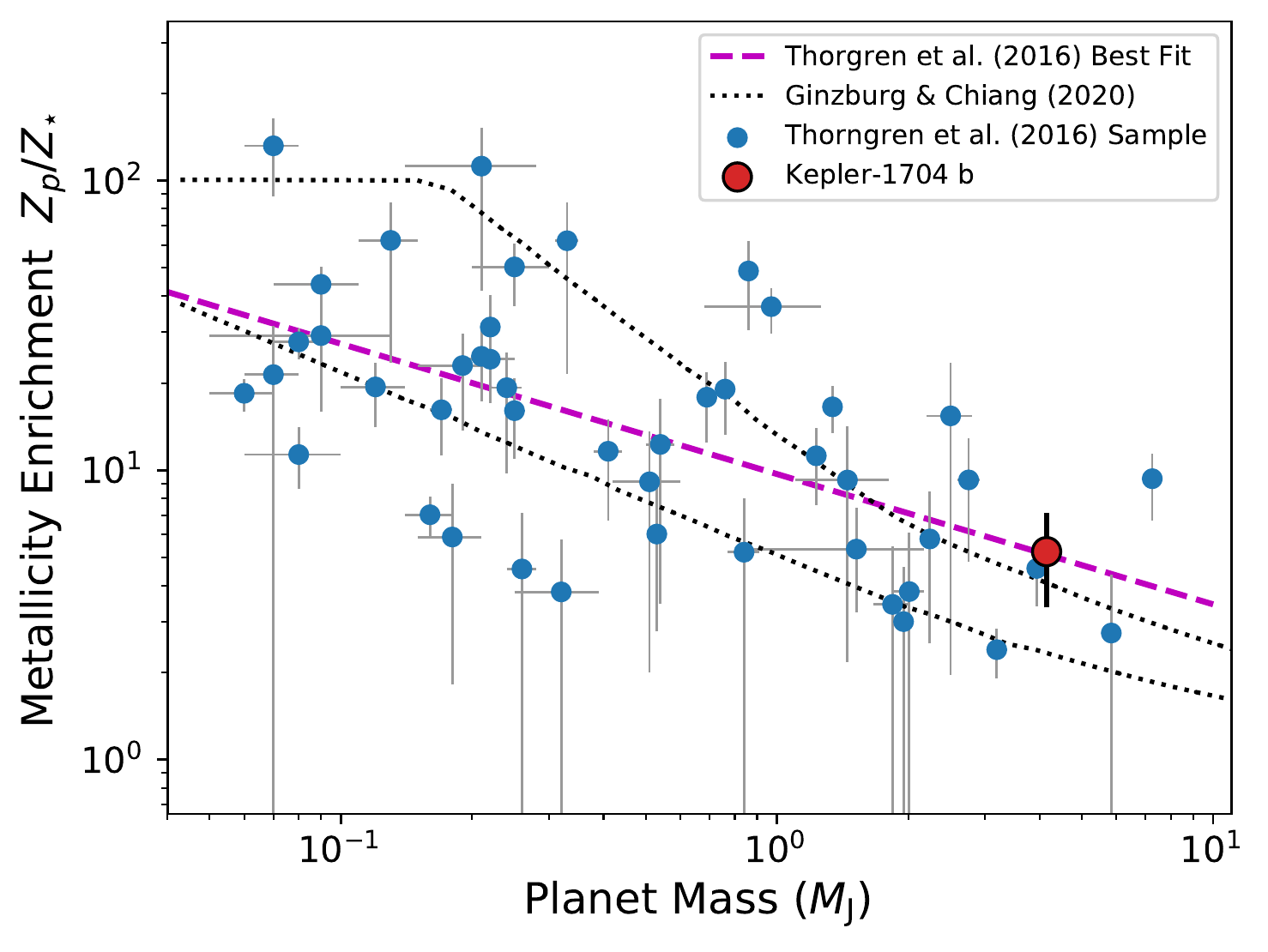} 
    \end{tabular}
    \caption{\textit{Left:} Heavy element mass of the weakly irradiated giant exoplanets from \citet{Thorngren2016} as well as \planet. \textit{Right:} Metallicity enrichment of the weakly irradiated giant exoplanets from \citet{Thorngren2016} as well as \planet. The dotted black lines show the scatter that can be accounted for by concurrent gas accretion and mergers assuming a critical core mass of 10~$M_{\earth}$ at the onset of runaway gas accretion \citep{Ginzburg2020}. The position of \planet\ in these panels is consistent with substantial late-stage accretion of heavy elements or core growth through mergers during gas accretion.}
    \label{fig:zpzs}
\end{figure*}

We place the bulk metal mass and metallicity enrichment in context of other cool ($T_{\rm eq} \lesssim 1000$~K), weakly irradiated ($\langle F \rangle < 2\times10^8$~erg~s$^{-1}$~cm$^{-2}$) giant exoplanets from the \citet{Thorngren2016} sample\footnote{We exclude Kepler-75~b in all related figures and analyses since \citet{Thorngren2016} only derived an upper limit on its metal mass.} in Figure~\ref{fig:zpzs}. By metal mass and enrichment, \planet\ is entirely consistent with the known trends. \planet\ contains more metal mass than its lower-mass counterparts, but it is broadly less enriched in metals relative to its host star. These findings are consistent with the theory of core accretion as its formation scenario, followed by a period of late-stage heavy element accretion \citep[e.g.,][]{Mousis2009,Mordasini2014}. \planet\ is similar to the other high-mass ($M_p \gtrsim 2$~$M_{\rm J}$) giant planets in that it orbits a metal-rich star, something that has been predicted by population synthesis models \citep[e.g.,][]{Mordasini2012} and likely relates to the correlation between host star metallicity and giant planet occurrence \citep[e.g.,][]{Gonzalez1997,Santos2004,Fischer2005}. 

In Figure~\ref{fig:zpzs} (right panel), we also include a prediction from \citet{Ginzburg2020}. They model concurrent gas accretion and mergers during giant planet formation as an alternate means of explaining the heavy metal content of giant planets. The scatter in the data enclosed by the dotted black lines can be explained by the intrinsically chaotic nature of mergers, even if all systems evolve from nearly identical conditions as quantified by a critical core mass of 10~$M_{\earth}$. We find that the mass and bulk metallicity enrichment of \planet\ are also consistent with the theory of concurrent gas accretion and mergers.

\begin{figure}
    \centering
    \includegraphics[width=\columnwidth]{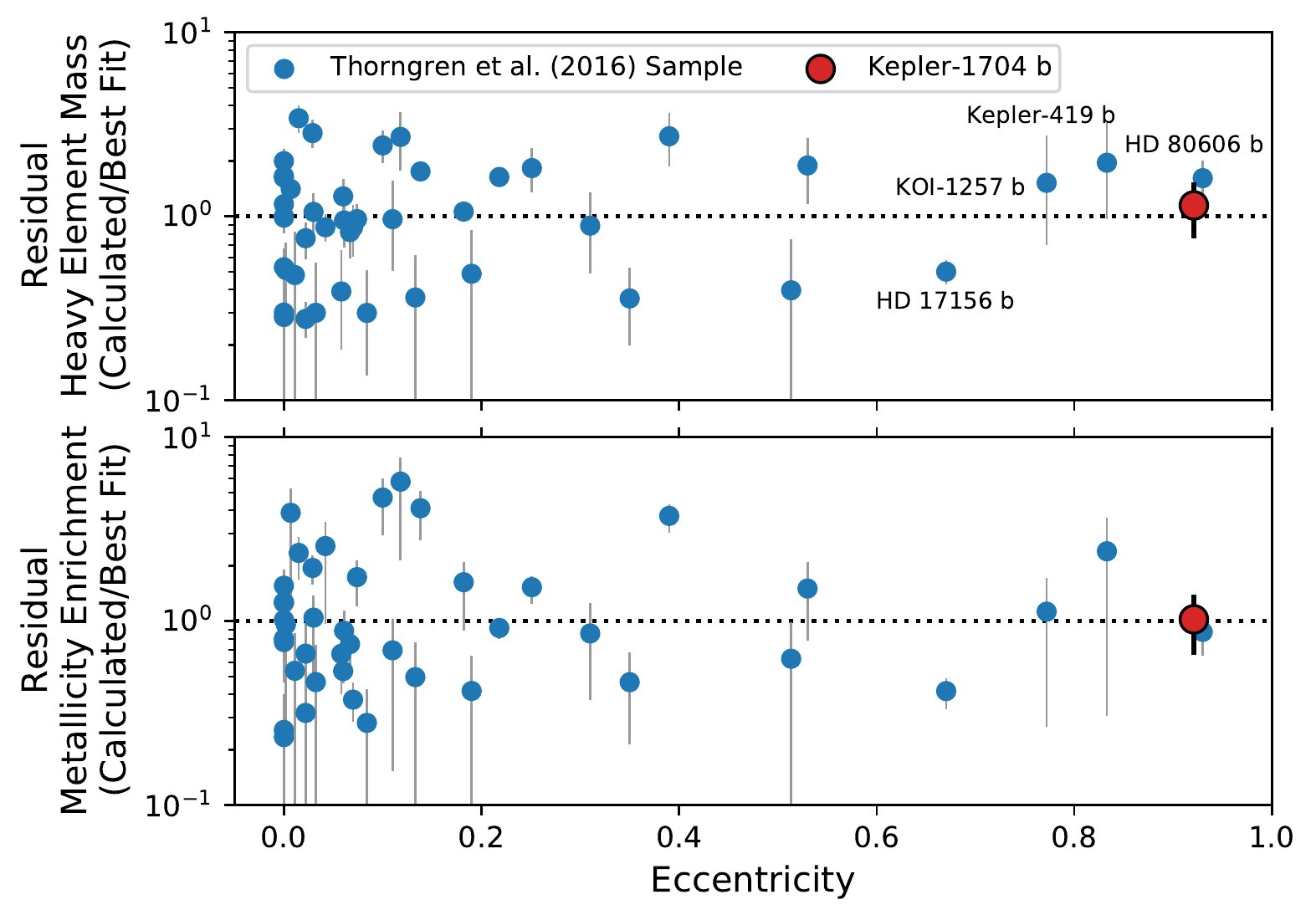}
    \caption{Relative residuals (calculated/best fit) of the heavy element mass (top) and the metallicity enrichment relative to stellar (bottom) as a function of eccentricity for the \citet{Thorngren2016} sample of weakly irradiated giant exoplanets and \planet.}
    \label{fig:ecc_resid}
\end{figure}

It is interesting to consider how trends in heavy element mass, metal enrichment, and total planet mass relate to other orbital and stellar properties. In Figure~\ref{fig:ecc_resid}, we show the relative residuals (calculated/best fit) of heavy element mass and metallicity enrichment relative to stellar as a function of eccentricity for the \citet{Thorngren2016} sample of weakly irradiated giant exoplanets and \planet. As noted by \citet{Thorngren2016}, there is no discernible trend in either quantity. However, given how sparsely populated the high-eccentricity region is, it is worthwhile to consider the (now) five systems with $e>0.6$ individually. The residual heavy element mass and metallicity enrichment of \planet\ and HD~80606~b are nearly identical, as are their orbital eccentricity and planet mass. However, HD~80606~b likely migrated via secular perturbations with HD~80607 \citep[e.g.,][]{Wu2003,Fabrycky2007,Winn2009} whereas we are unsure if similar interactions with a planetary or stellar companion have influenced the migration history of \planet. If \planet\ and HD~80606~b followed different migration pathways, there is no evidence in their bulk metallicity to distinguish them. The residual heavy element mass and the metallicity enrichment of these two planets are significantly different than those of HD~17156~b, which has $e\approx 0.67$ \citep{Fischer2007,Bonomo2017}. Unlike HD~80606~b, HD~17156~b has no stellar companion and its orbit is nearly aligned with its host star \citep{Cochran2008,Narita2008,Barbieri2009}. Also, HD~17156~b's orbital period is almost two orders of magnitude shorter than that of \planet. Therefore, it is perhaps not surprising that these planets experienced different formation histories that could account for the metallicity differences. The final two high-eccentricity planets in Figure~\ref{fig:ecc_resid} (KOI-1257~b and Kepler-419~b) have relatively imprecise residual heavy element masses and metallicity enrichments. KOI-1257~b is thought to be in a binary star system, possibly pointing to Kozai migration \citep{Santerne2014}. On the other hand, Kepler-419~b is joined by a massive outer giant planet that has a low mutual inclination, such that Kozai migration is likely not a viable migration theory \citep{Dawson2014}. The overall lack of a clear trend between heavy element mass or metallicity enrichment and the presence of a companion and/or high stellar obliquity is likely in part a result of the small number of data points. However, it could also suggest that the heavy element accretion occurs before or independently from the various channels of eccentricity excitement. 

\subsection{Atmospheric Characterization Prospects for \planet}

The bulk heavy element mass retrieval suggested that {$\sim$150} Earth-masses of metals should exist within \planet. The distribution of those heavy elements within the planet can possibly affect the composition of its atmosphere. Specifically, \citet{Thorngren2019a} showed that the bulk metallicity places an upper limit on atmospheric metallicity. For \planet, the core-free $2\sigma$ upper limit (${Z_p = 0.2}$) for atmospheric metallicity is ${35.7}\times$ solar. A natural next step in the exploration of \planet\ is to test this prediction via atmospheric characterization.

Considering only orbital period or semi-major axis, \planet\ is a rare opportunity for transmission spectroscopy \citep{Seager2000}. Although, long-period exoplanets pose specific challenges to this kind of technique. Not only are transits of such planets geometrically rare, but their timing is often uncertain. Since only two transits of \planet\ have been observed, the presence of extreme transit timing variations \citep[TTVs;][]{Wang2015b} cannot be ruled out \citep[e.g.,][see Section~\ref{sec:discTTV}]{Dalba2016,Dalba2019c}. Furthermore, atmospheric temperature will (to first order) decrease with increasing orbital distance. As a result, atmospheres will be cooler and scale heights and transmission spectrum features will be smaller. Surprisingly, this can be balanced by low surface gravity, as would be the case if Saturn was subject to transmission spectroscopy \citep{Dalba2015}. The transiting geometry of the long-period \planet\ also makes it a unique candidate for testing theories of atmospheric refraction \citep[e.g.,][]{Sidis2010,Dalba2017b,Alp2018} that have not yet been observationally tested \citep{Sheets2018}. 

However, considering the large radius of the subgiant \host\ and the high mass of \planet, this system is a challenging target for transmission spectroscopy. With a surface gravity of {86~m~s$^{-2}$} and the average equilibrium temperature of {254~K} from Table~\ref{tab:planet} (assuming no albedo), the atmospheric scale height is only {$\sim$12~km}, which corresponds to {1}~part-per-million (ppm) in the transmission spectrum. The out-of-transit stellar mirage caused by refraction also scales with the atmospheric scale height, making such a detection similarly difficult \citep[e.g.,][]{Dalba2017b}. 

On the other hand, we used Equation~\ref{eq:teq} to estimate that $T_{\rm eq}$ at periastron, which is within several days of transit, is {$\sim$900~K}. This suggests a {3.6$\times$} increase in the atmospheric scale height and transmission spectrum feature size. Although {4~ppm} is still beyond the reach of current and future facilities, we caution that our intuition for predicting favorable transmission spectroscopy targets is largely based on our current understanding of hot, close-in exoplanet atmospheres. This possibly warrants skepticism. If Saturn were a transiting exoplanet, its warm stratosphere and active photochemistry would produce a $\sim$90~ppm absorption feature in its transmission spectrum at 3.4~$\mu$m \citep{Dalba2015}. Considering only Saturn's $T_{\rm eq}$ as defined in Equation~\ref{eq:teq} would under predict its amenability to transmission spectroscopy. Other long-period giant exoplanets may prove surprising as well. 

Even if transmission spectroscopy is not a viable atmospheric characterization technique, the {0.16~au} periastron distance of \planet\ caused by its extreme eccentricity possible qualifies it for an IR phase curve analysis.

\subsubsection{IR Phase Curve Analysis}\label{sec:atm}

\begin{figure*}
    \centering
    \includegraphics[width=0.9\textwidth]{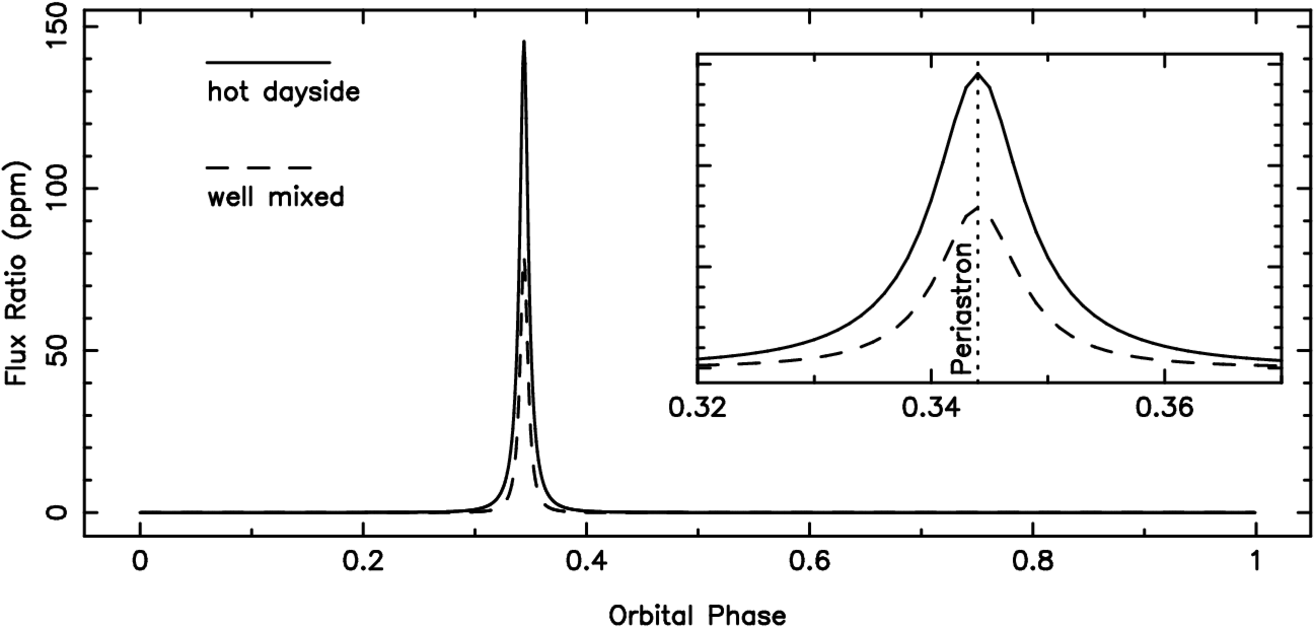}
    \caption{Simulated 4.5~$\mu$m phase curve of \planet\ following \citet{Kane2011e}. The ``hot dayside'' and ``well mixed'' models correspond to atmospheric heat redistribution efficiencies of 0 and 1, respectively. The $\sim$100~ppm amplitude of this variation is favorable for \jwst\ observation. This simulation assumed a pseudo-synchronous rotation of \planet. If the planet's rotation is not synchronized, an oscillation in flux at the frequency of the planets effective rotation rate may also be detectable.}
    \label{fig:phase}
\end{figure*}

To predict the expected thermal signature of the planet during periastron passage, we calculated the IR phase curve for \planet\ during one complete orbital period. These calculations followed the methodology of \citet{Kane2011e} using the stellar and planetary parameters provided in Tables~\ref{tab:stellar} and \ref{tab:planet}, respectively. We assumed a passband of 4.5~$\mu$m, a Bond albedo of zero, and we calculated the flux ratio of planet to star using the ``hot dayside'' and ``well mixed'' models. These models represent the extremes of heat redistributions as they assume re-radiated energy over 2$\pi$ and 4$\pi$~sr, respectively. The full IR phase curve for both models are shown in Figure~\ref{fig:phase}, along with a zoomed panel that shows the location of the periastron passage. 

There are several caveats to this calculation. We assumed an instantaneous response of the planetary absorption and IR emission, whereas the radiative and advective time scales will determine the nature of phase lags in thermal emission profiles \citep{Langton2008,Cowan2011b}. This, combined with the blackbody emission and zero albedo assumptions, means that the calculations presented in Figure~\ref{fig:phase} may be considered as an upper limit on the expected IR emission. Furthermore, the variation in temperature would also alter the atmospheric composition. Some of the energy would be converted into latent heat to dissociate larger molecules or particulates. There would also be an interconversion between CO and CH$_4$ \citep[e.g.,][]{Visscher2012}. The timescale of this reaction, and also the vertical mixing timescale, should be considered to produce a more accurate model of the phase curve. We leave these considerations for a future work and instead derive a first-order, upper limit on the phase curve emission.

Despite the various assumptions that apply to this phase curve modeling, the order of magnitude ($\mathcal{O}$(10$^2$)~ppm) of the thermal flux increase is likely accurate. Several instruments onboard \jwst\ will have sensitivity in the near- to thermal-IR and, based on preliminary noise floor expectations \citep{Greene2016}, should be capable of detecting the \planet\ phase variation. Borrowing from solar system intuition, 4--6~$\mu$m is likely a promising wavelength for such an observation. Jupiter and Saturn's atmospheres have low opacity in this wavelength region that exists between bands of methane and phosphine where radiation from 5--8~bars can escape \citep{Irwin2014}. Jupiter's radiance near 5~$\mu$m even exceeds that at mid-IR wavelengths \citep[e.g.,][]{Irwin1998,Fletcher2009b,Irwin2014}. 

The periastron passage of \planet\ occurs over $\sim$10~days and includes the transit. Low cadence time series observations from the F444W filter of NIRCam, for example, could detect the peak flux ratio and the width of the feature assuming that the visit-to-visit photometric variability does not overwhelm the astrophysical signal. Including at least one high cadence, longer visit at or following periastron would also be valuable because the phase curve may exhibit ``ringing'' as the hot spot from periastron rotates in and out of view \citep[e.g.,][]{Cowan2011a}. This effect is not featured in our simulation of \planet, which assumed pseudo-synchronous rotation \citep{Hut1981}. However, given the inefficiency of tides at the periastron distance of \planet, this assumption may be an oversimplification. The detection of a ringing oscillation in the IR phase curve would test this assumption and possibly directly yield the effective planetary rotation rate.

The detectability of the thermal phase curve for \planet\ should be explored in more detail with an atmospheric structure code \citep[e.g.,][]{Mayorga2021}. Such an effort is beyond the scope of this paper and should likely wait until \jwst\ is launched and commissioned. In addition to a broadband detection of phase variability, the prospects for spectroscopic detection should also be investigated. Transmission spectroscopy may not be an effective tool to measure atmospheric composition (e.g., metallicity), so any any other possible method would be extremely useful. Atmospheric metal enrichment (relative to stellar) is a specifically valuable property to measure because it can be compared to the planet's bulk metallicity enhancement (Section~\ref{sec:zpzs}). One prediction would be an atmospheric metallicity less than the bulk metallicity if some heavy elements comprise a planetary core or there is otherwise an increasing gradient in metals with depth. However, for Jupiter and Saturn, recent high-precision gravity data and well established atmospheric composition results suggest a more complicated picture \citep{Niemann1998,Wong2004,Fletcher2009a,Wahl2017,Guillot2018,Iess2019,Muller2020a}. More elaborate theories including inverse compositional gradients \citep[e.g.,][]{Debras2019} are needed to explain Jupiter and Saturn and could possibly be refined through atmospheric characterization of exoplanets like \planet.

Based on the optimistic prospect of \jwst\ observations, we determined the timing of transits and periastron passages of \planet\ occurring in the next 10 years (Table~\ref{tab:transits}). For each event, we checked for visibility from \jwst\ using the General Target Visibility Tool\footnote{Accessed 2021 February 11 (\url{https://github.com/spacetelescope/jwst_gtvt}).}. This tool only predicts visibility through the end of 2023, but we assumed the same visibility of \host\ in later years. The 2023 transit of \planet\ will not be visible to \jwst\ but the 2028 transit will be visible. The 2025 transit will occur within 24~hr after the visibility window closes and the 2031 transit will occur roughly six days after the visibility window opens. If the solar avoidance restrictions change after launch, these transits may or may not be visible to \jwst. The periastron passage of \planet\ occurs several hours before transit, so its visibility is similar. However, as shown in Figure~\ref{fig:phase}, the peak of the thermal flux ratio spans $\sim$10~days. Even if the exact moment of periastron is (or is not visible), some portion of the event is expected to be visible to \jwst. 

\begin{deluxetable*}{cccccl}
\tablecaption{Future Transit and Periastron Timing Predictions \label{tab:transits}}
\tablehead{
  \colhead{} & 
  \multicolumn{2}{c}{\underline{Conjunction (Transit) Time}} &
  \multicolumn{2}{c}{\underline{Periastron Time}} &
  \colhead{\jwst} \\
  \colhead{Epoch$^a$} & 
  \colhead{~~~~~~~BJD$_{\rm TDB}$~~~~~~~~} &
  \colhead{UTC} &
  \colhead{~~~~~~~BJD$_{\rm TDB}$~~~~~~~~} &
  \colhead{UTC} & 
  \colhead{Visibility$^b$}}
  \startdata
    5 & $2460016.6290\pm$0.0046 & 2023-03-13 03:06 & $2460016.29\pm0.19$ & 2023-03-12 18:51 & None \\
    6 & $2461005.5109\pm$0.0055 & 2025-11-26 00:15 & $2461005.17\pm0.19$ & 2025-11-25 16:00 & Partial\\
    7 & $2461994.3912\pm$0.0064 & 2028-08-10 21:23 & $2461994.05\pm0.19$ & 2028-08-10 13:09 & Full\\
    8 & $2462983.2723\pm$0.0073 & 2031-04-26 18:32 & $2462982.93\pm0.19$ & 2031-04-26 10:18 & Partial\\
  \enddata
\tablenotetext{}{The times listed here do not account for possible uncertainty owing to yet undiscovered TTVs (see Section~\ref{sec:discTTV}).}
\tablenotetext{a}{Epoch=0 is defined as the first transit observed by the \kepler\ spacecraft.}
\tablenotetext{b}{\jwst\ visibility after 2023 December 31 is based on previous years' visibility. Epochs for which the full periastron passage of \planet\ partially falls outside of the predicted visibility windows are labeled as ``Partial'' (see the text).} 
\end{deluxetable*}

\subsubsection{Radio Emission}

Unlike for transmission spectroscopy, the relatively high mass of \planet\ is beneficial to attempts to measure planetary radio emission. \citet{Lazio2010} searched for radio emission from HD~80606~b during a periastron passage but measured only an upper limit. That experiment was based on the expectation that the variation in planet--star distance over an eccentric orbit would lead to dramatic increase in magnetospheric emission. Assuming that luminosity scales with the planet--star distance as $L \propto d^{-1.6}$ \citep[e.g.,][]{Farrell1999}, then the factor of 24.3 change in distance for \planet\ would produce a 165x increase in luminosity. While this is slightly smaller than the 200x increase expected for HD~80606~b, a future radio search may be aided by the fact that \planet\ can possibly emit at higher frequencies. We estimate that the upper limit emission frequency as determined by the local plasma frequency in the emission region for \planet\ is 
\begin{equation}\label{eq:radio}
    \nu = 24 \; {\rm MHz} \left ( \frac{\Omega}{\Omega_{\rm J}} \right ) \left ( \frac{M_p}{M_{\rm J}} \right )^{5/3} \left ( \frac{R_p}{R_{\rm J}} \right )^3 
\end{equation}

\noindent where $\Omega$ is the angular rotation rate \citep{Farrell1999,Lazio2004,Lazio2010}. In this Equation, all values are scaled to those of Jupiter. For HD~80606~b, tidal forces are expected to force the planet into pseudo-synchronous rotation with a period of 39.9~hr \citep{Hut1981,Lazio2010}. It is unlikely that this would also apply to \planet, for which the larger periastron distance renders tides inefficient. Therefore, the assumption of a Jupiter-like rotation period ($\sim$9.9~hr) is reasonable. In that case, evaluating Equation~\ref{eq:radio} gives {310~MHz}. \citet{Lazio2010} argued that this equation may actually under-predict the cutoff frequency of exoplanets, as it does for Jupiter, and suggested that the upper limit may be 60\% larger. In that case, the cutoff frequency for \planet\ would be {497~MHz}, which is more accessible to existing radio observatories than HD~80606~b's 55--90~MHz.

A full simulation of the potential for radio emission from \planet\ is beyond the scope of this paper, and the ability to make such a detection, at least relative to previous attempts for HD~80606~b, will likely be hindered somewhat by the greater distance to the \host\ system. However, even our approximate calculation suggests that \planet\ is one of the best systems to investigate magnetospheric response to a rapidly changing planet--star distance. Such an observation stands to extend the study of giant exoplanet magnetic fields beyond the inner most hot Jupiters \citep[e.g.,][]{Cauley2019} and explore magnetic field generation in planets akin to Jupiter and Saturn.


\section{Discussion}\label{sec:disc}

\begin{figure}
    \centering
    \includegraphics[width=\columnwidth]{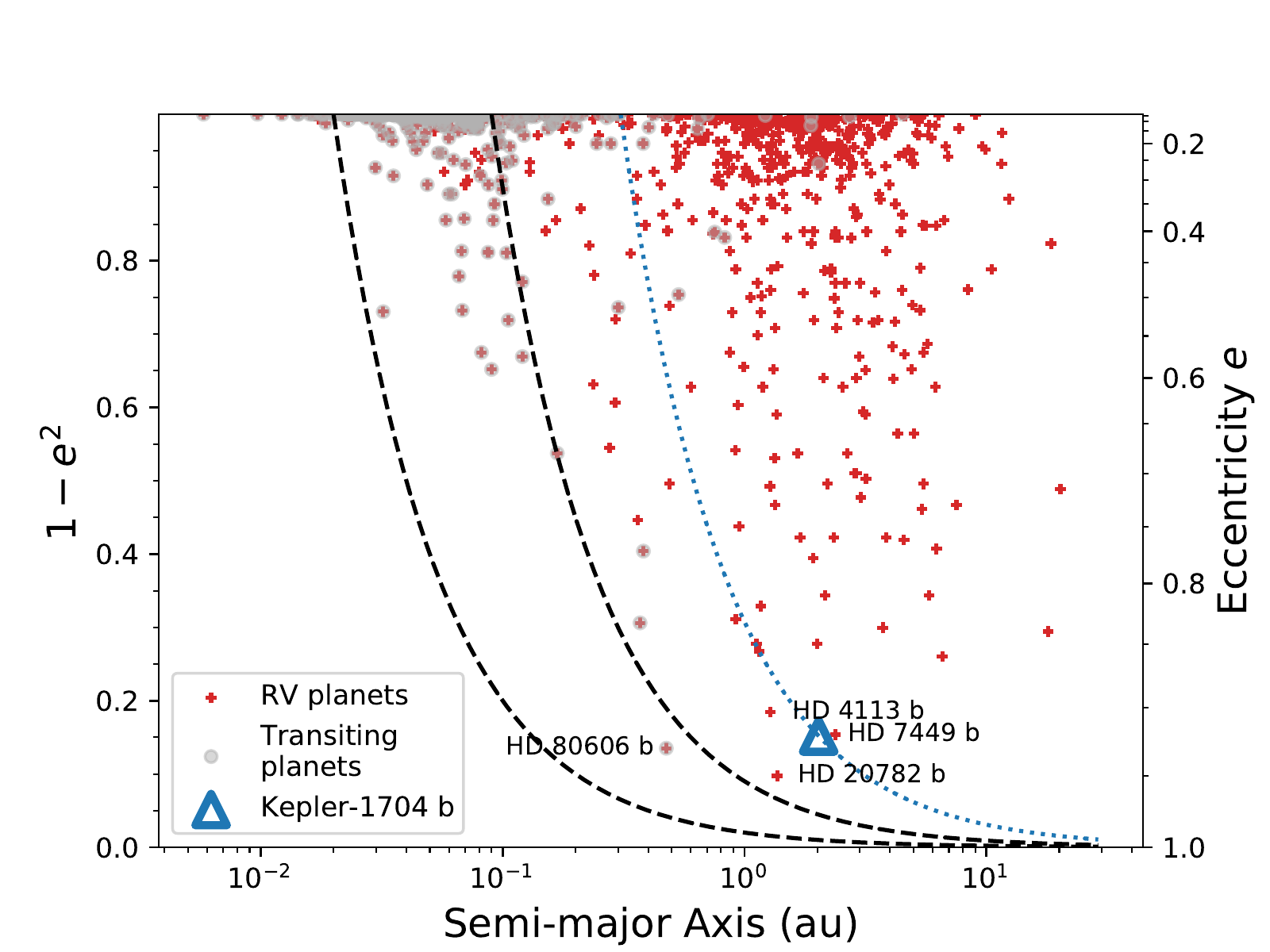}
    \caption{The eccentricity for all non-controversial exoplanets with known $a$ (or with the necessary parameters to calculate $a$) and with (minimum) mass greater than 0.3~$M_{\rm J}$ as listed in the NASA Exoplanet Archive (accessed 2021 February 17). The marker style indicates whether a planet has been detected by transits and/or RVs. The dashed black lines indicate tracks of constant angular momentum with final orbital periods of 1 and 10~days. The dotted blue line indicates the track for \planet.}
    \label{fig:ecc_vs_a}
\end{figure}

Much of the previous analysis has focused on key pieces of information that inform the formation and migration history of \planet. Orbital period and eccentricity are two of the most notable properties in this respect. As shown in Figure~\ref{fig:ecc_vs_a}, these properties place \planet\ among a small group of known exoplanets on long-period, highly eccentric orbits that are useful for testing the extremes of planetary formation theories. More remarkable, though, is the transiting geometry of the orbit of \planet. Relative to other transiting exoplanets, the position of \planet\ in $a$--$e$ space is unrivaled (Figure~\ref{fig:ecc_vs_a}). \planet\ thereby offers its radius and bulk composition, as well as its orbital properties, as clues to its formation and migration history.

Here, we assemble all of this information into a coherent narrative describing the history of this interesting planet.

\subsection{\planet: The Failed Hot Jupiter}\label{sec:failedHJ}

Based solely on the measured orbital eccentricity, we discard disk-migration as the explanation for the orbital properties of \planet. \citet{Papaloizou2001} showed that eccentricities up to $\sim$0.25 could be achieved through disk interactions for a variety of planet masses. Although, eccentricity is generally damped by the disk for giant planets with $M_p < 5$~$M_{\rm J}$ \citep{Bitsch2013}. Recent work revisiting disk cavity migration argued for eccentricities up to 0.4 for giant planets \citep{Debras2021}, which is possibly a viable theory for other outer giant planets like Kepler-1514~b ($e=0.401^{+0.013}_{-0.014}$), which also harbors an inner Earth-sized companion \citep{Dalba2021a}. Explaining the current orbit of \planet, however, requires excitation to high eccentricity by another body.

Through multiple analyses, we rule out stellar companions with mass greater than $\sim$700~$M_{\rm J}$ at most orbital separations in the \host\ system. As described in Section~\ref{sec:companions}, undetected less massive companions may still be present at a variety of separations and could have driven \planet\ to its high eccentricity through secular Kozai--Lidov perturbations \citep[e.g.,][]{Wu2003,Naoz2011}. Also, star-planet Kozai migration from a stellar companion that was present when \planet\ formed but subsequently lost due three-body interactions also remains a possible explanation. However, motivated by our nondetection of a companion and only a tentative detection of acceleration in $\sim$10~years of RV measurements, we discard secular perturbations as being the most likely explanation for the high eccentricity of \planet. We recommend that future dynamical simulations explore the areas of parameter space that we have not ruled out to see if a hidden companion could theoretically explain the properties of \planet\ \citep[c.f.,][]{Jackson2019}. 

This brings us to eccentricity excitation theories involving close, fast dynamical interactions. Specifically, could planet-planet scattering \citep[e.g.,][]{Rasio1996} provide an explanation for the eccentricity of \planet? Many aspects of the observed eccentricity distribution of giant exoplanets can be explained by planet-planet scattering \citep[e.g.,][]{Moorhead2005,Chatterjee2008,Raymond2010,Bitsch2020}, including planets with eccentricities above 0.99 \citep{Carrera2019}. For \planet, we find planet-planet scattering is consistent with its orbital properties, its host star, and its bulk interior properties. \host\ is metal-rich ({[Fe/H]~$=0.196\pm0.057$} from Table~\ref{tab:stellar}). \citet{Dawson2013} demonstrated that metal-rich stars tend to host high-eccentricity hot Jupiters, which they interpreted as evidence support HEM by planet-planet scattering owing to the well know correlation between stellar metallicity and giant planet occurrence \citep[e.g.,][]{Santos2004,Fischer2005}. Even though \planet\ is not a hot Jupiter, it is reasonable that it could have formed alongside other giant planets that were subsequently scattered. After many close encounters, possibly even tens of thousands \citep{Carrera2019}, \planet\ could have been driven to its current eccentricity. However, its final periastron distance was too far for tides to efficiently circularize the orbit, leaving \planet\ as the failed hot Jupiter that we have characterized here. 

Now, we consider this migration pathway in the context of the bulk heavy metal mass, which we found to be {$\sim$150~$M_{\earth}$}. This enrichment could have been acquired through late-stage accretion of planetesimals or core mergers with concurrent gas accretion (Section~\ref{sec:zpzs}). For the former explanation, \citet{Shibata2020} found that migrating giant planets can capture tens of Earths masses worth of planetesimals that may otherwise not be available \textit{in situ} \citep[e.g.,][]{Shibata2019}. The amount of heavy elements accreted scales with increasing migration distance and decreasing migration timescale, both of which are expected for eccentricity excitement through planet-planet scattering. However, \citet{Shibata2020} also found that the most enriched giant exoplanets, containing more than $\sim$100~$M_{\earth}$ of heavy elements likely require an additional source of enrichment. 

The latter explanation mentioned above for how \planet\ acquired its heavy metal mass involves the merger of cores during gas accretion \citep{Ginzburg2020}. The interaction of cores during this stage of giant planet formation, while gas is still present in the disk, is broadly consistent with the high eccentricity of \planet. \citet{Bitsch2020} found that such scattering events at this stage are common, and systems that start with more planetary embryos create giant planets with higher eccentricities so long as the damping rates for inclination and eccentricity are slow. Indeed, slow rates are required to reproduce the eccentricity distribution of the known giant planets \citep{Bitsch2020}. It is therefore possible that the same processes that led to the accretion of heavy elements for \planet\ also contributed to exciting its eccentricity. In reality, owing to the fact that planet mergers (or collisions) are less efficient at producing high eccentricities than scattering events \citep[e.g.,][]{Ford2008b,Juric2008,Anderson2020}, some combination of the aforementioned theories along with planet-planet scattering after the dispersal of the gas disk likely produced the \host\ system as seen today.

Moving forward, it would be useful to compare this proposed formation history to other well characterized long-period transiting giant planets. It will be particularly interesting to compare the bulk interior properties of giant planets in systems with and without outer companions that could have induced Kozai migration. If bulk heavy element composition and migration mechanisms are linked, as may be the case for \planet, we might expect to find a correlation between interior properties, orbital properties, and the existence of companions.

\subsection{Stellar Obliquity}\label{sec:discRM}

A critical missing piece in our discussion of the migration of \planet\ is the stellar obliquity. A substantially misaligned orbit of \planet\ would warrant a reexamination of Kozai eccentricity oscillations, although planet-planet scattering can also cause misaligned orbits \citep[e.g.,][]{Chatterjee2008,Naoz2012}. Moreover, the effective temperature of \host\ makes this system a perfect laboratory for testing the theory that hot Jupiters preferentially realign cool stars \citep[e.g.,][]{Winn2010a,Schlaufman2010}. The effective temperature of \host\ is {$5745^{+88}_{-89}$~K}, which is well below the $\sim$6200~K Kraft break \citep{Kraft1967} that has been implicated by hot Jupiter obliquity observations. Since tidal forces are inefficient for failed hot Jupiters like \planet, we would expect that these planets would show a variety of obliquities and would not be preferentially aligned like hot Jupiters orbiting similarly cool stars. 

In theory, the obliquity between \planet\ and its host star could be measured through the Rossiter--McLaughlin (RM) effect \citep{Rossiter1924,McLaughlin1924}. If successful, it would stand as the longest-period planet, by far, to have an obliquity measurement. In practice, an RM experiment will be challenging. Our \textsf{SpecMatch} analysis (Section~\ref{sec:hires}) inferred a low stellar rotational velocity of 2.74$\pm$1.0~km~s$^{-1}$. By Equation~40 of \citet{Winn2010b}, the maximum expected amplitude of the RM effect is only 11~m~s$^{-1}$. Assuming 30~minute exposure times (as used for the current RV data), this would only allow 12 data points across the entire transit. With the $\sim$7~m~s$^{-1}$ RV precision achieved using the best-match template (see Table~\ref{tab:rvs} and Section~\ref{sec:hires}), any detection of obliquity would likely be marginal. We recommend that any future effort to observe the spectroscopic transit of \planet\ should first acquire a high-S/N spectral template of \host\ to reduce the internal RV precision by several m~s$^{-1}$. Owing to the extreme eccentricity and the argument of periastron, the transit duration is short enough that a fortunately timed transit could be observed from a single site. For the Keck~I telescope, only the second half of the 2023 transit (Table~\ref{tab:transits}) is visible. \host\ will rise above the Nasmyth deck at Keck~I at a favorable airmass of $\sim$1.5 around 14:30~UTC on 2023 March 12. Again assuming 30~minute exposure times, that would place roughly 6 data points across the second half of RM signal. Even with the actual template and improved internal precision, a detection of obliquity would likely be moderate at best. It is not until 2028 that the Keck~I telescope has the optimal position for an RM detection. The mid-transit time of the August 2028 transit is almost perfect timed with \host\ crossing the meridian, and the full transit (plus post-transit baseline) is observable. However, in the coming years, new precise RV facilities with the capability of achieving few m~s$^{-1}$ precision on faint ($V=13.4$) stars such as MAROON-X \citep{Seifahrt2018} or the Keck Planet Finder \citep{Gibson2016} should consider conducting RM measurements of long-period \kepler\ planets like \planet.

\subsection{A Third Transit of \planet\ to Explore TTVs}\label{sec:discTTV}

With only two transit events detected by \kepler, we cannot rule out large TTVs that could possibly preclude future transit observations \citep[e.g.,][]{Wang2015b}. TTVs of this nature would require a massive perturber with an orbit that is sufficiently close to \planet\ to allow for gravitational interaction. The RV observations presented here largely rule out such a companion on orbits interior to \planet\ (Figure~\ref{fig:rvsearch}). However, giant planet companions on wider orbits could be present. One avenue of future work would be to apply the companion limits and stability results described here to a dynamical analysis of the \host\ system to set limits on TTV magnitudes. 

Another avenue toward addressing this issue would be to observe a third transit. As has been done with Kepler-421~b \citep{Dalba2016}, Kepler-167~e \citep{Dalba2019c}, and HIP~41378~f \citep{Bryant2021}, \host\ can be observed for a window of time surrounding the expected third transit according to a linear ephemeris. For this experiment, a missed transit places a lower limit on transit-to-transit timing variations. The ground-based observability of future \planet\ transits is not as restrictive as that described for \jwst\ in Section~\ref{sec:atm}. For example, in March 2023, \host\ rises to reasonable elevations in the last few hours of the night in the Northern hemisphere. Any single site will likely struggle to detect the full transit. However, detections of ingress and egress from multiple sites spread out in longitude would refine the ephemeris of \planet\ and identify any TTVs.

\subsection{Comparison to the Kepler-167 System}

\host\ represents an interesting comparison for the Kepler-167 system, in which an early K dwarf star hosts three inner super-Earth-sized planets and an outer transiting Jupiter-analog on a $P=1071$~day orbit \citep{Kipping2016b}. Although the mass of Kepler-167~e---the outer giant planet---has not been measured, its orbital eccentricity has been constrained to $\sim0.06$ by the transit shape and duration. This low eccentricity combined with the presence of multiple inner super-Earth planets suggests that the migration mechanism for Kepler-167~e was likely gentle and driven by interactions with the disk. Kepler-167 is solar metallicity, if not slightly metal poor, so it is possible that Kepler-167~e was the only giant planet formed in the outer disk, so scattering events never occurred. \citet{Dalba2019c} ruled out the existence of TTVs in the ephemeris of Kepler-167~e, which further implied a lack of an outer massive companion. A mass and bulk metallicity measurement for Kepler-167~e would provide an interesting comparison with \planet, which likely experienced dynamical interactions with other bodies during and/or after its formation. 

\subsection{Could \planet\ Host Exomoons?}

Giant transiting exoplanets with multi-year orbital periods are possibly exciting targets for dedicated exomoons searches \citep[e.g.,][]{Kipping2012b,Heller2014b,Teachey2018b}. Now that we have measured the mass and orbital properties of \planet, the plausibility of this planet hosting a system of exomoons should be investigated in more detail. Given the suspected active dynamical formation history of \planet, its ability to have maintained a system of exomoons is perhaps questionable. Indeed, the investigation of exomoon stability under tidal forces \citep[e.g.,][]{Barnes2002,Adams2016,Sucerguia2020}, planet-planet scattering \citep[e.g.,][]{Nesvorny2007,Gong2013,Hong2018}, disk torques \citep[e.g.,][]{Namouni2010,Spalding2016}, and secular migration owing to a stellar companion \citep[e.g.,][]{Martinez2019,Trani2020} are active areas of theoretical research. Although any such study is beyond the scope of this work, we can approximate the Hill radius of \planet\ at periastron (where it is smallest):
\begin{equation}\label{eq:hill}
r_{\rm H, peri} \approx a (1 - e) \left (\frac{M_p}{3M_{\star}} \right )^{1/3} \; .    
\end{equation}
For \planet, we find that {$r_{\rm H, peri} \approx 2.6\times10^6$~km}. Any exomoon on a circular prograde orbit around \planet\ would need a semi-major axis less than (roughly) half of this value to survive the close periastron passages \citep{Hamilton1991}. For perspective, the semi-major axis of Callisto, Jupiter's most distant Galilean moon, is roughly 1.9$\times$10$^6$~km. Of course, this calculation neglects all of the processes that led to \planet\ reaching its current orbital configuration. We offer \planet\ as a potentially interesting case study for more detailed investigations of exomoon formation and stability in the future. 

The fact that \planet\ swings through its host star's habitable zone on its eccentric orbit is also potentially interesting from an exomoon standpoint \citep[e.g.,][]{Heller2012,Heller2013,Hill2018}. However, the plausibility of life developing on an exomoon that experiences such intense variation in stellar irradiation should be thoroughly scrutinized.

\subsection{One Path Forward for Giant, Long-period Transiting Exoplanets}

The vast majority of known giant planets on au-scale orbits have unknown radii because they either do not transit or they are not known to transit. Without a radius, subsequent investigations of atmospheres and interiors are uncertain if not altogether impossible. Measuring the masses of the modest sample of known transiting giant planets with au-scale orbits and discovering more such planets will be important to advancing our understanding of giant planet formation and migration. These discoveries will also drive new theoretical advances in giant planet interiors, which are needed given that changing model assumptions can substantially alter our conclusions about the interior structures of giant exoplanets \citep[e.g.,][]{Muller2020b}.   

Only a handful of outer giant exoplanets like \planet\ exist within the \kepler\ sample and they all orbit relatively faint stars. This creates two problems. Firstly, their limited number means that unfortunate transit timing (see Table~\ref{tab:transits} and also \citet{Dalba2019c}) can drastically slow progress to obtain new observations and advance our theoretical understanding. Secondly, their faintness must be overcome (if at all possible) by larger investments of highly competitive telescope time.

The Transiting Exoplanet Survey Satellite \citep[\tess;][]{Ricker2015}, which is actively searching for transits of bright stars around the entire sky, presents solutions to both problems. The only drawback is the tendency of \tess's observing strategy to yield single transit events for most planets with orbital periods greater than a couple dozen days \citep[e.g.,][]{Gill2020b,Dalba2020a,Diaz2020,Lendl2020}. If the \kepler\ mission had adopted the \tess\ mission's observing strategy, not only would \planet\ have been identified through a single transit, but its 6~hr transit duration could have easily been misconstrued as corresponding to a relatively short orbital period. This suggests \citep[and more quantitative efforts have shown;][]{Cooke2018,Villanueva2019} that given enough time and targets, \tess\ will identify transits from a unprecedented sample of long-period giant planets. Yet, the advancement of giant planet theory and understanding will rely on continued challenging follow-up efforts to characterize these planets masses, orbits, interiors, and atmospheres.


\section{Summary}\label{sec:concl}

We obtained nearly 10~years of RV observations of the $\sim5750$~K subgiant star \host, which was found to host a transiting giant planet candidate (\koi, now \planet) by the primary \kepler\ mission. Our observations and analyses confirmed the genuine nature of this exoplanet, now known as \planet, which is a {4.15}-Jupiter-mass planet on a {988.88}~day orbit with an extreme {0.921$^{+0.010}_{-0.015}$} eccentricity. We performed an AO imaging analysis, interior and atmosphere modeling, and dynamical simulations to characterize this system and make predictions for future observations. The primary results of this work are as follows. 

\begin{enumerate}
    \item We collected 14 RV measurements (Table~\ref{tab:rvs}) of \host\ from Keck-HIRES spanning 9.6~years that confirm the {988.88~day} orbital period for \planet, thereby ruling out the possibility of a third transit occurring in a \kepler\ data gap (Section~\ref{sec:kepler}). The RVs also confirmed the extremely high orbital eccentricity ({$e=0.921^{+0.010}_{-0.015}$}) that was suspected from our photoeccentric effect modeling (Section~\ref{sec:photo}) and measured the planet mass to be {$4.15\pm0.29$~$M_{\rm J}$}. \planet\ has the longest apastron distance ({3.9~au}) of any confirmed transiting exoplanet with a precisely known orbital period. Moreover, we found that between periastron and apastron, the equilibrium temperature of \planet\ varies from {$\sim$180~K} to {$\sim$900~K}.
    \item Archival AO imaging of \host\ from the PHARO instrument identified three possible stellar companions within $\sim$10$\arcsec$, two of which were previously published \citep{Wang2015a}. We found that two of the companions are spurious sources and the third is not gravitationally associated (Section~\ref{sec:imaging}). Additional archival AO imaging from the NIRC2 instrument \citep{Furlan2017a} yielded a nondetection of stellar companions within 2$\arcsec$ and placed upper limits on the mass of any undetected companion within 1,000~au of \host\ (Figure~\ref{fig:mass_limit}).
    \item The joint analysis of transit, RV, and broadband photometry (Section~\ref{sec:model}) identified a bimodality in stellar properties due to the evolutionary state of \host\ (Figure~\ref{fig:bimod}). We split the solutions based on stellar mass and publish the favored set of stellar and planetary parameters in Tables \ref{tab:stellar} and \ref{tab:planet}, respectively.
    \item We conducted three investigations of companions to \planet\ (Section~\ref{sec:companions}). Firstly, an injection-recovery analysis demonstrated that the RVs of \host\ are sensitive enough to have detected planetary companions within the orbit of \planet\ down to $\sim$100~$M_{\earth}$ and companions out to a few au with a few Jupiter masses (Figure~\ref{fig:rvsearch}). Secondly, we synthesized RV time series to determine the region of mass--semimajor-axis parameter space that is consistent with the subtle acceleration of \host. Although this analysis does not conclusively rule out any portion of companion parameter space, it identifies a preference for those with separations greater than $\sim$30~au (Figure~\ref{fig:trend_map}, right panel). Thirdly, we conducted a dynamical simulation using the MEGNO chaos indicator that failed to substantially rule out any other regions of parameter space for additional companions (Figure~\ref{fig:megno}). Based on these three analyses, we disfavor, although fail to entirely rule out, Kozai migration and secular chaos as the primary scenario to explain the orbital properties of \planet. 
    \item Using the mass and radius of \planet\ and the bimodal age of \host, we retrieved the bulk heavy element mass and metal enrichment relative to stellar for \planet\ (Figure~\ref{fig:metal_post}). This planet likely contains {$\sim$150~$M_{\earth}$} of heavy elements, making it enriched relative to \host\ by a factor of $\sim$5. These finding suggest that \planet\ is consistent with the mass--metallicity trends of \citet{Thorngren2016} and theories of core accretion with late-stage heavy element accretion (Figure~\ref{fig:zpzs}). \planet\ could have also acquired its heavy elements through core mergers during the gas accretion phase \citep{Ginzburg2020}. 
    \item Based on the aforementioned analyses, we hypothesized that \planet\ is a failed hot Jupiter \citep[e.g.,][]{Dawson2014} that reached its high eccentricity through planet-planet scattering events, but its periastron distance was too large for efficient tidal circularization (Section~\ref{sec:failedHJ}). We speculated that it may have ejected companions through these events. Furthermore, based on the stellar metallicity of \host\ and the bulk composition of \planet, the same processes that excited this planet's eccentricity may have also contributed to its heavy element accretion. 
    \item A critical missing piece of the discussion on the migration of \planet\ is the stellar obliquity (Section~\ref{sec:discRM}). Given the {5750~K} effective temperature of \host, this system can provide a valuable test of the theory that hot Jupiters preferentially align the spins of cool stars \citep[e.g.,][]{Winn2010a}. A detection of the RM effect for this system is feasible, however the timing of the future transits of \planet\ (Table~\ref{tab:transits}) will make this a challenging endeavour.
    \item Finally, we consider prospects for characterizing the atmosphere of \planet\ (Section~\ref{sec:atm}). While the large stellar radius and high planet mass may impede transmission spectroscopy, the IR phase curve of \planet\ near periastron is expected to be detectable from \jwst\ (Figure~\ref{fig:phase}). Such a detection would reveal the heat redistribution properties of this cold ({$T_{\rm eq} = 254$~K}, assuming no albedo) Jovian planet. Furthermore, since tidal forces are inefficient, the rotation of \planet\ is likely not pseudo-synchronized with its orbit, and its rotation period is possibly measurable via a ``ringing'' in thermal phase curve \citep[e.g.,][]{Cowan2011a}.
\end{enumerate}

The GOT `EM survey aims to characterize systems of long-period transiting giant planets, which serve as stepping stones between many exoplanet systems and the solar system \citep{Dalba2021a}. \planet\ is an extraordinary system owing to its high eccentricity and transiting geometry. Much like HD~80606~b, \planet\ provides a laboratory for testing the extremes of planetary migration scenarios. Continued observation and characterization of this system stands to refine the theories underlying the formation and evolution of all planetary systems. 


\acknowledgements
The authors are grateful to the anonymous referee for a thorough review that improved the quality of this research. The authors thank all of the observers in the California Planet Search team for their many hours of hard work. The authors thank Ji Wang for a helpful discussion about PHARO AO imaging. The authors are grateful to Daniel Foreman-Mackey for helpful discussion regarding \textsf{exoplanet} and photoeccentric modeling. The authors thank all of the members of the Unistellar Citizen Science campaign to observe \host. 

P. D. is supported by a National Science Foundation (NSF) Astronomy and Astrophysics Postdoctoral Fellowship under award AST-1903811. E. W. S. acknowledges support from the NASA Astrobiology Institute's Alternative Earths team funded under Cooperative Agreement Number NNA15BB03A and the Virtual Planetary Laboratory, which is a member of the NASA Nexus for Exoplanet System Science, and funded via NASA Astrobiology Program Grant No. 80NSSC18K0829.

This research has made use of the NASA Exoplanet Archive, which is operated by the California Institute of Technology, under contract with the National Aeronautics and Space Administration under the Exoplanet Exploration Program. This paper includes data collected by the \kepler\ mission and obtained from the MAST data archive at the Space Telescope Science Institute (STScI). Funding for the Kepler mission is provided by the NASA Science Mission Directorate. STScI is operated by the Association of Universities for Research in Astronomy, Inc., under NASA contract NAS 5–26555. This research has made use of the Exoplanet Follow-up Observation Program website, which is operated by the California Institute of Technology, under contract with the National Aeronautics and Space Administration under the Exoplanet Exploration Program. 

Some of the data presented herein were obtained at the W. M. Keck Observatory, which is operated as a scientific partnership among the California Institute of Technology, the University of California, and NASA. The Observatory was made possible by the generous financial support of the W. M. Keck Foundation. Some of the Keck data were obtained under PI Data awards 2013A and 2013B (M. Payne). Finally, the authors recognize and acknowledge the cultural role and reverence that the summit of Maunakea has within the indigenous Hawaiian community. We are deeply grateful to have the opportunity to conduct observations from this mountain.\\

\vspace{5mm}
\facilities{Keck:I (HIRES), Keck:II (NIRC2), Kepler, Hale (PHARO)}\\

\vspace{5mm}
\software{   \textsf{astropy} \citep{astropy2013,astropy2018},
                \textsf{corner} \citep{ForemanMackey2016a},
                \textsf{EXOFASTv2} \citep{Eastman2013,Eastman2017,Eastman2019}, 
                \textsf{lightkurve} \citep{Lightkurve2018}, 
                \textsf{SpecMatch} \citep{Petigura2015,Petigura2017b}, 
                \textsf{SpecMatch--Emp} \citep{Yee2017}, 
                \textsf{exoplanet} \citep{exoplanet},
                \textsf{pymc3} \citep{pymc3},
                \textsf{theano} \citep{theano},
                \textsf{REBOUND} \citep{Rein2012},
                \textsf{RVSearch} \citep{Rosenthal2021}
                } \\


\begin{thebibliography}{}
\expandafter\ifx\csname natexlab\endcsname\relax\def\natexlab#1{#1}\fi
\providecommand{\url}[1]{\href{#1}{#1}}
\providecommand{\dodoi}[1]{doi:~\href{http://doi.org/#1}{\nolinkurl{#1}}}
\providecommand{\doeprint}[1]{\href{http://ascl.net/#1}{\nolinkurl{http://ascl.net/#1}}}
\providecommand{\doarXiv}[1]{\href{https://arxiv.org/abs/#1}{\nolinkurl{https://arxiv.org/abs/#1}}}

\bibitem[{{Adams} \& {Bloch}(2016)}]{Adams2016}
{Adams}, F.~C., \& {Bloch}, A.~M. 2016, \mnras, 462, 2527,
  \dodoi{10.1093/mnras/stw1883}

\bibitem[{{Adams} \& {Laughlin}(2006)}]{Adams2006}
{Adams}, F.~C., \& {Laughlin}, G. 2006, \apj, 649, 1004, \dodoi{10.1086/506145}

\bibitem[{{Alibert} {et~al.}(2005){Alibert}, {Mousis}, {Mordasini}, \&
  {Benz}}]{Alibert2005}
{Alibert}, Y., {Mousis}, O., {Mordasini}, C., \& {Benz}, W. 2005, \apjl, 626,
  L57, \dodoi{10.1086/431325}

\bibitem[{{Alp} \& {Demory}(2018)}]{Alp2018}
{Alp}, D., \& {Demory}, B.~O. 2018, \aap, 609, A90,
  \dodoi{10.1051/0004-6361/201731484}

\bibitem[{{Anderson} {et~al.}(2020){Anderson}, {Lai}, \& {Pu}}]{Anderson2020}
{Anderson}, K.~R., {Lai}, D., \& {Pu}, B. 2020, \mnras, 491, 1369,
  \dodoi{10.1093/mnras/stz3119}

\bibitem[{{Asplund} {et~al.}(2009){Asplund}, {Grevesse}, {Sauval}, \&
  {Scott}}]{Asplund2009}
{Asplund}, M., {Grevesse}, N., {Sauval}, A.~J., \& {Scott}, P. 2009, \araa, 47,
  481, \dodoi{10.1146/annurev.astro.46.060407.145222}

\bibitem[{{Astropy Collaboration} {et~al.}(2013){Astropy Collaboration},
  {Robitaille}, {Tollerud}, {Greenfield}, {Droettboom}, {Bray}, {Aldcroft},
  {Davis}, {Ginsburg}, {Price-Whelan}, {Kerzendorf}, {Conley}, {Crighton},
  {Barbary}, {Muna}, {Ferguson}, {Grollier}, {Parikh}, {Nair}, {Unther},
  {Deil}, {Woillez}, {Conseil}, {Kramer}, {Turner}, {Singer}, {Fox}, {Weaver},
  {Zabalza}, {Edwards}, {Azalee Bostroem}, {Burke}, {Casey}, {Crawford},
  {Dencheva}, {Ely}, {Jenness}, {Labrie}, {Lim}, {Pierfederici}, {Pontzen},
  {Ptak}, {Refsdal}, {Servillat}, \& {Streicher}}]{astropy2013}
{Astropy Collaboration}, {Robitaille}, T.~P., {Tollerud}, E.~J., {et~al.} 2013,
  \aap, 558, A33, \dodoi{10.1051/0004-6361/201322068}

\bibitem[{{Astropy Collaboration} {et~al.}(2018){Astropy Collaboration},
  {Price-Whelan}, {Sip{\H o}cz}, {G{\"u}nther}, {Lim}, {Crawford}, {Conseil},
  {Shupe}, {Craig}, {Dencheva}, {Ginsburg}, {VanderPlas}, {Bradley},
  {P{\'e}rez-Su{\'a}rez}, {de Val-Borro}, {Aldcroft}, {Cruz}, {Robitaille},
  {Tollerud}, {Ardelean}, {Babej}, {Bach}, {Bachetti}, {Bakanov}, {Bamford},
  {Barentsen}, {Barmby}, {Baumbach}, {Berry}, {Biscani}, {Boquien}, {Bostroem},
  {Bouma}, {Brammer}, {Bray}, {Breytenbach}, {Buddelmeijer}, {Burke},
  {Calderone}, {Cano Rodr{\'{\i}}guez}, {Cara}, {Cardoso}, {Cheedella},
  {Copin}, {Corrales}, {Crichton}, {D'Avella}, {Deil}, {Depagne}, {Dietrich},
  {Donath}, {Droettboom}, {Earl}, {Erben}, {Fabbro}, {Ferreira}, {Finethy},
  {Fox}, {Garrison}, {Gibbons}, {Goldstein}, {Gommers}, {Greco}, {Greenfield},
  {Groener}, {Grollier}, {Hagen}, {Hirst}, {Homeier}, {Horton}, {Hosseinzadeh},
  {Hu}, {Hunkeler}, {Ivezi{\'c}}, {Jain}, {Jenness}, {Kanarek}, {Kendrew},
  {Kern}, {Kerzendorf}, {Khvalko}, {King}, {Kirkby}, {Kulkarni}, {Kumar},
  {Lee}, {Lenz}, {Littlefair}, {Ma}, {Macleod}, {Mastropietro}, {McCully},
  {Montagnac}, {Morris}, {Mueller}, {Mumford}, {Muna}, {Murphy}, {Nelson},
  {Nguyen}, {Ninan}, {N{\"o}the}, {Ogaz}, {Oh}, {Parejko}, {Parley}, {Pascual},
  {Patil}, {Patil}, {Plunkett}, {Prochaska}, {Rastogi}, {Reddy Janga},
  {Sabater}, {Sakurikar}, {Seifert}, {Sherbert}, {Sherwood-Taylor}, {Shih},
  {Sick}, {Silbiger}, {Singanamalla}, {Singer}, {Sladen}, {Sooley},
  {Sornarajah}, {Streicher}, {Teuben}, {Thomas}, {Tremblay}, {Turner},
  {Terr{\'o}n}, {van Kerkwijk}, {de la Vega}, {Watkins}, {Weaver}, {Whitmore},
  {Woillez}, {Zabalza}, \& {Astropy Contributors}}]{astropy2018}
{Astropy Collaboration}, {Price-Whelan}, A.~M., {Sip{\H o}cz}, B.~M., {et~al.}
  2018, \aj, 156, 123, \dodoi{10.3847/1538-3881/aabc4f}

\bibitem[{{Baraffe} {et~al.}(2003){Baraffe}, {Chabrier}, {Barman}, {Allard}, \&
  {Hauschildt}}]{Baraffe2003}
{Baraffe}, I., {Chabrier}, G., {Barman}, T.~S., {Allard}, F., \& {Hauschildt},
  P.~H. 2003, \aap, 402, 701, \dodoi{10.1051/0004-6361:20030252}

\bibitem[{{Barbieri} {et~al.}(2009){Barbieri}, {Alonso}, {Desidera},
  {Sozzetti}, {Martinez Fiorenzano}, {Almenara}, {Cecconi}, {Claudi},
  {Charbonneau}, {Endl}, {Granata}, {Gratton}, {Laughlin}, {Loeillet}, \&
  {EXOPLANET Amateur Consortium}}]{Barbieri2009}
{Barbieri}, M., {Alonso}, R., {Desidera}, S., {et~al.} 2009, \aap, 503, 601,
  \dodoi{10.1051/0004-6361/200811466}

\bibitem[{{Barnes} \& {O'Brien}(2002)}]{Barnes2002}
{Barnes}, J.~W., \& {O'Brien}, D.~P. 2002, \apj, 575, 1087,
  \dodoi{10.1086/341477}

\bibitem[{{Baruteau} {et~al.}(2014){Baruteau}, {Crida}, {Paardekooper},
  {Masset}, {Guilet}, {Bitsch}, {Nelson}, {Kley}, \&
  {Papaloizou}}]{Baruteau2014}
{Baruteau}, C., {Crida}, A., {Paardekooper}, S.-J., {et~al.} 2014, in
  Protostars and Planets VI, ed. H.~{Beuther}, R.~S. {Klessen}, C.~P.
  {Dullemond}, \& T.~Henning (Tucson: Univ. of Arizona Press), 667--689,
  \dodoi{10.2458/azu_uapress_9780816531240-ch029}

\bibitem[{{Beatty} \& {Gaudi}(2008)}]{Beatty2008}
{Beatty}, T.~G., \& {Gaudi}, B.~S. 2008, \apj, 686, 1302,
  \dodoi{10.1086/591441}

\bibitem[{{Bitsch} {et~al.}(2013){Bitsch}, {Crida}, {Libert}, \&
  {Lega}}]{Bitsch2013}
{Bitsch}, B., {Crida}, A., {Libert}, A.~S., \& {Lega}, E. 2013, \aap, 555,
  A124, \dodoi{10.1051/0004-6361/201220310}

\bibitem[{{Bitsch} {et~al.}(2020){Bitsch}, {Trifonov}, \&
  {Izidoro}}]{Bitsch2020}
{Bitsch}, B., {Trifonov}, T., \& {Izidoro}, A. 2020, \aap, 643, A66,
  \dodoi{10.1051/0004-6361/202038856}

\bibitem[{{Bonomo} {et~al.}(2017){Bonomo}, {Desidera}, {Benatti}, {Borsa},
  {Crespi}, {Damasso}, {Lanza}, {Sozzetti}, {Lodato}, {Marzari}, {Boccato},
  {Claudi}, {Cosentino}, {Covino}, {Gratton}, {Maggio}, {Micela}, {Molinari},
  {Pagano}, {Piotto}, {Poretti}, {Smareglia}, {Affer}, {Biazzo}, {Bignamini},
  {Esposito}, {Giacobbe}, {H{\'e}brard}, {Malavolta}, {Maldonado}, {Mancini},
  {Martinez Fiorenzano}, {Masiero}, {Nascimbeni}, {Pedani}, {Rainer}, \& {Scand
  ariato}}]{Bonomo2017}
{Bonomo}, A.~S., {Desidera}, S., {Benatti}, S., {et~al.} 2017, \aap, 602, A107,
  \dodoi{10.1051/0004-6361/201629882}

\bibitem[{{Borucki} {et~al.}(2010){Borucki}, {Koch}, {Basri}, {Batalha},
  {Brown}, {Caldwell}, {Caldwell}, {Christensen-Dalsgaard}, {Cochran},
  {DeVore}, {Dunham}, {Dupree}, {Gautier}, {Geary}, {Gilliland}, {Gould},
  {Howell}, {Jenkins}, {Kondo}, {Latham}, {Marcy}, {Meibom}, {Kjeldsen},
  {Lissauer}, {Monet}, {Morrison}, {Sasselov}, {Tarter}, {Boss}, {Brownlee},
  {Owen}, {Buzasi}, {Charbonneau}, {Doyle}, {Fortney}, {Ford}, {Holman},
  {Seager}, {Steffen}, {Welsh}, {Rowe}, {Anderson}, {Buchhave}, {Ciardi},
  {Walkowicz}, {Sherry}, {Horch}, {Isaacson}, {Everett}, {Fischer}, {Torres},
  {Johnson}, {Endl}, {MacQueen}, {Bryson}, {Dotson}, {Haas}, {Kolodziejczak},
  {Van Cleve}, {Chandrasekaran}, {Twicken}, {Quintana}, {Clarke}, {Allen},
  {Li}, {Wu}, {Tenenbaum}, {Verner}, {Bruhweiler}, {Barnes}, \&
  {Prsa}}]{Borucki2010}
{Borucki}, W.~J., {Koch}, D., {Basri}, G., {et~al.} 2010, \sci, 327, 977,
  \dodoi{10.1126/science.1185402}

\bibitem[{{Brady} {et~al.}(2018){Brady}, {Petigura}, {Knutson}, {Sinukoff},
  {Isaacson}, {Hirsch}, {Fulton}, {Kosiarek}, \& {Howard}}]{Brady2018}
{Brady}, M.~T., {Petigura}, E.~A., {Knutson}, H.~A., {et~al.} 2018, \aj, 156,
  147, \dodoi{10.3847/1538-3881/aad773}

\bibitem[{{Brown}(2003)}]{Brown2003}
{Brown}, T.~M. 2003, \apjl, 593, L125, \dodoi{10.1086/378310}

\bibitem[{{Bryant} {et~al.}(2021){Bryant}, {Bayliss}, {Santerne}, {Wheatley},
  {Nascimbeni}, {Ducrot}, {Burdanov}, {Acton}, {Alves}, {Anderson},
  {Armstrong}, {Awiphan}, {Cooke}, {Burleigh}, {Casewell}, {Delrez}, {Demory},
  {Eigm{\"u}ller}, {Fukui}, {Gan}, {Gill}, {Gillon}, {Goad}, {Tan},
  {G{\"u}nther}, {Hardee}, {Henderson}, {Jehin}, {Jenkins}, {Kosiarek},
  {Lendl}, {Moyano}, {Murray}, {Narita}, {Niraula}, {Odden}, {Palle},
  {Parviainen}, {Pedersen}, {Pozuelos}, {Rackham}, {Sebastian}, {Stockdale},
  {Tilbrook}, {Thompson}, {Triaud}, {Udry}, {Vines}, {West}, \& {de
  Wit}}]{Bryant2021}
{Bryant}, E.~M., {Bayliss}, D., {Santerne}, A., {et~al.} 2021, \mnras,
  \dodoi{10.1093/mnrasl/slab037}

\bibitem[{{Buchhave} {et~al.}(2018){Buchhave}, {Bitsch}, {Johansen}, {Latham},
  {Bizzarro}, {Bieryla}, \& {Kipping}}]{Buchhave2018}
{Buchhave}, L.~A., {Bitsch}, B., {Johansen}, A., {et~al.} 2018, \apj, 856, 37,
  \dodoi{10.3847/1538-4357/aaafca}

\bibitem[{{Cameron}(2012)}]{Cameron2012}
{Cameron}, A.~C. 2012, \nat, 492, 48, \dodoi{10.1038/492048a}

\bibitem[{{Carrera} {et~al.}(2019){Carrera}, {Raymond}, \&
  {Davies}}]{Carrera2019}
{Carrera}, D., {Raymond}, S.~N., \& {Davies}, M.~B. 2019, \aap, 629, L7,
  \dodoi{10.1051/0004-6361/201935744}

\bibitem[{{Cauley} {et~al.}(2019){Cauley}, {Shkolnik}, {Llama}, \&
  {Lanza}}]{Cauley2019}
{Cauley}, P.~W., {Shkolnik}, E.~L., {Llama}, J., \& {Lanza}, A.~F. 2019, Nature
  Astronomy, 3, 1128, \dodoi{10.1038/s41550-019-0840-x}

\bibitem[{{Chatterjee} {et~al.}(2008){Chatterjee}, {Ford}, {Matsumura}, \&
  {Rasio}}]{Chatterjee2008}
{Chatterjee}, S., {Ford}, E.~B., {Matsumura}, S., \& {Rasio}, F.~A. 2008, \apj,
  686, 580, \dodoi{10.1086/590227}

\bibitem[{{Choi} {et~al.}(2016){Choi}, {Dotter}, {Conroy}, {Cantiello},
  {Paxton}, \& {Johnson}}]{Choi2016}
{Choi}, J., {Dotter}, A., {Conroy}, C., {et~al.} 2016, \apj, 823, 102,
  \dodoi{10.3847/0004-637X/823/2/102}

\bibitem[{{Christiansen} {et~al.}(2020){Christiansen}, {Clarke}, {Burke},
  {Jenkins}, {Bryson}, {Coughlin}, {Mullally}, {Twicken}, {Batalha},
  {Catanzarite}, {Uddin}, {Zamudio}, {Smith}, {Henze}, \&
  {Campbell}}]{Christiansen2020}
{Christiansen}, J.~L., {Clarke}, B.~D., {Burke}, C.~J., {et~al.} 2020, \aj,
  160, 159, \dodoi{10.3847/1538-3881/abab0b}

\bibitem[{{Cincotta} \& {Sim{\'o}}(2000)}]{Cincotta2000}
{Cincotta}, P.~M., \& {Sim{\'o}}, C. 2000, \aaps, 147, 205,
  \dodoi{10.1051/aas:2000108}

\bibitem[{{Cochran} {et~al.}(2008){Cochran}, {Redfield}, {Endl}, \&
  {Cochran}}]{Cochran2008}
{Cochran}, W.~D., {Redfield}, S., {Endl}, M., \& {Cochran}, A.~L. 2008, \apjl,
  683, L59, \dodoi{10.1086/591317}

\bibitem[{{Cooke} {et~al.}(2018){Cooke}, {Pollacco}, {West}, {McCormac}, \&
  {Wheatley}}]{Cooke2018}
{Cooke}, B.~F., {Pollacco}, D., {West}, R., {McCormac}, J., \& {Wheatley},
  P.~J. 2018, \aap, 619, A175, \dodoi{10.1051/0004-6361/201834014}

\bibitem[{{Cowan} \& {Agol}(2011{\natexlab{a}})}]{Cowan2011b}
{Cowan}, N.~B., \& {Agol}, E. 2011{\natexlab{a}}, \apj, 729, 54,
  \dodoi{10.1088/0004-637X/729/1/54}

\bibitem[{{Cowan} \& {Agol}(2011{\natexlab{b}})}]{Cowan2011a}
---. 2011{\natexlab{b}}, \apj, 726, 82, \dodoi{10.1088/0004-637X/726/2/82}

\bibitem[{{Crepp} {et~al.}(2012){Crepp}, {Johnson}, {Howard}, {Marcy},
  {Fischer}, {Hillenbrand}, {Yantek}, {Delaney}, {Wright}, {Isaacson}, \&
  {Montet}}]{Crepp2012}
{Crepp}, J.~R., {Johnson}, J.~A., {Howard}, A.~W., {et~al.} 2012, \apj, 761,
  39, \dodoi{10.1088/0004-637X/761/1/39}

\bibitem[{{Cutri} {et~al.}(2003){Cutri}, {Skrutskie}, {van Dyk}, {Beichman},
  {Carpenter}, {Chester}, {Cambresy}, {Evans}, {Fowler}, {Gizis}, {Howard},
  {Huchra}, {Jarrett}, {Kopan}, {Kirkpatrick}, {Light}, {Marsh}, {McCallon},
  {Schneider}, {Stiening}, {Sykes}, {Weinberg}, {Wheaton}, {Wheelock}, \&
  {Zacarias}}]{Cutri2003}
{Cutri}, R.~M., {Skrutskie}, M.~F., {van Dyk}, S., {et~al.} 2003, VizieR Online
  Data Catalog, II/246

\bibitem[{{Cutri} {et~al.}(2014){Cutri}, {Wright}, {Conrow}, {Fowler},
  {Eisenhardt}, {Grillmair}, {Kirkpatrick}, {Masci}, {McCallon}, {Wheelock},
  {Fajardo-Acosta}, {Yan}, {Benford}, {Harbut}, {Jarrett}, {Lake}, {Leisawitz},
  {Ressler}, {Stanford}, {Tsai}, {Liu}, {Helou}, {Mainzer}, {Gettngs},
  {Gonzalez}, {Hoffman}, {Marsh}, {Padgett}, {Skrutskie}, {Beck}, {Papin}, \&
  {Wittman}}]{Cutri2014}
{Cutri}, R.~M., {Wright}, E.~L., {Conrow}, T., {et~al.} 2014, VizieR Online
  Data Catalog, II/328

\bibitem[{{Dalba}(2017)}]{Dalba2017b}
{Dalba}, P.~A. 2017, \apj, 848, 91, \dodoi{10.3847/1538-4357/aa8e47}

\bibitem[{{Dalba} {et~al.}(2020{\natexlab{a}}){Dalba}, {Fulton}, {Isaacson},
  {Kane}, \& {Howard}}]{Dalba2020b}
{Dalba}, P.~A., {Fulton}, B., {Isaacson}, H., {Kane}, S.~R., \& {Howard}, A.~W.
  2020{\natexlab{a}}, \aj, 160, 149, \dodoi{10.3847/1538-3881/abad27}

\bibitem[{{Dalba} {et~al.}(2019){Dalba}, {Kane}, {Barclay}, {Bean}, {Campante},
  {Pepper}, {Ragozzine}, \& {Turnbull}}]{Dalba2019a}
{Dalba}, P.~A., {Kane}, S.~R., {Barclay}, T., {et~al.} 2019, \pasp, 131,
  034401, \dodoi{10.1088/1538-3873/aaf183}

\bibitem[{{Dalba} \& {Muirhead}(2016)}]{Dalba2016}
{Dalba}, P.~A., \& {Muirhead}, P.~S. 2016, \apjl, 826, L7,
  \dodoi{10.3847/2041-8205/826/1/L7}

\bibitem[{{Dalba} {et~al.}(2015){Dalba}, {Muirhead}, {Fortney}, {Hedman},
  {Nicholson}, \& {Veyette}}]{Dalba2015}
{Dalba}, P.~A., {Muirhead}, P.~S., {Fortney}, J.~J., {et~al.} 2015, \apj, 814,
  154, \dodoi{10.1088/0004-637X/814/2/154}

\bibitem[{{Dalba} \& {Tamburo}(2019)}]{Dalba2019c}
{Dalba}, P.~A., \& {Tamburo}, P. 2019, \apjl, 873, L17,
  \dodoi{10.3847/2041-8213/ab0bb4}

\bibitem[{{Dalba} {et~al.}(2020{\natexlab{b}}){Dalba}, {Gupta}, {Rodriguez},
  {Dragomir}, {Huang}, {Kane}, {Quinn}, {Bieryla}, {Esquerdo}, {Fulton},
  {Scarsdale}, {Batalha}, {Beard}, {Behmard}, {Chontos}, {Crossfield},
  {Dressing}, {Giacalone}, {Hill}, {Hirsch}, {Howard}, {Huber}, {Isaacson},
  {Kosiarek}, {Lubin}, {Mayo}, {Mocnik}, {Akana Murphy}, {Petigura},
  {Robertson}, {Rosenthal}, {Roy}, {Rubenzahl}, {Van Zandt}, {Weiss},
  {Knudstrup}, {Andersen}, {Grundahl}, {Yao}, {Pepper}, {Villanueva}, {Ciardi},
  {Cloutier}, {Jacobs}, {Kristiansen}, {LaCourse}, {Lendl}, {Osborn}, {Palle},
  {Stassun}, {Stevens}, {Ricker}, {Vanderspek}, {Latham}, {Seager}, {Winn},
  {Jenkins}, {Caldwell}, {Daylan}, {Fong}, {Goeke}, {Rose}, {Rowden},
  {Schlieder}, {Smith}, \& {Vanderburg}}]{Dalba2020a}
{Dalba}, P.~A., {Gupta}, A.~F., {Rodriguez}, J.~E., {et~al.}
  2020{\natexlab{b}}, \aj, 159, 241, \dodoi{10.3847/1538-3881/ab84e3}

\bibitem[{{Dalba} {et~al.}(2021{\natexlab{a}}){Dalba}, {Kane}, {Isaacson},
  {Giacalone}, {Howard}, {Rodriguez}, {Vanderburg}, {Eastman}, {Kraus},
  {Dupuy}, {Weiss}, \& {Schwieterman}}]{Dalba2021a}
{Dalba}, P.~A., {Kane}, S.~R., {Isaacson}, H., {et~al.} 2021{\natexlab{a}},
  \aj, 161, 103, \dodoi{10.3847/1538-3881/abd408}

\bibitem[{{Dalba} {et~al.}(2021{\natexlab{b}}){Dalba}, {Kane}, {Howell},
  {Horch}, {Li}, {Hirsch}, {Burt}, {Brandt}, {Mo{\v{c}}nik}, {Henry},
  {Everett}, {Rosenthal}, \& {Howard}}]{Dalba2021b}
{Dalba}, P.~A., {Kane}, S.~R., {Howell}, S.~B., {et~al.} 2021{\natexlab{b}},
  \aj, 161, 123, \dodoi{10.3847/1538-3881/abd6ed}

\bibitem[{{Dawson} \& {Johnson}(2012)}]{Dawson2012a}
{Dawson}, R.~I., \& {Johnson}, J.~A. 2012, \apj, 756, 122,
  \dodoi{10.1088/0004-637X/756/2/122}

\bibitem[{{Dawson} \& {Johnson}(2018)}]{Dawson2018}
---. 2018, \araa, 56, 175, \dodoi{10.1146/annurev-astro-081817-051853}

\bibitem[{{Dawson} {et~al.}(2012){Dawson}, {Johnson}, {Morton}, {Crepp},
  {Fabrycky}, {Murray-Clay}, \& {Howard}}]{Dawson2012b}
{Dawson}, R.~I., {Johnson}, J.~A., {Morton}, T.~D., {et~al.} 2012, \apj, 761,
  163, \dodoi{10.1088/0004-637X/761/2/163}

\bibitem[{{Dawson} \& {Murray-Clay}(2013)}]{Dawson2013}
{Dawson}, R.~I., \& {Murray-Clay}, R.~A. 2013, \apjl, 767, L24,
  \dodoi{10.1088/2041-8205/767/2/L24}

\bibitem[{{Dawson} {et~al.}(2015){Dawson}, {Murray-Clay}, \&
  {Johnson}}]{Dawson2015}
{Dawson}, R.~I., {Murray-Clay}, R.~A., \& {Johnson}, J.~A. 2015, \apj, 798, 66,
  \dodoi{10.1088/0004-637X/798/2/66}

\bibitem[{{Dawson} {et~al.}(2014){Dawson}, {Johnson}, {Fabrycky},
  {Foreman-Mackey}, {Murray-Clay}, {Buchhave}, {Cargile}, {Clubb}, {Fulton},
  {Hebb}, {Howard}, {Huber}, {Shporer}, \& {Valenti}}]{Dawson2014}
{Dawson}, R.~I., {Johnson}, J.~A., {Fabrycky}, D.~C., {et~al.} 2014, \apj, 791,
  89, \dodoi{10.1088/0004-637X/791/2/89}

\bibitem[{{Debras} {et~al.}(2021){Debras}, {Baruteau}, \&
  {Donati}}]{Debras2021}
{Debras}, F., {Baruteau}, C., \& {Donati}, J.-F. 2021, \mnras, 500, 1621,
  \dodoi{10.1093/mnras/staa3397}

\bibitem[{{Debras} \& {Chabrier}(2019)}]{Debras2019}
{Debras}, F., \& {Chabrier}, G. 2019, \apj, 872, 100,
  \dodoi{10.3847/1538-4357/aaff65}

\bibitem[{{Demory} \& {Seager}(2011)}]{Demory2011b}
{Demory}, B.-O., \& {Seager}, S. 2011, \apjs, 197, 12,
  \dodoi{10.1088/0067-0049/197/1/12}

\bibitem[{{D{\'\i}az} {et~al.}(2020){D{\'\i}az}, {Jenkins}, {Feng}, {Butler},
  {Tuomi}, {Shectman}, {Thorngren}, {Soto}, {Vines}, {Teske}, {Dragomir},
  {Villanueva}, {Kane}, {Berdi{\~n}as}, {Crane}, {Wang}, \&
  {Arriagada}}]{Diaz2020}
{D{\'\i}az}, M.~R., {Jenkins}, J.~S., {Feng}, F., {et~al.} 2020, \mnras, 496,
  4330, \dodoi{10.1093/mnras/staa1724}

\bibitem[{{Dong} {et~al.}(2013){Dong}, {Katz}, \& {Socrates}}]{Dong2013}
{Dong}, S., {Katz}, B., \& {Socrates}, A. 2013, \apjl, 763, L2,
  \dodoi{10.1088/2041-8205/763/1/L2}

\bibitem[{{Dong} {et~al.}(2014){Dong}, {Katz}, \& {Socrates}}]{Dong2014}
---. 2014, \apjl, 781, L5, \dodoi{10.1088/2041-8205/781/1/L5}

\bibitem[{{Dotter}(2016)}]{Dotter2016}
{Dotter}, A. 2016, \apjs, 222, 8, \dodoi{10.3847/0067-0049/222/1/8}

\bibitem[{{Duncan} {et~al.}(1998){Duncan}, {Levison}, \& {Lee}}]{Duncan1998}
{Duncan}, M.~J., {Levison}, H.~F., \& {Lee}, M.~H. 1998, \aj, 116, 2067,
  \dodoi{10.1086/300541}

\bibitem[{{Eastman}(2017)}]{Eastman2017}
{Eastman}, J. 2017, {EXOFASTv2: Generalized publication-quality exoplanet
  modeling code}, v2, Astrophysics Source Code Library.
\newblock \doeprint{1710.003}

\bibitem[{{Eastman} {et~al.}(2013){Eastman}, {Gaudi}, \& {Agol}}]{Eastman2013}
{Eastman}, J., {Gaudi}, B.~S., \& {Agol}, E. 2013, \pasp, 125, 83,
  \dodoi{10.1086/669497}

\bibitem[{{Eastman} {et~al.}(2019){Eastman}, {Rodriguez}, {Agol}, {Stassun},
  {Beatty}, {Vanderburg}, {Gaudi}, {Collins}, \& {Luger}}]{Eastman2019}
{Eastman}, J.~D., {Rodriguez}, J.~E., {Agol}, E., {et~al.} 2019, arXiv
  e-prints.
\newblock \doarXiv{1907.09480}

\bibitem[{{Fabrycky} \& {Tremaine}(2007)}]{Fabrycky2007}
{Fabrycky}, D., \& {Tremaine}, S. 2007, \apj, 669, 1298, \dodoi{10.1086/521702}

\bibitem[{{Fabrycky} \& {Winn}(2009)}]{Fabrycky2009}
{Fabrycky}, D.~C., \& {Winn}, J.~N. 2009, \apj, 696, 1230,
  \dodoi{10.1088/0004-637X/696/2/1230}

\bibitem[{{Farrell} {et~al.}(1999){Farrell}, {Desch}, \& {Zarka}}]{Farrell1999}
{Farrell}, W.~M., {Desch}, M.~D., \& {Zarka}, P. 1999, \jgr, 104, 14025,
  \dodoi{10.1029/1998JE900050}

\bibitem[{{Fischer} \& {Valenti}(2005)}]{Fischer2005}
{Fischer}, D.~A., \& {Valenti}, J. 2005, \apj, 622, 1102,
  \dodoi{10.1086/428383}

\bibitem[{{Fischer} {et~al.}(2007){Fischer}, {Vogt}, {Marcy}, {Butler}, {Sato},
  {Henry}, {Robinson}, {Laughlin}, {Ida}, {Toyota}, {Omiya}, {Driscoll},
  {Takeda}, {Wright}, \& {Johnson}}]{Fischer2007}
{Fischer}, D.~A., {Vogt}, S.~S., {Marcy}, G.~W., {et~al.} 2007, \apj, 669,
  1336, \dodoi{10.1086/521869}

\bibitem[{{Fletcher} {et~al.}(2009{\natexlab{a}}){Fletcher}, {Orton}, {Teanby},
  \& {Irwin}}]{Fletcher2009b}
{Fletcher}, L.~N., {Orton}, G.~S., {Teanby}, N.~A., \& {Irwin}, P.~G.~J.
  2009{\natexlab{a}}, \icarus, 202, 543, \dodoi{10.1016/j.icarus.2009.03.023}

\bibitem[{{Fletcher} {et~al.}(2009{\natexlab{b}}){Fletcher}, {Orton}, {Teanby},
  {Irwin}, \& {Bjoraker}}]{Fletcher2009a}
{Fletcher}, L.~N., {Orton}, G.~S., {Teanby}, N.~A., {Irwin}, P.~G.~J., \&
  {Bjoraker}, G.~L. 2009{\natexlab{b}}, \icarus, 199, 351,
  \dodoi{10.1016/j.icarus.2008.09.019}

\bibitem[{{Ford}(2006)}]{Ford2006b}
{Ford}, E.~B. 2006, \apj, 642, 505, \dodoi{10.1086/500802}

\bibitem[{{Ford} {et~al.}(2008){Ford}, {Quinn}, \& {Veras}}]{Ford2008a}
{Ford}, E.~B., {Quinn}, S.~N., \& {Veras}, D. 2008, \apj, 678, 1407,
  \dodoi{10.1086/587046}

\bibitem[{{Ford} \& {Rasio}(2006)}]{Ford2006a}
{Ford}, E.~B., \& {Rasio}, F.~A. 2006, \apjl, 638, L45, \dodoi{10.1086/500734}

\bibitem[{{Ford} \& {Rasio}(2008)}]{Ford2008b}
---. 2008, \apj, 686, 621, \dodoi{10.1086/590926}

\bibitem[{Foreman-Mackey(2016)}]{ForemanMackey2016a}
Foreman-Mackey, D. 2016, The Journal of Open Source Software, 24,
  \dodoi{10.21105/joss.00024}

\bibitem[{Foreman-Mackey {et~al.}(2020)Foreman-Mackey, Luger, Czekala, Agol,
  Price-Whelan, \& Barclay}]{exoplanet}
Foreman-Mackey, D., Luger, R., Czekala, I., {et~al.} 2020,
  exoplanet-dev/exoplanet v0.3.2, \dodoi{10.5281/zenodo.1998447}

\bibitem[{{Foreman-Mackey} {et~al.}(2016){Foreman-Mackey}, {Morton}, {Hogg},
  {Agol}, \& {Sch{\"o}lkopf}}]{ForemanMackey2016b}
{Foreman-Mackey}, D., {Morton}, T.~D., {Hogg}, D.~W., {Agol}, E., \&
  {Sch{\"o}lkopf}, B. 2016, \aj, 152, 206, \dodoi{10.3847/0004-6256/152/6/206}

\bibitem[{{Fortney} {et~al.}(2007){Fortney}, {Marley}, \&
  {Barnes}}]{Fortney2007}
{Fortney}, J.~J., {Marley}, M.~S., \& {Barnes}, J.~W. 2007, \apj, 659, 1661,
  \dodoi{10.1086/512120}

\bibitem[{{Furlan} {et~al.}(2017){Furlan}, {Ciardi}, {Everett}, {Saylors},
  {Teske}, {Horch}, {Howell}, {van Belle}, {Hirsch}, {Gautier}, {Adams},
  {Barrado}, {Cartier}, {Dressing}, {Dupree}, {Gilliland}, {Lillo-Box},
  {Lucas}, \& {Wang}}]{Furlan2017a}
{Furlan}, E., {Ciardi}, D.~R., {Everett}, M.~E., {et~al.} 2017, \aj, 153, 71,
  \dodoi{10.3847/1538-3881/153/2/71}

\bibitem[{{Gaia Collaboration} {et~al.}(2020){Gaia Collaboration}, {Brown},
  {Vallenari}, {Prusti}, {de Bruijne}, {Babusiaux}, \& {Biermann}}]{Gaia2020}
{Gaia Collaboration}, {Brown}, A.~G.~A., {Vallenari}, A., {et~al.} 2020, arXiv
  e-prints, arXiv:2012.01533.
\newblock \doarXiv{2012.01533}

\bibitem[{{Gaia Collaboration} {et~al.}(2016){Gaia Collaboration}, {Prusti},
  {de Bruijne}, {Brown}, {Vallenari}, {Babusiaux}, {Bailer-Jones}, {Bastian},
  {Biermann}, {Evans}, {Eyer}, {Jansen}, {Jordi}, {Klioner}, {Lammers},
  {Lindegren}, {Luri}, {Mignard}, {Milligan}, {Panem}, {Poinsignon},
  {Pourbaix}, {Randich}, {Sarri}, {Sartoretti}, {Siddiqui}, {Soubiran},
  {Valette}, {van Leeuwen}, {Walton}, {Aerts}, {Arenou}, {Cropper}, {Drimmel},
  {H{\o}g}, {Katz}, {Lattanzi}, {O'Mullane}, {Grebel}, {Holland}, {Huc},
  {Passot}, {Bramante}, {Cacciari}, {Casta{\~n}eda}, {Chaoul}, {Cheek}, {De
  Angeli}, {Fabricius}, {Guerra}, {Hern{\'a}ndez}, {Jean-Antoine-Piccolo},
  {Masana}, {Messineo}, {Mowlavi}, {Nienartowicz}, {Ord{\'o}{\~n}ez-Blanco},
  {Panuzzo}, {Portell}, {Richards}, {Riello}, {Seabroke}, {Tanga},
  {Th{\'e}venin}, {Torra}, {Els}, {Gracia-Abril}, {Comoretto},
  {Garcia-Reinaldos}, {Lock}, {Mercier}, {Altmann}, {Andrae}, {Astraatmadja},
  {Bellas-Velidis}, {Benson}, {Berthier}, {Blomme}, {Busso}, {Carry},
  {Cellino}, {Clementini}, {Cowell}, {Creevey}, {Cuypers}, {Davidson}, {De
  Ridder}, {de Torres}, {Delchambre}, {Dell'Oro}, {Ducourant}, {Fr{\'e}mat},
  {Garc{\'\i}a-Torres}, {Gosset}, {Halbwachs}, {Hambly}, {Harrison}, {Hauser},
  {Hestroffer}, {Hodgkin}, {Huckle}, {Hutton}, {Jasniewicz}, {Jordan},
  {Kontizas}, {Korn}, {Lanzafame}, {Manteiga}, {Moitinho}, {Muinonen},
  {Osinde}, {Pancino}, {Pauwels}, {Petit}, {Recio-Blanco}, {Robin}, {Sarro},
  {Siopis}, {Smith}, {Smith}, {Sozzetti}, {Thuillot}, {van Reeven}, {Viala},
  {Abbas}, {Abreu Aramburu}, {Accart}, {Aguado}, {Allan}, {Allasia},
  {Altavilla}, {{\'A}lvarez}, {Alves}, {Anderson}, {Andrei}, {Anglada Varela},
  {Antiche}, {Antoja}, {Ant{\'o}n}, {Arcay}, {Atzei}, {Ayache}, {Bach},
  {Baker}, {Balaguer-N{\'u}{\~n}ez}, {Barache}, {Barata}, {Barbier}, {Barblan},
  {Baroni}, {Barrado y Navascu{\'e}s}, {Barros}, {Barstow}, {Becciani},
  {Bellazzini}, {Bellei}, {Bello Garc{\'\i}a}, {Belokurov}, {Bendjoya},
  {Berihuete}, {Bianchi}, {Bienaym{\'e}}, {Billebaud}, {Blagorodnova},
  {Blanco-Cuaresma}, {Boch}, {Bombrun}, {Borrachero}, {Bouquillon}, {Bourda},
  {Bouy}, {Bragaglia}, {Breddels}, {Brouillet}, {Br{\"u}semeister},
  {Bucciarelli}, {Budnik}, {Burgess}, {Burgon}, {Burlacu}, {Busonero}, {Buzzi},
  {Caffau}, {Cambras}, {Campbell}, {Cancelliere}, {Cantat-Gaudin}, {Carlucci},
  {Carrasco}, {Castellani}, {Charlot}, {Charnas}, {Charvet}, {Chassat},
  {Chiavassa}, {Clotet}, {Cocozza}, {Collins}, {Collins}, {Costigan}, {Crifo},
  {Cross}, {Crosta}, {Crowley}, {Dafonte}, {Damerdji}, {Dapergolas}, {David},
  {David}, {De Cat}, {de Felice}, {de Laverny}, {De Luise}, {De March}, {de
  Martino}, {de Souza}, {Debosscher}, {del Pozo}, {Delbo}, {Delgado},
  {Delgado}, {di Marco}, {Di Matteo}, {Diakite}, {Distefano}, {Dolding}, {Dos
  Anjos}, {Drazinos}, {Dur{\'a}n}, {Dzigan}, {Ecale}, {Edvardsson}, {Enke},
  {Erdmann}, {Escolar}, {Espina}, {Evans}, {Eynard Bontemps}, {Fabre},
  {Fabrizio}, {Faigler}, {Falc{\~a}o}, {Farr{\`a}s Casas}, {Faye}, {Federici},
  {Fedorets}, {Fern{\'a}ndez-Hern{\'a}ndez}, {Fernique}, {Fienga}, {Figueras},
  {Filippi}, {Findeisen}, {Fonti}, {Fouesneau}, {Fraile}, {Fraser}, {Fuchs},
  {Furnell}, {Gai}, {Galleti}, {Galluccio}, {Garabato}, {Garc{\'\i}a-Sedano},
  {Gar{\'e}}, {Garofalo}, {Garralda}, {Gavras}, {Gerssen}, {Geyer}, {Gilmore},
  {Girona}, {Giuffrida}, {Gomes}, {Gonz{\'a}lez-Marcos},
  {Gonz{\'a}lez-N{\'u}{\~n}ez}, {Gonz{\'a}lez-Vidal}, {Granvik}, {Guerrier},
  {Guillout}, {Guiraud}, {G{\'u}rpide}, {Guti{\'e}rrez-S{\'a}nchez}, {Guy},
  {Haigron}, {Hatzidimitriou}, {Haywood}, {Heiter}, {Helmi}, {Hobbs},
  {Hofmann}, {Holl}, {Holland}, {Hunt}, {Hypki}, {Icardi}, {Irwin}, {Jevardat
  de Fombelle}, {Jofr{\'e}}, {Jonker}, {Jorissen}, {Julbe}, {Karampelas},
  {Kochoska}, {Kohley}, {Kolenberg}, {Kontizas}, {Koposov}, {Kordopatis},
  {Koubsky}, {Kowalczyk}, {Krone-Martins}, {Kudryashova}, {Kull}, {Bachchan},
  {Lacoste-Seris}, {Lanza}, {Lavigne}, {Le Poncin-Lafitte}, {Lebreton},
  {Lebzelter}, {Leccia}, {Leclerc}, {Lecoeur-Taibi}, {Lemaitre}, {Lenhardt},
  {Leroux}, {Liao}, {Licata}, {Lindstr{\o}m}, {Lister}, {Livanou}, {Lobel},
  {L{\"o}ffler}, {L{\'o}pez}, {Lopez-Lozano}, {Lorenz}, {Loureiro},
  {MacDonald}, {Magalh{\~a}es Fernandes}, {Managau}, {Mann}, {Mantelet},
  {Marchal}, {Marchant}, {Marconi}, {Marie}, {Marinoni}, {Marrese},
  {Marschalk{\'o}}, {Marshall}, {Mart{\'\i}n-Fleitas}, {Martino}, {Mary},
  {Matijevi{\v{c}}}, {Mazeh}, {McMillan}, {Messina}, {Mestre}, {Michalik},
  {Millar}, {Miranda}, {Molina}, {Molinaro}, {Molinaro}, {Moln{\'a}r},
  {Moniez}, {Montegriffo}, {Monteiro}, {Mor}, {Mora}, {Morbidelli}, {Morel},
  {Morgenthaler}, {Morley}, {Morris}, {Mulone}, {Muraveva}, {Musella},
  {Narbonne}, {Nelemans}, {Nicastro}, {Noval}, {Ord{\'e}novic},
  {Ordieres-Mer{\'e}}, {Osborne}, {Pagani}, {Pagano}, {Pailler}, {Palacin},
  {Palaversa}, {Parsons}, {Paulsen}, {Pecoraro}, {Pedrosa}, {Pentik{\"a}inen},
  {Pereira}, {Pichon}, {Piersimoni}, {Pineau}, {Plachy}, {Plum}, {Poujoulet},
  {Pr{\v{s}}a}, {Pulone}, {Ragaini}, {Rago}, {Rambaux}, {Ramos-Lerate},
  {Ranalli}, {Rauw}, {Read}, {Regibo}, {Renk}, {Reyl{\'e}}, {Ribeiro},
  {Rimoldini}, {Ripepi}, {Riva}, {Rixon}, {Roelens}, {Romero-G{\'o}mez},
  {Rowell}, {Royer}, {Rudolph}, {Ruiz-Dern}, {Sadowski}, {Sagrist{\`a}
  Sell{\'e}s}, {Sahlmann}, {Salgado}, {Salguero}, {Sarasso}, {Savietto},
  {Schnorhk}, {Schultheis}, {Sciacca}, {Segol}, {Segovia}, {Segransan},
  {Serpell}, {Shih}, {Smareglia}, {Smart}, {Smith}, {Solano}, {Solitro},
  {Sordo}, {Soria Nieto}, {Souchay}, {Spagna}, {Spoto}, {Stampa}, {Steele},
  {Steidelm{\"u}ller}, {Stephenson}, {Stoev}, {Suess}, {S{\"u}veges}, {Surdej},
  {Szabados}, {Szegedi-Elek}, {Tapiador}, {Taris}, {Tauran}, {Taylor},
  {Teixeira}, {Terrett}, {Tingley}, {Trager}, {Turon}, {Ulla}, {Utrilla},
  {Valentini}, {van Elteren}, {Van Hemelryck}, {van Leeuwen}, {Varadi},
  {Vecchiato}, {Veljanoski}, {Via}, {Vicente}, {Vogt}, {Voss}, {Votruba},
  {Voutsinas}, {Walmsley}, {Weiler}, {Weingrill}, {Werner}, {Wevers},
  {Whitehead}, {Wyrzykowski}, {Yoldas}, {{\v{Z}}erjal}, {Zucker}, {Zurbach},
  {Zwitter}, {Alecu}, {Allen}, {Allende Prieto}, {Amorim},
  {Anglada-Escud{\'e}}, {Arsenijevic}, {Azaz}, {Balm}, {Beck}, {Bernstein},
  {Bigot}, {Bijaoui}, {Blasco}, {Bonfigli}, {Bono}, {Boudreault}, {Bressan},
  {Brown}, {Brunet}, {Bunclark}, {Buonanno}, {Butkevich}, {Carret}, {Carrion},
  {Chemin}, {Ch{\'e}reau}, {Corcione}, {Darmigny}, {de Boer}, {de Teodoro}, {de
  Zeeuw}, {Delle Luche}, {Domingues}, {Dubath}, {Fodor}, {Fr{\'e}zouls},
  {Fries}, {Fustes}, {Fyfe}, {Gallardo}, {Gallegos}, {Gardiol}, {Gebran},
  {Gomboc}, {G{\'o}mez}, {Grux}, {Gueguen}, {Heyrovsky}, {Hoar}, {Iannicola},
  {Isasi Parache}, {Janotto}, {Joliet}, {Jonckheere}, {Keil}, {Kim},
  {Klagyivik}, {Klar}, {Knude}, {Kochukhov}, {Kolka}, {Kos}, {Kutka}, {Lainey},
  {LeBouquin}, {Liu}, {Loreggia}, {Makarov}, {Marseille}, {Martayan},
  {Martinez-Rubi}, {Massart}, {Meynadier}, {Mignot}, {Munari}, {Nguyen},
  {Nordlander}, {Ocvirk}, {O'Flaherty}, {Olias Sanz}, {Ortiz}, {Osorio},
  {Oszkiewicz}, {Ouzounis}, {Palmer}, {Park}, {Pasquato}, {Peltzer}, {Peralta},
  {P{\'e}turaud}, {Pieniluoma}, {Pigozzi}, {Poels}, {Prat}, {Prod'homme},
  {Raison}, {Rebordao}, {Risquez}, {Rocca-Volmerange}, {Rosen}, {Ruiz-Fuertes},
  {Russo}, {Sembay}, {Serraller Vizcaino}, {Short}, {Siebert}, {Silva},
  {Sinachopoulos}, {Slezak}, {Soffel}, {Sosnowska}, {Strai{\v{z}}ys}, {ter
  Linden}, {Terrell}, {Theil}, {Tiede}, {Troisi}, {Tsalmantza}, {Tur},
  {Vaccari}, {Vachier}, {Valles}, {Van Hamme}, {Veltz}, {Virtanen}, {Wallut},
  {Wichmann}, {Wilkinson}, {Ziaeepour}, \& {Zschocke}}]{Gaia2016}
{Gaia Collaboration}, {Prusti}, T., {de Bruijne}, J.~H.~J., {et~al.} 2016,
  \aap, 595, A1, \dodoi{10.1051/0004-6361/201629272}

\bibitem[{{Gaia Collaboration} {et~al.}(2018){Gaia Collaboration}, {Brown},
  {Vallenari}, {Prusti}, {de Bruijne}, {Babusiaux}, {Bailer-Jones}, {Biermann},
  {Evans}, {Eyer}, \& et~al.}]{Gaia2018}
{Gaia Collaboration}, {Brown}, A.~G.~A., {Vallenari}, A., {et~al.} 2018, \aap,
  616, A1, \dodoi{10.1051/0004-6361/201833051}

\bibitem[{{Gelman} \& {Rubin}(1992)}]{Gelman1992}
{Gelman}, A., \& {Rubin}, D.~B. 1992, Statistical Science, 7, 457,
  \dodoi{10.1214/ss/1177011136}

\bibitem[{{Gibson} {et~al.}(2016){Gibson}, {Howard}, {Marcy}, {Edelstein},
  {Wishnow}, \& {Poppett}}]{Gibson2016}
{Gibson}, S.~R., {Howard}, A.~W., {Marcy}, G.~W., {et~al.} 2016, in Society of
  Photo-Optical Instrumentation Engineers (SPIE) Conference Series, Vol. 9908,
  Ground-based and Airborne Instrumentation for Astronomy VI, ed. C.~J.
  {Evans}, L.~{Simard}, \& H.~{Takami}, 990870, \dodoi{10.1117/12.2233334}

\bibitem[{{Gill} {et~al.}(2020){Gill}, {Wheatley}, {Cooke}, {Jord{\'a}n},
  {Nielsen}, {Bayliss}, {Anderson}, {Vines}, {Lendl}, {Acton}, {Armstrong},
  {Bouchy}, {Brahm}, {Bryant}, {Burleigh}, {Casewell}, {Eigm{\"u}ller},
  {Espinoza}, {Gillen}, {Goad}, {Grieves}, {G{\"u}nther}, {Henning}, {Hobson},
  {Hogan}, {Jenkins}, {McCormac}, {Moyano}, {Osborn}, {Pollacco}, {Queloz},
  {Rauer}, {Raynard}, {Rojas}, {Sarkis}, {Smith}, {Pinto}, {Tilbrook}, {Udry},
  {Watson}, \& {West}}]{Gill2020b}
{Gill}, S., {Wheatley}, P.~J., {Cooke}, B.~F., {et~al.} 2020, \apjl, 898, L11,
  \dodoi{10.3847/2041-8213/ab9eb9}

\bibitem[{{Ginzburg} \& {Chiang}(2020)}]{Ginzburg2020}
{Ginzburg}, S., \& {Chiang}, E. 2020, \mnras, 498, 680,
  \dodoi{10.1093/mnras/staa2500}

\bibitem[{{Goldreich} \& {Tremaine}(1980)}]{Goldreich1980}
{Goldreich}, P., \& {Tremaine}, S. 1980, \apj, 241, 425, \dodoi{10.1086/158356}

\bibitem[{{Gong} {et~al.}(2013){Gong}, {Zhou}, {Xie}, \& {Wu}}]{Gong2013}
{Gong}, Y.-X., {Zhou}, J.-L., {Xie}, J.-W., \& {Wu}, X.-M. 2013, \apjl, 769,
  L14, \dodoi{10.1088/2041-8205/769/1/L14}

\bibitem[{{Gonzalez}(1997)}]{Gonzalez1997}
{Gonzalez}, G. 1997, \mnras, 285, 403, \dodoi{10.1093/mnras/285.2.403}

\bibitem[{{Greene} {et~al.}(2016){Greene}, {Line}, {Montero}, {Fortney},
  {Lustig-Yaeger}, \& {Luther}}]{Greene2016}
{Greene}, T.~P., {Line}, M.~R., {Montero}, C., {et~al.} 2016, \apj, 817, 17,
  \dodoi{10.3847/0004-637X/817/1/17}

\bibitem[{{Guillot} {et~al.}(2018){Guillot}, {Miguel}, {Militzer}, {Hubbard},
  {Kaspi}, {Galanti}, {Cao}, {Helled}, {Wahl}, {Iess}, {Folkner}, {Stevenson},
  {Lunine}, {Reese}, {Biekman}, {Parisi}, {Durante}, {Connerney}, {Levin}, \&
  {Bolton}}]{Guillot2018}
{Guillot}, T., {Miguel}, Y., {Militzer}, B., {et~al.} 2018, \nat, 555, 227,
  \dodoi{10.1038/nature25775}

\bibitem[{{Hamers} {et~al.}(2017){Hamers}, {Antonini}, {Lithwick}, {Perets}, \&
  {Portegies Zwart}}]{Hamers2017a}
{Hamers}, A.~S., {Antonini}, F., {Lithwick}, Y., {Perets}, H.~B., \& {Portegies
  Zwart}, S.~F. 2017, \mnras, 464, 688, \dodoi{10.1093/mnras/stw2370}

\bibitem[{{Hamilton} \& {Burns}(1991)}]{Hamilton1991}
{Hamilton}, D.~P., \& {Burns}, J.~A. 1991, \icarus, 92, 118,
  \dodoi{10.1016/0019-1035(91)90039-V}

\bibitem[{{Hansen} \& {Barman}(2007)}]{Hansen2007}
{Hansen}, B. M.~S., \& {Barman}, T. 2007, \apj, 671, 861,
  \dodoi{10.1086/523038}

\bibitem[{{Hayward} {et~al.}(2001){Hayward}, {Brandl}, {Pirger}, {Blacken},
  {Gull}, {Schoenwald}, \& {Houck}}]{Hayward2001}
{Hayward}, T.~L., {Brandl}, B., {Pirger}, B., {et~al.} 2001, \pasp, 113, 105,
  \dodoi{10.1086/317969}

\bibitem[{{Heller}(2012)}]{Heller2012}
{Heller}, R. 2012, \aap, 545, L8, \dodoi{10.1051/0004-6361/201220003}

\bibitem[{{Heller} \& {Barnes}(2013)}]{Heller2013}
{Heller}, R., \& {Barnes}, R. 2013, Astrobiology, 13, 18,
  \dodoi{10.1089/ast.2012.0859}

\bibitem[{{Heller} {et~al.}(2014){Heller}, {Williams}, {Kipping}, {Limbach},
  {Turner}, {Greenberg}, {Sasaki}, {Bolmont}, {Grasset}, {Lewis}, {Barnes}, \&
  {Zuluaga}}]{Heller2014b}
{Heller}, R., {Williams}, D., {Kipping}, D., {et~al.} 2014, Astrobiology, 14,
  798, \dodoi{10.1089/ast.2014.1147}

\bibitem[{{Hill} {et~al.}(2018){Hill}, {Kane}, {Seperuelo Duarte}, {Kopparapu},
  {Gelino}, \& {Wittenmyer}}]{Hill2018}
{Hill}, M.~L., {Kane}, S.~R., {Seperuelo Duarte}, E., {et~al.} 2018, \apj, 860,
  67, \dodoi{10.3847/1538-4357/aac384}

\bibitem[{{Hinse} {et~al.}(2010){Hinse}, {Christou}, {Alvarellos}, \&
  {Go{\'z}dziewski}}]{Hinse2010}
{Hinse}, T.~C., {Christou}, A.~A., {Alvarellos}, J.~L.~A., \&
  {Go{\'z}dziewski}, K. 2010, \mnras, 404, 837,
  \dodoi{10.1111/j.1365-2966.2010.16307.x}

\bibitem[{{Hong} {et~al.}(2018){Hong}, {Raymond}, {Nicholson}, \&
  {Lunine}}]{Hong2018}
{Hong}, Y.-C., {Raymond}, S.~N., {Nicholson}, P.~D., \& {Lunine}, J.~I. 2018,
  \apj, 852, 85, \dodoi{10.3847/1538-4357/aaa0db}

\bibitem[{{Howard} \& {Fulton}(2016)}]{Howard2016}
{Howard}, A.~W., \& {Fulton}, B.~J. 2016, \pasp, 128, 114401,
  \dodoi{10.1088/1538-3873/128/969/114401}

\bibitem[{{Howard} {et~al.}(2010){Howard}, {Johnson}, {Marcy}, {Fischer},
  {Wright}, {Bernat}, {Henry}, {Peek}, {Isaacson}, {Apps}, {Endl}, {Cochran},
  {Valenti}, {Anderson}, \& {Piskunov}}]{Howard2010}
{Howard}, A.~W., {Johnson}, J.~A., {Marcy}, G.~W., {et~al.} 2010, \apj, 721,
  1467, \dodoi{10.1088/0004-637X/721/2/1467}

\bibitem[{{Hut}(1981)}]{Hut1981}
{Hut}, P. 1981, \aap, 99, 126

\bibitem[{{Iess} {et~al.}(2019){Iess}, {Militzer}, {Kaspi}, {Nicholson},
  {Durante}, {Racioppa}, {Anabtawi}, {Galanti}, {Hubbard}, {Mariani},
  {Tortora}, {Wahl}, \& {Zannoni}}]{Iess2019}
{Iess}, L., {Militzer}, B., {Kaspi}, Y., {et~al.} 2019, Science, 364, aat2965,
  \dodoi{10.1126/science.aat2965}

\bibitem[{{Ikwut-Ukwa} {et~al.}(2020){Ikwut-Ukwa}, {Rodriguez}, {Bieryla},
  {Vanderburg}, {Mocnik}, {Kane}, {Quinn}, {Col{\'o}n}, {Zhou}, {Eastman},
  {Huang}, {Latham}, {Dotson}, {Jenkins}, {Ricker}, {Seager}, {Vanderspek},
  {Winn}, {Barclay}, {Barentsen}, {Berta-Thompson}, {Charbonneau}, {Dragomir},
  {Daylan}, {G{\"u}nther}, {Hedges}, {Henze}, {McDermott}, {Schlieder},
  {Quintana}, {Smith}, {Twicken}, \& {Yahalomi}}]{IkwutUkwa2020}
{Ikwut-Ukwa}, M., {Rodriguez}, J.~E., {Bieryla}, A., {et~al.} 2020, \aj, 160,
  209, \dodoi{10.3847/1538-3881/aba964}

\bibitem[{{Irwin} {et~al.}(2014){Irwin}, {Barstow}, {Bowles}, {Fletcher},
  {Aigrain}, \& {Lee}}]{Irwin2014}
{Irwin}, P.~G.~J., {Barstow}, J.~K., {Bowles}, N.~E., {et~al.} 2014, \icarus,
  242, 172, \dodoi{10.1016/j.icarus.2014.08.005}

\bibitem[{{Irwin} {et~al.}(1998){Irwin}, {Weir}, {Smith}, {Taylor}, {Lambert},
  {Calcutt}, {Cameron-Smith}, {Carlson}, {Baines}, {Orton}, {Drossart},
  {Encrenaz}, \& {Roos-Serote}}]{Irwin1998}
{Irwin}, P.~G.~J., {Weir}, A.~L., {Smith}, S.~E., {et~al.} 1998, \jgr, 103,
  23001, \dodoi{10.1029/98JE00948}

\bibitem[{{Isaacson} \& {Fischer}(2010)}]{Isaacson2010}
{Isaacson}, H., \& {Fischer}, D. 2010, \apj, 725, 875,
  \dodoi{10.1088/0004-637X/725/1/875}

\bibitem[{{Jackson} {et~al.}(2019){Jackson}, {Dawson}, \&
  {Zalesky}}]{Jackson2019}
{Jackson}, J.~M., {Dawson}, R.~I., \& {Zalesky}, J. 2019, \aj, 157, 166,
  \dodoi{10.3847/1538-3881/ab09eb}

\bibitem[{{Jenkins} {et~al.}(2010){Jenkins}, {Caldwell}, {Chandrasekaran},
  {Twicken}, {Bryson}, {Quintana}, {Clarke}, {Li}, {Allen}, {Tenenbaum}, {Wu},
  {Klaus}, {Middour}, {Cote}, {McCauliff}, {Girouard}, {Gunter}, {Wohler},
  {Sommers}, {Hall}, {Uddin}, {Wu}, {Bhavsar}, {Van Cleve}, {Pletcher},
  {Dotson}, {Haas}, {Gilliland}, {Koch}, \& {Borucki}}]{Jenkins2010a}
{Jenkins}, J.~M., {Caldwell}, D.~A., {Chandrasekaran}, H., {et~al.} 2010,
  \apjl, 713, L87, \dodoi{10.1088/2041-8205/713/2/L87}

\bibitem[{{Juri{\'c}} \& {Tremaine}(2008)}]{Juric2008}
{Juri{\'c}}, M., \& {Tremaine}, S. 2008, \apj, 686, 603, \dodoi{10.1086/590047}

\bibitem[{{Kane}(2007)}]{Kane2007}
{Kane}, S.~R. 2007, \mnras, 380, 1488, \dodoi{10.1111/j.1365-2966.2007.12144.x}

\bibitem[{{Kane} \& {Gelino}(2011)}]{Kane2011e}
{Kane}, S.~R., \& {Gelino}, D.~M. 2011, \apj, 741, 52,
  \dodoi{10.1088/0004-637X/741/1/52}

\bibitem[{{Kane} {et~al.}(2019){Kane}, {Dalba}, {Li}, {Horch}, {Hirsch},
  {Horner}, {Wittenmyer}, {Howell}, {Everett}, {Butler}, {Tinney}, {Carter},
  {Wright}, {Jones}, {Bailey}, \& {O{\textquoteright}Toole}}]{Kane2019b}
{Kane}, S.~R., {Dalba}, P.~A., {Li}, Z., {et~al.} 2019, \aj, 157, 252,
  \dodoi{10.3847/1538-3881/ab1ddf}

\bibitem[{{Kipping}(2010)}]{Kipping2010a}
{Kipping}, D.~M. 2010, \mnras, 407, 301,
  \dodoi{10.1111/j.1365-2966.2010.16894.x}

\bibitem[{{Kipping}(2013)}]{Kipping2013a}
---. 2013, \mnras, 434, L51, \dodoi{10.1093/mnrasl/slt075}

\bibitem[{{Kipping} {et~al.}(2012{\natexlab{a}}){Kipping}, {Bakos}, {Buchhave},
  {Nesvorn{\'y}}, \& {Schmitt}}]{Kipping2012b}
{Kipping}, D.~M., {Bakos}, G.~{\'A}., {Buchhave}, L., {Nesvorn{\'y}}, D., \&
  {Schmitt}, A. 2012{\natexlab{a}}, \apj, 750, 115,
  \dodoi{10.1088/0004-637X/750/2/115}

\bibitem[{{Kipping} {et~al.}(2012{\natexlab{b}}){Kipping}, {Dunn}, {Jasinski},
  \& {Manthri}}]{Kipping2012a}
{Kipping}, D.~M., {Dunn}, W.~R., {Jasinski}, J.~M., \& {Manthri}, V.~P.
  2012{\natexlab{b}}, \mnras, 421, 1166,
  \dodoi{10.1111/j.1365-2966.2011.20376.x}

\bibitem[{{Kipping} {et~al.}(2016){Kipping}, {Torres}, {Henze}, {Teachey},
  {Isaacson}, {Petigura}, {Marcy}, {Buchhave}, {Chen}, {Bryson}, \&
  {Sandford}}]{Kipping2016b}
{Kipping}, D.~M., {Torres}, G., {Henze}, C., {et~al.} 2016, \apj, 820, 112,
  \dodoi{10.3847/0004-637X/820/2/112}

\bibitem[{{Kov{\'a}cs} {et~al.}(2002){Kov{\'a}cs}, {Zucker}, \&
  {Mazeh}}]{Kovacs2002}
{Kov{\'a}cs}, G., {Zucker}, S., \& {Mazeh}, T. 2002, \aap, 391, 369,
  \dodoi{10.1051/0004-6361:20020802}

\bibitem[{{Kozai}(1962)}]{Kozai1962}
{Kozai}, Y. 1962, \aj, 67, 591, \dodoi{10.1086/108790}

\bibitem[{{Kraft}(1967)}]{Kraft1967}
{Kraft}, R.~P. 1967, \apj, 150, 551, \dodoi{10.1086/149359}

\bibitem[{{Langton} \& {Laughlin}(2008)}]{Langton2008}
{Langton}, J., \& {Laughlin}, G. 2008, \apj, 674, 1106, \dodoi{10.1086/523957}

\bibitem[{{Laughlin}(2018)}]{Laughlin2018}
{Laughlin}, G. 2018, {Mass-Radius Relations of Giant Planets: The Radius
  Anomaly and Interior Models}, ed. H.~J. {Deeg} \& J.~A. {Belmonte} (Cham:
  Springer), 1, \dodoi{10.1007/978-3-319-55333-7_1}

\bibitem[{{Lazio} {et~al.}(2010){Lazio}, {Shankland}, {Farrell}, \&
  {Blank}}]{Lazio2010}
{Lazio}, T. J.~W., {Shankland}, P.~D., {Farrell}, W.~M., \& {Blank}, D.~L.
  2010, \aj, 140, 1929, \dodoi{10.1088/0004-6256/140/6/1929}

\bibitem[{{Lazio} {et~al.}(2004){Lazio}, {Farrell}, {Dietrick}, {Greenlees},
  {Hogan}, {Jones}, \& {Hennig}}]{Lazio2004}
{Lazio}, T.~Joseph, W., {Farrell}, W.~M., {Dietrick}, J., {et~al.} 2004, \apj,
  612, 511, \dodoi{10.1086/422449}

\bibitem[{{Lendl} {et~al.}(2020){Lendl}, {Bouchy}, {Gill}, {Nielsen}, {Turner},
  {Stassun}, {Acton}, {Anderson}, {Armstrong}, {Bayliss}, {Belardi}, {Bryant},
  {Burleigh}, {Chaushev}, {Casewell}, {Cooke}, {Eigm{\"u}ller}, {Gillen},
  {Goad}, {G{\"u}nther}, {Hagelberg}, {Jenkins}, {Louden}, {Marmier},
  {McCormac}, {Moyano}, {Pollacco}, {Raynard}, {Tilbrook}, {Udry}, {Vines},
  {West}, {Wheatley}, {Ricker}, {Vanderspek}, {Latham}, {Seager}, {Winn},
  {Jenkins}, {Addison}, {Brice{\~n}o}, {Brahm}, {Caldwell}, {Doty}, {Espinoza},
  {Goeke}, {Henning}, {Jord{\'a}n}, {Krishnamurthy}, {Law}, {Morris},
  {Okumura}, {Mann}, {Rodriguez}, {Sarkis}, {Schlieder}, {Twicken},
  {Villanueva}, {Wittenmyer}, {Wright}, \& {Ziegler}}]{Lendl2020}
{Lendl}, M., {Bouchy}, F., {Gill}, S., {et~al.} 2020, \mnras, 492, 1761,
  \dodoi{10.1093/mnras/stz3545}

\bibitem[{{Lidov}(1962)}]{Lidov1962}
{Lidov}, M.~L. 1962, \planss, 9, 719, \dodoi{10.1016/0032-0633(62)90129-0}

\bibitem[{{Lightkurve Collaboration} {et~al.}(2018){Lightkurve Collaboration},
  {Cardoso}, {Hedges}, {Gully-Santiago}, {Saunders}, {Cody}, {Barclay}, {Hall},
  {Sagear}, {Turtelboom}, {Zhang}, {Tzanidakis}, {Mighell}, {Coughlin}, {Bell},
  {Berta-Thompson}, {Williams}, {Dotson}, \& {Barentsen}}]{Lightkurve2018}
{Lightkurve Collaboration}, {Cardoso}, J.~V.~d.~M., {Hedges}, C., {et~al.}
  2018, {Lightkurve: Kepler and TESS time series analysis in Python}, 1.11,
  Astrophysics Source Code Library.
\newblock \doeprint{1812.013}

\bibitem[{{Lin} \& {Papaloizou}(1986)}]{Lin1986}
{Lin}, D.~N.~C., \& {Papaloizou}, J. 1986, \apj, 309, 846,
  \dodoi{10.1086/164653}

\bibitem[{{Lindegren} {et~al.}(2020){Lindegren}, {Klioner}, {Hern{\'a}ndez},
  {Bombrun}, {Ramos-Lerate}, {Steidelm{\"u}ller}, {Bastian}, {Biermann}, {de
  Torres}, {Gerlach}, {Geyer}, {Hilger}, {Hobbs}, {Lammers}, {McMillan},
  {Stephenson}, {Casta{\~n}eda}, {Davidson}, {Fabricius}, {Gracia-Abril},
  {Portell}, {Rowell}, {Teyssier}, {Torra}, {Bartolom{\'e}}, {Clotet},
  {Garralda}, {Gonz{\'a}lez-Vidal}, {Torra}, {Abbas}, {Altmann}, {Anglada
  Varela}, {Balaguer-N{\'u}{\~n}ez}, {Balog}, {Barache}, {Becciani}, {Bernet},
  {Bertone}, {Bianchi}, {Bouquillon}, {Brown}, {Bucciarelli}, {Busonero},
  {Butkevich}, {Buzzi}, {Cancelliere}, {Carlucci}, {Charlot}, {Cioni},
  {Crosta}, {Crowley}, {del Peloso}, {del Pozo}, {Drimmel}, {Esquej}, {Fienga},
  {Fraile}, {Gai}, {Garcia-Reinaldos}, {Guerra}, {Hambly}, {Hauser},
  {Jan{\ss}en}, {Jordan}, {Kostrzewa-Rutkowska}, {Lattanzi}, {Liao}, {Licata},
  {Lister}, {L{\"o}ffler}, {Marchant}, {Masip}, {Mignard}, {Mints}, {Molina},
  {Mora}, {Morbidelli}, {Murphy}, {Pagani}, {Panuzzo}, {Pe{\~n}alosa Esteller},
  {Poggio}, {Re Fiorentin}, {Riva}, {Sagrist{\`a} Sell{\'e}s}, {Sanchez
  Gimenez}, {Sarasso}, {Sciacca}, {Siddiqui}, {Smart}, {Souami}, {Spagna},
  {Steele}, {Taris}, {Utrilla}, {van Reeven}, \& {Vecchiato}}]{Lindegren2020}
{Lindegren}, L., {Klioner}, S.~A., {Hern{\'a}ndez}, J., {et~al.} 2020, arXiv
  e-prints, arXiv:2012.03380.
\newblock \doarXiv{2012.03380}

\bibitem[{{Lithwick} \& {Naoz}(2011)}]{Lithwick2011b}
{Lithwick}, Y., \& {Naoz}, S. 2011, \apj, 742, 94,
  \dodoi{10.1088/0004-637X/742/2/94}

\bibitem[{{Lithwick} \& {Wu}(2011)}]{Lithwick2011a}
{Lithwick}, Y., \& {Wu}, Y. 2011, \apj, 739, 31,
  \dodoi{10.1088/0004-637X/739/1/31}

\bibitem[{{Martinez} {et~al.}(2019){Martinez}, {Stone}, \&
  {Metzger}}]{Martinez2019}
{Martinez}, M. A.~S., {Stone}, N.~C., \& {Metzger}, B.~D. 2019, \mnras, 489,
  5119, \dodoi{10.1093/mnras/stz2464}

\bibitem[{{Mayorga} {et~al.}(2021){Mayorga}, {Robinson}, {Marley}, {May}, \&
  {Stevenson}}]{Mayorga2021}
{Mayorga}, L.~C., {Robinson}, T.~D., {Marley}, M.~S., {May}, E.~M., \&
  {Stevenson}, K.~B. 2021, arXiv e-prints, arXiv:2105.08009.
\newblock \doarXiv{2105.08009}

\bibitem[{{McLaughlin}(1924)}]{McLaughlin1924}
{McLaughlin}, D.~B. 1924, \apj, 60, 22, \dodoi{10.1086/142826}

\bibitem[{{Miller} \& {Fortney}(2011)}]{Miller2011}
{Miller}, N., \& {Fortney}, J.~J. 2011, \apjl, 736, L29,
  \dodoi{10.1088/2041-8205/736/2/L29}

\bibitem[{{Montet} {et~al.}(2014){Montet}, {Crepp}, {Johnson}, {Howard}, \&
  {Marcy}}]{Montet2014}
{Montet}, B.~T., {Crepp}, J.~R., {Johnson}, J.~A., {Howard}, A.~W., \& {Marcy},
  G.~W. 2014, \apj, 781, 28, \dodoi{10.1088/0004-637X/781/1/28}

\bibitem[{{Moorhead} \& {Adams}(2005)}]{Moorhead2005}
{Moorhead}, A.~V., \& {Adams}, F.~C. 2005, \icarus, 178, 517,
  \dodoi{10.1016/j.icarus.2005.05.005}

\bibitem[{{Mordasini} {et~al.}(2012){Mordasini}, {Alibert}, {Georgy},
  {Dittkrist}, {Klahr}, \& {Henning}}]{Mordasini2012}
{Mordasini}, C., {Alibert}, Y., {Georgy}, C., {et~al.} 2012, \aap, 547, A112,
  \dodoi{10.1051/0004-6361/201118464}

\bibitem[{{Mordasini} {et~al.}(2014){Mordasini}, {Klahr}, {Alibert}, {Miller},
  \& {Henning}}]{Mordasini2014}
{Mordasini}, C., {Klahr}, H., {Alibert}, Y., {Miller}, N., \& {Henning}, T.
  2014, \aap, 566, A141, \dodoi{10.1051/0004-6361/201321479}

\bibitem[{{Mousis} {et~al.}(2009){Mousis}, {Marboeuf}, {Lunine}, {Alibert},
  {Fletcher}, {Orton}, {Pauzat}, \& {Ellinger}}]{Mousis2009}
{Mousis}, O., {Marboeuf}, U., {Lunine}, J.~I., {et~al.} 2009, \apj, 696, 1348,
  \dodoi{10.1088/0004-637X/696/2/1348}

\bibitem[{{Moutou} {et~al.}(2009){Moutou}, {H{\'e}brard}, {Bouchy},
  {Eggenberger}, {Boisse}, {Bonfils}, {Gravallon}, {Ehrenreich}, {Forveille},
  {Delfosse}, {Desort}, {Lagrange}, {Lovis}, {Mayor}, {Pepe}, {Perrier},
  {Pont}, {Queloz}, {Santos}, {S{\'e}gransan}, {Udry}, \&
  {Vidal-Madjar}}]{Moutou2009}
{Moutou}, C., {H{\'e}brard}, G., {Bouchy}, F., {et~al.} 2009, \aap, 498, L5,
  \dodoi{10.1051/0004-6361/200911954}

\bibitem[{{M{\"u}ller} {et~al.}(2020{\natexlab{a}}){M{\"u}ller}, {Ben-Yami}, \&
  {Helled}}]{Muller2020b}
{M{\"u}ller}, S., {Ben-Yami}, M., \& {Helled}, R. 2020{\natexlab{a}}, \apj,
  903, 147, \dodoi{10.3847/1538-4357/abba19}

\bibitem[{{M{\"u}ller} {et~al.}(2020{\natexlab{b}}){M{\"u}ller}, {Helled}, \&
  {Cumming}}]{Muller2020a}
{M{\"u}ller}, S., {Helled}, R., \& {Cumming}, A. 2020{\natexlab{b}}, \aap, 638,
  A121, \dodoi{10.1051/0004-6361/201937376}

\bibitem[{{Naef} {et~al.}(2001){Naef}, {Latham}, {Mayor}, {Mazeh}, {Beuzit},
  {Drukier}, {Perrier-Bellet}, {Queloz}, {Sivan}, {Torres}, {Udry}, \&
  {Zucker}}]{Naef2001}
{Naef}, D., {Latham}, D.~W., {Mayor}, M., {et~al.} 2001, \aap, 375, L27,
  \dodoi{10.1051/0004-6361:20010853}

\bibitem[{{Nagasawa} \& {Ida}(2011)}]{Nagasawa2011}
{Nagasawa}, M., \& {Ida}, S. 2011, \apj, 742, 72,
  \dodoi{10.1088/0004-637X/742/2/72}

\bibitem[{{Nagasawa} {et~al.}(2008){Nagasawa}, {Ida}, \&
  {Bessho}}]{Nagasawa2008}
{Nagasawa}, M., {Ida}, S., \& {Bessho}, T. 2008, \apj, 678, 498,
  \dodoi{10.1086/529369}

\bibitem[{{Namouni}(2010)}]{Namouni2010}
{Namouni}, F. 2010, \apjl, 719, L145, \dodoi{10.1088/2041-8205/719/2/L145}

\bibitem[{{Naoz}(2016)}]{Naoz2016}
{Naoz}, S. 2016, \araa, 54, 441, \dodoi{10.1146/annurev-astro-081915-023315}

\bibitem[{{Naoz} {et~al.}(2011){Naoz}, {Farr}, {Lithwick}, {Rasio}, \&
  {Teyssandier}}]{Naoz2011}
{Naoz}, S., {Farr}, W.~M., {Lithwick}, Y., {Rasio}, F.~A., \& {Teyssandier}, J.
  2011, Nature, 473, 187, \dodoi{10.1038/nature10076}

\bibitem[{{Naoz} {et~al.}(2012){Naoz}, {Farr}, \& {Rasio}}]{Naoz2012}
{Naoz}, S., {Farr}, W.~M., \& {Rasio}, F.~A. 2012, \apjl, 754, L36,
  \dodoi{10.1088/2041-8205/754/2/L36}

\bibitem[{{Narita} {et~al.}(2008){Narita}, {Sato}, {Ohshima}, \&
  {Winn}}]{Narita2008}
{Narita}, N., {Sato}, B., {Ohshima}, O., \& {Winn}, J.~N. 2008, \pasj, 60, L1,
  \dodoi{10.1093/pasj/60.2.L1}

\bibitem[{{Nesvorn{\'y}} {et~al.}(2007){Nesvorn{\'y}}, {Vokrouhlick{\'y}}, \&
  {Morbidelli}}]{Nesvorny2007}
{Nesvorn{\'y}}, D., {Vokrouhlick{\'y}}, D., \& {Morbidelli}, A. 2007, \aj, 133,
  1962, \dodoi{10.1086/512850}

\bibitem[{{Niemann} {et~al.}(1998){Niemann}, {Atreya}, {Carignan}, {Donahue},
  {Haberman}, {Harpold}, {Hartle}, {Hunten}, {Kasprzak}, {Mahaffy}, {Owen}, \&
  {Way}}]{Niemann1998}
{Niemann}, H.~B., {Atreya}, S.~K., {Carignan}, G.~R., {et~al.} 1998, \jgr, 103,
  22831, \dodoi{10.1029/98JE01050}

\bibitem[{{Papaloizou} {et~al.}(2001){Papaloizou}, {Nelson}, \&
  {Masset}}]{Papaloizou2001}
{Papaloizou}, J.~C.~B., {Nelson}, R.~P., \& {Masset}, F. 2001, \aap, 366, 263,
  \dodoi{10.1051/0004-6361:20000011}

\bibitem[{{Paxton} {et~al.}(2011){Paxton}, {Bildsten}, {Dotter}, {Herwig},
  {Lesaffre}, \& {Timmes}}]{Paxton2011}
{Paxton}, B., {Bildsten}, L., {Dotter}, A., {et~al.} 2011, \apjs, 192, 3,
  \dodoi{10.1088/0067-0049/192/1/3}

\bibitem[{{Paxton} {et~al.}(2013){Paxton}, {Cantiello}, {Arras}, {Bildsten},
  {Brown}, {Dotter}, {Mankovich}, {Montgomery}, {Stello}, {Timmes}, \&
  {Townsend}}]{Paxton2013}
{Paxton}, B., {Cantiello}, M., {Arras}, P., {et~al.} 2013, \apjs, 208, 4,
  \dodoi{10.1088/0067-0049/208/1/4}

\bibitem[{{Paxton} {et~al.}(2015){Paxton}, {Marchant}, {Schwab}, {Bauer},
  {Bildsten}, {Cantiello}, {Dessart}, {Farmer}, {Hu}, {Langer}, {Townsend},
  {Townsley}, \& {Timmes}}]{Paxton2015}
{Paxton}, B., {Marchant}, P., {Schwab}, J., {et~al.} 2015, \apjs, 220, 15,
  \dodoi{10.1088/0067-0049/220/1/15}

\bibitem[{{Petigura}(2015)}]{Petigura2015}
{Petigura}, E.~A. 2015, PhD thesis, University of California, Berkeley

\bibitem[{{Petigura} {et~al.}(2017){Petigura}, {Howard}, {Marcy}, {Johnson},
  {Isaacson}, {Cargile}, {Hebb}, {Fulton}, {Weiss}, {Morton}, {Winn}, {Rogers},
  {Sinukoff}, {Hirsch}, \& {Crossfield}}]{Petigura2017b}
{Petigura}, E.~A., {Howard}, A.~W., {Marcy}, G.~W., {et~al.} 2017, \aj, 154,
  107, \dodoi{10.3847/1538-3881/aa80de}

\bibitem[{{Rasio} \& {Ford}(1996)}]{Rasio1996}
{Rasio}, F.~A., \& {Ford}, E.~B. 1996, \sci, 274, 954,
  \dodoi{10.1126/science.274.5289.954}

\bibitem[{{Raymond} {et~al.}(2010){Raymond}, {Armitage}, \&
  {Gorelick}}]{Raymond2010}
{Raymond}, S.~N., {Armitage}, P.~J., \& {Gorelick}, N. 2010, \apj, 711, 772,
  \dodoi{10.1088/0004-637X/711/2/772}

\bibitem[{{Rein} \& {Liu}(2012)}]{Rein2012}
{Rein}, H., \& {Liu}, S.~F. 2012, \aap, 537, A128,
  \dodoi{10.1051/0004-6361/201118085}

\bibitem[{{Rein} \& {Tamayo}(2015)}]{Rein2015}
{Rein}, H., \& {Tamayo}, D. 2015, \mnras, 452, 376,
  \dodoi{10.1093/mnras/stv1257}

\bibitem[{{Ricker} {et~al.}(2015){Ricker}, {Winn}, {Vanderspek}, {Latham},
  {Bakos}, {Bean}, {Berta-Thompson}, {Brown}, {Buchhave}, {Butler}, {Butler},
  {Chaplin}, {Charbonneau}, {Christensen-Dalsgaard}, {Clampin}, {Deming},
  {Doty}, {De Lee}, {Dressing}, {Dunham}, {Endl}, {Fressin}, {Ge}, {Henning},
  {Holman}, {Howard}, {Ida}, {Jenkins}, {Jernigan}, {Johnson}, {Kaltenegger},
  {Kawai}, {Kjeldsen}, {Laughlin}, {Levine}, {Lin}, {Lissauer}, {MacQueen},
  {Marcy}, {McCullough}, {Morton}, {Narita}, {Paegert}, {Palle}, {Pepe},
  {Pepper}, {Quirrenbach}, {Rinehart}, {Sasselov}, {Sato}, {Seager},
  {Sozzetti}, {Stassun}, {Sullivan}, {Szentgyorgyi}, {Torres}, {Udry}, \&
  {Villasenor}}]{Ricker2015}
{Ricker}, G.~R., {Winn}, J.~N., {Vanderspek}, R., {et~al.} 2015, \jatis, 1,
  014003, \dodoi{10.1117/1.JATIS.1.1.014003}

\bibitem[{{Rosenthal} {et~al.}(2021){Rosenthal}, {Fulton}, {Hirsch},
  {Isaacson}, {Howard}, {Dedrick}, {Sherstyuk}, {Blunt}, {Petigura}, {Knutson},
  {Behmard}, {Chontos}, {Crepp}, {Crossfield}, {Dalba}, {Fischer}, {Henry},
  {Kane}, {Kosiarek}, {Marcy}, {Rubenzahl}, {Weiss}, \&
  {Wright}}]{Rosenthal2021}
{Rosenthal}, L.~J., {Fulton}, B.~J., {Hirsch}, L.~A., {et~al.} 2021, arXiv
  e-prints, arXiv:2105.11583.
\newblock \doarXiv{2105.11583}

\bibitem[{{Rossiter}(1924)}]{Rossiter1924}
{Rossiter}, R.~A. 1924, \apj, 60, 15, \dodoi{10.1086/142825}

\bibitem[{Salvatier {et~al.}(2016)Salvatier, Wiecki, \& Fonnesbeck}]{pymc3}
Salvatier, J., Wiecki, T.~V., \& Fonnesbeck, C. 2016, PeerJ Computer Science,
  2, e55

\bibitem[{{Santerne} {et~al.}(2014){Santerne}, {H{\'e}brard}, {Deleuil},
  {Havel}, {Correia}, {Almenara}, {Alonso}, {Arnold}, {Barros}, {Behrend},
  {Bernasconi}, {Boisse}, {Bonomo}, {Bouchy}, {Bruno}, {Damiani}, {D{\'\i}az},
  {Gravallon}, {Guillot}, {Labrevoir}, {Montagnier}, {Moutou}, {Rinner},
  {Santos}, {Abe}, {Audejean}, {Bendjoya}, {Gillier}, {Gregorio}, {Martinez},
  {Michelet}, {Montaigut}, {Poncy}, {Rivet}, {Rousseau}, {Roy}, {Suarez},
  {Vanhuysse}, \& {Verilhac}}]{Santerne2014}
{Santerne}, A., {H{\'e}brard}, G., {Deleuil}, M., {et~al.} 2014, \aap, 571,
  A37, \dodoi{10.1051/0004-6361/201424158}

\bibitem[{{Santerne} {et~al.}(2016){Santerne}, {Moutou}, {Tsantaki}, {Bouchy},
  {H{\'e}brard}, {Adibekyan}, {Almenara}, {Amard}, {Barros}, {Boisse},
  {Bonomo}, {Bruno}, {Courcol}, {Deleuil}, {Demangeon}, {D{\'{\i}}az},
  {Guillot}, {Havel}, {Montagnier}, {Rajpurohit}, {Rey}, \&
  {Santos}}]{Santerne2016}
{Santerne}, A., {Moutou}, C., {Tsantaki}, M., {et~al.} 2016, \aap, 587, A64,
  \dodoi{10.1051/0004-6361/201527329}

\bibitem[{{Santos} {et~al.}(2004){Santos}, {Israelian}, \&
  {Mayor}}]{Santos2004}
{Santos}, N.~C., {Israelian}, G., \& {Mayor}, M. 2004, \aap, 415, 1153,
  \dodoi{10.1051/0004-6361:20034469}

\bibitem[{{Schlafly} \& {Finkbeiner}(2011)}]{Schlafly2011}
{Schlafly}, E.~F., \& {Finkbeiner}, D.~P. 2011, \apj, 737, 103,
  \dodoi{10.1088/0004-637X/737/2/103}

\bibitem[{{Schlaufman}(2010)}]{Schlaufman2010}
{Schlaufman}, K.~C. 2010, \apj, 719, 602, \dodoi{10.1088/0004-637X/719/1/602}

\bibitem[{{Seager} \& {Sasselov}(2000)}]{Seager2000}
{Seager}, S., \& {Sasselov}, D.~D. 2000, \apj, 537, 916, \dodoi{10.1086/309088}

\bibitem[{{Seifahrt} {et~al.}(2018){Seifahrt}, {St{\"u}rmer}, {Bean}, \&
  {Schwab}}]{Seifahrt2018}
{Seifahrt}, A., {St{\"u}rmer}, J., {Bean}, J.~L., \& {Schwab}, C. 2018, in
  Society of Photo-Optical Instrumentation Engineers (SPIE) Conference Series,
  Vol. 10702, Ground-based and Airborne Instrumentation for Astronomy VII, ed.
  C.~J. {Evans}, L.~{Simard}, \& H.~{Takami}, 107026D,
  \dodoi{10.1117/12.2312936}

\bibitem[{{Sestovic} {et~al.}(2018){Sestovic}, {Demory}, \&
  {Queloz}}]{Sestovic2018}
{Sestovic}, M., {Demory}, B.-O., \& {Queloz}, D. 2018, \aap, 616, A76,
  \dodoi{10.1051/0004-6361/201731454}

\bibitem[{{Sheets} {et~al.}(2018){Sheets}, {Jacob}, {Cowan}, \&
  {Deming}}]{Sheets2018}
{Sheets}, H.~A., {Jacob}, L., {Cowan}, N.~B., \& {Deming}, D. 2018, Research
  Notes of the American Astronomical Society, 2, 153,
  \dodoi{10.3847/2515-5172/aadcb1}

\bibitem[{{Shibata} {et~al.}(2020){Shibata}, {Helled}, \&
  {Ikoma}}]{Shibata2020}
{Shibata}, S., {Helled}, R., \& {Ikoma}, M. 2020, \aap, 633, A33,
  \dodoi{10.1051/0004-6361/201936700}

\bibitem[{{Shibata} \& {Ikoma}(2019)}]{Shibata2019}
{Shibata}, S., \& {Ikoma}, M. 2019, \mnras, 487, 4510,
  \dodoi{10.1093/mnras/stz1629}

\bibitem[{{Sidis} \& {Sari}(2010)}]{Sidis2010}
{Sidis}, O., \& {Sari}, R. 2010, \apj, 720, 904,
  \dodoi{10.1088/0004-637X/720/1/904}

\bibitem[{{Smith} {et~al.}(2012){Smith}, {Stumpe}, {Van Cleve}, {Jenkins},
  {Barclay}, {Fanelli}, {Girouard}, {Kolodziejczak}, {McCauliff}, {Morris}, \&
  {Twicken}}]{Smith2012}
{Smith}, J.~C., {Stumpe}, M.~C., {Van Cleve}, J.~E., {et~al.} 2012, \pasp, 124,
  1000, \dodoi{10.1086/667697}

\bibitem[{{Socrates} {et~al.}(2012){Socrates}, {Katz}, {Dong}, \&
  {Tremaine}}]{Socrates2012}
{Socrates}, A., {Katz}, B., {Dong}, S., \& {Tremaine}, S. 2012, \apj, 750, 106,
  \dodoi{10.1088/0004-637X/750/2/106}

\bibitem[{{Spalding} {et~al.}(2016){Spalding}, {Batygin}, \&
  {Adams}}]{Spalding2016}
{Spalding}, C., {Batygin}, K., \& {Adams}, F.~C. 2016, \apj, 817, 18,
  \dodoi{10.3847/0004-637X/817/1/18}

\bibitem[{{Stumpe} {et~al.}(2012){Stumpe}, {Smith}, {Van Cleve}, {Twicken},
  {Barclay}, {Fanelli}, {Girouard}, {Jenkins}, {Kolodziejczak}, {McCauliff}, \&
  {Morris}}]{Stumpe2012}
{Stumpe}, M.~C., {Smith}, J.~C., {Van Cleve}, J.~E., {et~al.} 2012, \pasp, 124,
  985, \dodoi{10.1086/667698}

\bibitem[{{Sucerquia} {et~al.}(2020){Sucerquia}, {Ram{\'\i}rez},
  {Alvarado-Montes}, \& {Zuluaga}}]{Sucerguia2020}
{Sucerquia}, M., {Ram{\'\i}rez}, V., {Alvarado-Montes}, J.~A., \& {Zuluaga},
  J.~I. 2020, \mnras, 492, 3499, \dodoi{10.1093/mnras/stz3548}

\bibitem[{{Tayar} {et~al.}(2020){Tayar}, {Claytor}, {Huber}, \& {van
  Saders}}]{Tayar2020}
{Tayar}, J., {Claytor}, Z.~R., {Huber}, D., \& {van Saders}, J. 2020, arXiv
  e-prints, arXiv:2012.07957.
\newblock \doarXiv{2012.07957}

\bibitem[{{Teachey} \& {Kipping}(2018)}]{Teachey2018b}
{Teachey}, A., \& {Kipping}, D.~M. 2018, Science Advances, 4, eaav1784,
  \dodoi{10.1126/sciadv.aav1784}

\bibitem[{{Teske} {et~al.}(2019){Teske}, {Thorngren}, {Fortney}, {Hinkel}, \&
  {Brewer}}]{Teske2019}
{Teske}, J.~K., {Thorngren}, D., {Fortney}, J.~J., {Hinkel}, N., \& {Brewer},
  J.~M. 2019, \aj, 158, 239, \dodoi{10.3847/1538-3881/ab4f79}

\bibitem[{{Theano Development Team}(2016)}]{theano}
{Theano Development Team}. 2016, arXiv e-prints, abs/1605.02688.
\newblock \url{http://arxiv.org/abs/1605.02688}

\bibitem[{{Thompson} {et~al.}(2018){Thompson}, {Coughlin}, {Hoffman},
  {Mullally}, {Christiansen}, {Burke}, {Bryson}, {Batalha}, {Haas},
  {Catanzarite}, {Rowe}, {Barentsen}, {Caldwell}, {Clarke}, {Jenkins}, {Li},
  {Latham}, {Lissauer}, {Mathur}, {Morris}, {Seader}, {Smith}, {Klaus},
  {Twicken}, {Van Cleve}, {Wohler}, {Akeson}, {Ciardi}, {Cochran}, {Henze},
  {Howell}, {Huber}, {Pr{\v s}a}, {Ram{\'{\i}}rez}, {Morton}, {Barclay},
  {Campbell}, {Chaplin}, {Charbonneau}, {Christensen-Dalsgaard}, {Dotson},
  {Doyle}, {Dunham}, {Dupree}, {Ford}, {Geary}, {Girouard}, {Isaacson},
  {Kjeldsen}, {Quintana}, {Ragozzine}, {Shabram}, {Shporer}, {Silva Aguirre},
  {Steffen}, {Still}, {Tenenbaum}, {Welsh}, {Wolfgang}, {Zamudio}, {Koch}, \&
  {Borucki}}]{Thompson2018}
{Thompson}, S.~E., {Coughlin}, J.~L., {Hoffman}, K., {et~al.} 2018, \apjs, 235,
  38, \dodoi{10.3847/1538-4365/aab4f9}

\bibitem[{{Thorngren} \& {Fortney}(2019)}]{Thorngren2019a}
{Thorngren}, D., \& {Fortney}, J.~J. 2019, \apjl, 874, L31,
  \dodoi{10.3847/2041-8213/ab1137}

\bibitem[{{Thorngren} {et~al.}(2016){Thorngren}, {Fortney}, {Murray-Clay}, \&
  {Lopez}}]{Thorngren2016}
{Thorngren}, D.~P., {Fortney}, J.~J., {Murray-Clay}, R.~A., \& {Lopez}, E.~D.
  2016, \apj, 831, 64, \dodoi{10.3847/0004-637X/831/1/64}

\bibitem[{{Torres} {et~al.}(2005){Torres}, {Konacki}, {Sasselov}, \&
  {Jha}}]{Torres2005}
{Torres}, G., {Konacki}, M., {Sasselov}, D.~D., \& {Jha}, S. 2005, \apj, 619,
  558, \dodoi{10.1086/426496}

\bibitem[{{Trani} {et~al.}(2020){Trani}, {Hamers}, {Geller}, \&
  {Spera}}]{Trani2020}
{Trani}, A.~A., {Hamers}, A.~S., {Geller}, A., \& {Spera}, M. 2020, \mnras,
  499, 4195, \dodoi{10.1093/mnras/staa3098}

\bibitem[{{Villanueva} {et~al.}(2019){Villanueva}, {Dragomir}, \&
  {Gaudi}}]{Villanueva2019}
{Villanueva}, Steven, J., {Dragomir}, D., \& {Gaudi}, B.~S. 2019, \aj, 157, 84,
  \dodoi{10.3847/1538-3881/aaf85e}

\bibitem[{{Virtanen} {et~al.}(2020){Virtanen}, {Gommers}, {Oliphant},
  {Haberland}, {Reddy}, {Cournapeau}, {Burovski}, {Peterson}, {Weckesser},
  {Bright}, {van der Walt}, {Brett}, {Wilson}, {Millman}, {Mayorov}, {Nelson},
  {Jones}, {Kern}, {Larson}, {Carey}, {Polat}, {Feng}, {Moore}, {Vand erPlas},
  {Laxalde}, {Perktold}, {Cimrman}, {Henriksen}, {Quintero}, {Harris},
  {Archibald}, {Ribeiro}, {Pedregosa}, {van Mulbregt}, \& {SciPy 1. 0
  Contributors}}]{SciPy2020}
{Virtanen}, P., {Gommers}, R., {Oliphant}, T.~E., {et~al.} 2020, Nature
  Methods, 17, 261, \dodoi{10.1038/s41592-019-0686-2}

\bibitem[{{Visscher}(2012)}]{Visscher2012}
{Visscher}, C. 2012, \apj, 757, 5, \dodoi{10.1088/0004-637X/757/1/5}

\bibitem[{{Vogt} {et~al.}(1994){Vogt}, {Allen}, {Bigelow}, {Bresee}, {Brown},
  {Cantrall}, {Conrad}, {Couture}, {Delaney}, {Epps}, {Hilyard}, {Hilyard},
  {Horn}, {Jern}, {Kanto}, {Keane}, {Kibrick}, {Lewis}, {Osborne},
  {Pardeilhan}, {Pfister}, {Ricketts}, {Robinson}, {Stover}, {Tucker}, {Ward},
  \& {Wei}}]{Vogt1994}
{Vogt}, S.~S., {Allen}, S.~L., {Bigelow}, B.~C., {et~al.} 1994, in \procspie,
  Vol. 2198, Instrumentation in Astronomy VIII, ed. D.~L. {Crawford} \& E.~R.
  {Craine}, 362, \dodoi{10.1117/12.176725}

\bibitem[{{Wahl} {et~al.}(2017){Wahl}, {Hubbard}, {Militzer}, {Guillot},
  {Miguel}, {Movshovitz}, {Kaspi}, {Helled}, {Reese}, {Galanti}, {Levin},
  {Connerney}, \& {Bolton}}]{Wahl2017}
{Wahl}, S.~M., {Hubbard}, W.~B., {Militzer}, B., {et~al.} 2017, \grl, 44, 4649,
  \dodoi{10.1002/2017GL073160}

\bibitem[{{Wang} {et~al.}(2015{\natexlab{a}}){Wang}, {Fischer}, {Horch}, \&
  {Xie}}]{Wang2015a}
{Wang}, J., {Fischer}, D.~A., {Horch}, E.~P., \& {Xie}, J.-W.
  2015{\natexlab{a}}, \apj, 806, 248, \dodoi{10.1088/0004-637X/806/2/248}

\bibitem[{{Wang} {et~al.}(2015{\natexlab{b}}){Wang}, {Fischer}, {Barclay},
  {Picard}, {Ma}, {Bowler}, {Schmitt}, {Boyajian}, {Jek}, {LaCourse},
  {Baranec}, {Riddle}, {Law}, {Lintott}, {Schawinski}, {Simister},
  {Gr{\'e}goire}, {Babin}, {Poile}, {Jacobs}, {Jebson}, {Omohundro},
  {Schwengeler}, {Sejpka}, {Terentev}, {Gagliano}, {Paakkonen}, {Otnes Berge},
  {Winarski}, {Green}, {Schmitt}, {Kristiansen}, \& {Hoekstra}}]{Wang2015b}
{Wang}, J., {Fischer}, D.~A., {Barclay}, T., {et~al.} 2015{\natexlab{b}}, \apj,
  815, 127, \dodoi{10.1088/0004-637X/815/2/127}

\bibitem[{{Ward}(1997)}]{Ward1997}
{Ward}, W.~R. 1997, \icarus, 126, 261, \dodoi{10.1006/icar.1996.5647}

\bibitem[{{Winn}(2010)}]{Winn2010b}
{Winn}, J.~N. 2010, in Exoplanets, ed. S.~{Seager} (Tucson: Univ. of Arizona
  Press), 55--77

\bibitem[{{Winn} {et~al.}(2010){Winn}, {Fabrycky}, {Albrecht}, \&
  {Johnson}}]{Winn2010a}
{Winn}, J.~N., {Fabrycky}, D., {Albrecht}, S., \& {Johnson}, J.~A. 2010, \apjl,
  718, L145, \dodoi{10.1088/2041-8205/718/2/L145}

\bibitem[{{Winn} {et~al.}(2009){Winn}, {Howard}, {Johnson}, {Marcy}, {Gazak},
  {Starkey}, {Ford}, {Col{\'o}n}, {Reyes}, {Nortmann}, {Dreizler}, {Odewahn},
  {Welsh}, {Kadakia}, {Vanderbei}, {Adams}, {Lockhart}, {Crossfield},
  {Valenti}, {Dantowitz}, \& {Carter}}]{Winn2009}
{Winn}, J.~N., {Howard}, A.~W., {Johnson}, J.~A., {et~al.} 2009, \apj, 703,
  2091, \dodoi{10.1088/0004-637X/703/2/2091}

\bibitem[{{Wizinowich} {et~al.}(2000){Wizinowich}, {Acton}, {Shelton},
  {Stomski}, {Gathright}, {Ho}, {Lupton}, {Tsubota}, {Lai}, {Max}, {Brase},
  {An}, {Avicola}, {Olivier}, {Gavel}, {Macintosh}, {Ghez}, \&
  {Larkin}}]{Wizinowich2000}
{Wizinowich}, P., {Acton}, D.~S., {Shelton}, C., {et~al.} 2000, \pasp, 112,
  315, \dodoi{10.1086/316543}

\bibitem[{{Wong} {et~al.}(2004){Wong}, {Mahaffy}, {Atreya}, {Niemann}, \&
  {Owen}}]{Wong2004}
{Wong}, M.~H., {Mahaffy}, P.~R., {Atreya}, S.~K., {Niemann}, H.~B., \& {Owen},
  T.~C. 2004, \icarus, 171, 153, \dodoi{10.1016/j.icarus.2004.04.010}

\bibitem[{{Wright} {et~al.}(2004){Wright}, {Marcy}, {Butler}, \&
  {Vogt}}]{Wright2004}
{Wright}, J.~T., {Marcy}, G.~W., {Butler}, R.~P., \& {Vogt}, S.~S. 2004, \apjs,
  152, 261, \dodoi{10.1086/386283}

\bibitem[{{Wu} \& {Lithwick}(2011)}]{Wu2011}
{Wu}, Y., \& {Lithwick}, Y. 2011, \apj, 735, 109,
  \dodoi{10.1088/0004-637X/735/2/109}

\bibitem[{{Wu} \& {Murray}(2003)}]{Wu2003}
{Wu}, Y., \& {Murray}, N. 2003, \apj, 589, 605, \dodoi{10.1086/374598}

\bibitem[{{Yee} {et~al.}(2017){Yee}, {Petigura}, \& {von Braun}}]{Yee2017}
{Yee}, S.~W., {Petigura}, E.~A., \& {von Braun}, K. 2017, \apj, 836, 77,
  \dodoi{10.3847/1538-4357/836/1/77}

\end{thebibliography}
\end{document}